\documentclass[fleqn,usenatbib]{mnras}
   \usepackage[british]{babel}             % British English hyphenation
   \usepackage{newtxtext}                  % Good fonts
   \usepackage[slantedGreek]{newtxmath}    %   "    "   (slanted Greek)
   \let\la=\lesssim % for less than similar from newtxmath, not \la from mnras.cls
   \usepackage[T1]{fontenc}                % Good font encoding
   
\usepackage{graphics,graphicx,rotating}
\usepackage{natbib}
 \def\emerlin{$e$-MERLIN}
\usepackage{xspace}
\usepackage{amssymb}
\usepackage{amsmath}
\usepackage{epstopdf}
\usepackage{subfigure}
\usepackage{makecell} 
\usepackage{pdflscape}
\usepackage{geometry}
\usepackage{rotating}
\usepackage{xcolor}
\usepackage{xspace}
\usepackage{amssymb}
\usepackage{amsmath}
\usepackage{epstopdf}
\usepackage{subfigure}
\usepackage{xcolor}
\usepackage{hyperref}
\usepackage[percent]{overpic}
\usepackage{longtable,lscape}
\usepackage{xcolor}
\usepackage{pict2e}
\usepackage{float}
\let\la=\lesssim %
\let\ga=\gtrsim
 \def\emerlin{$e$-MERLIN}

\begin{document}
\title[Nuclear star clusters and AGN]{LeMMINGs. Multi-wavelength
  constraints on the co-existence of  nuclear star clusters and AGN  in nucleated galaxies  }  
  \author[Dullo et al.]{B.\ T.\  Dullo$,^{1,2}$\thanks{E-mail: bdullo@ucm.es}        
   J.\ H.\ Knapen$,^{3,4}$\
 R.\ D.\ Baldi$,^{5,6}$\
D.\ R.\  A.\ Williams$,^{7}$\
R.\ J.\  Beswick$,^{7}$\
I.\ M.\ McHardy$,^{6}$\
         \newauthor
D.\ A.\ Green$,^{8}$\
A.\ Gil de Paz$,^{2,9}$\
S.\ Aalto$,^{10}$
 A.\ Alberdi$,^{11}$\
M.\ K.\ Argo$,^{12}$\
J.\ S.\ Gallagher$,^{13}$\
 H.-R.\   Kl\"ockner$,^{14}$\
     \newauthor
  J.\ M.\ Marcaide$,^{15}$\
I.\ M.\ Mutie$,^{16,7}$\
    D.\ J.\ Saikia$,^{17}$\
    P.\ Saikia$,^{18}$\
I.\ R.\ Stevens$,^{19}$ \
            and
S.\ Torrej\'on$^{2}$\\
$^{1}$Department of Physical Sciences, Embry-Riddle Aeronautical University, Daytona Beach, FL 32114, USA\\
 $^{2}$Departamento de F\'isica de la Tierra y Astrof\'isica, Facultad de CC. Físicas, Universidad Complutense de Madrid, E-28040, Madrid, Spain\\
$^{3}$Instituto de Astrof\'isica de Canarias, V\'ia
  L\'actea S/N, E-38205, La Laguna, Tenerife, Spain\\
$^{4}$Departamento de Astrof\'isica, Universidad de
  La Laguna, E-38206, La Laguna, Tenerife, Spain\\
$^{5}$Istituto di Radioastronomia - INAF, Via P. Gobetti 101, I-40129 Bologna, Italy\\
$^{6}$School of Physics and Astronomy, University of Southampton, Southampton, SO17 1BJ, UK\\
$^{7}$Jodrell Bank Centre for Astrophysics, School of
  Physics and Astronomy, The University of Manchester, Alan Turing
  Building,\\ Oxford Road, Manchester, M13 9PL, UK\\
$^{8}$Astrophysics Group, Cavendish Laboratory, 19 J. J. Thomson Avenue, Cambridge CB3 0HE, UK\\
$^{9}$Instituto de Física de Partículas y del Cosmos IPARCOS, Facultad de CC. Físicas, Universidad Complutense de Madrid, E-28040, Madrid, Spain\\
 $^{10}$Department of Space, Earth and Environment, Chalmers University of Technology, 412 96 G\"oteborg,
Sweden\\
$^{11}$Instituto de Astrofísica de Andalucía, IAA-CSIC, Glorieta de la Astronomía s/n, E-18008 Granada, Spain\\
$^{12}$Jeremiah Horrocks Institute, University of Central Lancashire, Preston, Lancashire, PR1 2HE, UK\\
$^{13}$Department of Astronomy, University of Wisconsin-Madison, Madison, Wisconsin, USA\\
 $^{14}$Max-Planck-Institut für Radioastronomie, Auf dem Hügel 69, 53121 Bonn, Germany\\
 $^{15}$Real Academia de Ciencias, C/ Valverde 22, 28004, Madrid, Spain\\
$^{16}$Department of Astronomy and Space Science, Technical University of Kenya,  P.O. box 52428-00200, Nairobi, Kenya\\
$^{17}$Inter-University Centre for Astronomy and Astrophysics (IUCAA), P.O., Post Bag 4,  Ganeshkhind, Pune 411 007, India\\
$^{18}$Center for Astro, Particle and Planetary Physics, New York University Abu Dhabi, PO Box 129188, Abu Dhabi, UAE\\
 $^{19}$School of Physics and Astronomy, University of Birmingham, Birmingham B15 2TT, UK\\
  }

\maketitle
\label{firstpage}
\begin{abstract}
  The relation between nuclear star clusters (NSCs) and the growth of
  the central supermassive black holes (SMBHs), as well as their
  connection to the properties of the host galaxies, is crucial for
  understanding the evolution of galaxies.  Recent observations have
  revealed that about 10 per cent of nucleated galaxies host hybrid
  nuclei, consisting of both NSCs and accreting SMBHs that power
  active galactic nuclei (AGN).  Motivated by the potential of the
  recently published multi-wavelength data sets from LeMMINGs survey,
  here we present the most thorough investigation to date of the
  incidence of hybrid nuclei in a large sample of 100 nearby nucleated
  galaxies (10 E, 25 S0, 63 S, and 2 Irr), covering a wide range in
  stellar mass
  (\mbox{$M_{*,\rm gal} \sim 10^{8.7}-10^{12}~\rm M_{\sun}$}).  We
  identify the nuclei and derive their properties by performing
  detailed 1D and 2D multi-component decompositions of the optical and
  near-infrared {\it HST} stellar light distributions of the galaxies
  using S\'ersic and core-S\'ersic models. Our AGN diagnostics are
  based on homogeneously derived nuclear 1.5 GHz \emerlin\ radio, {\it
    Chandra} X-ray (0.3--10 keV) and optical emission-line data. We
  determine the nucleation fraction ($f_{\rm nuc} $) as the relative
  incidence of nuclei across the LeMMINGs {\it HST} sample and find
  \mbox{$f_{\rm nuc} =~ $100/149 (= 67 $\pm$ 7 per cent)}, confirming
  previous work, with a peak value of \mbox{$49/56~(= 88 \pm 13$ per
    cent}) at bulge masses
  \mbox{$M_{*,\rm bulge}\sim 10^{9.4}- 10^{10.8}~ \rm M_{\sun}$}.  We
  identify 30 nucleated LeMMINGs galaxies that are optically active,
  radio-detected and X-ray luminous ($L_\textnormal{X} > 10^{39}$ erg
  s$^{-1}$). This indicates that our nucleated sample has a lower
  limit $\sim$ 30 per cent occupancy of hybrid nuclei, which is a
  function of $M_{*,\rm bulge}$ and $M_{*,\rm gal}$. We find that
  hybrid nuclei have a number density of $(1.5 \pm 0.4)\times 10^{-5}$
  Mpc$^{-3}$, are more common at
  \mbox{$M_{*,\rm gal}\sim 10^{10.6}- 10^{11.8}~\rm M_{\sun}$} and
  occur, at least, three times more frequently than previously
  reported.
\end{abstract}


\section{Introduction}
Central massive objects appear to be ubiquitous at the centres of
galaxies, and may be a supermassive black hole (SMBH, with mass
\mbox{$M_{\rm BH} \sim 10^{5}-10^{10}~\rm M_{\sun}$}), a dense stellar
nucleus or a combination of both. All nearby galaxies with stellar
masses ($M_{*,\rm gal}$) above \mbox{$10^{10}~\rm M_{\sun}$} are
thought to host a SMBH at their centre
\citep{1998AJ....115.2285M,1998Natur.395A..14R,2005SSRv..116..523F,2013ARA&A..51..511K,2016ASSL..418..263G}.
Conversely, at the lower stellar masses
(\mbox{$M_{*,\rm gal} \sim 10^{8}-10^{10}~\rm M_{\sun}$}) evidence for
dynamically identified SMBHs is scarce and nuclear star clusters
(NSCs) are routinely observed. NSCs are compact, typically having
half-light radii of a few parsecs and stellar masses in the range
\mbox{$M_{*,\rm NSC} \sim 10^{5} - 10^{8}~\rm M_{\sun}$}
\citep[e.g.][]{2004AJ....127..105B,2006ApJS..165...57C}.  {\it Hubble
  Space Telescope (HST)} observations have revealed that as much as
$\sim$80 per cent of galaxies of low and intermediate stellar masses
host a dense stellar nucleus at their centres
\citep{1996AJ....111.1566P,2002AJ....123.1389B,2003AJ....125.2936G,
  2004AJ....127..105B,2004AJ....128.1124S,2006ApJS..165...57C,2007ApJ...665.1084B,2014MNRAS.441.3570G}. 
  We use the term `nucleated' when referring to galaxies that possess
  nuclei, which are bright and compact optical sources at or near the
  galaxies' photocenters
  \citep{2006ApJS..165...57C,2012ApJ...755..163D,2016MNRAS.462.3800D}. The fraction of
nucleated galaxies has been reported to increase 
systematically with galaxy stellar mass, reaching a peak at
\mbox{$M_{*,\rm gal} \sim 10^{9.5}~\rm M_{\sun}$} before decreasing
at higher galaxy masses (e.g.\
\citealt{2014MNRAS.445.2385D,2019ApJ...878...18S,2020A&ARv..28....4N,2021MNRAS.507.3246H,2021MNRAS.508..986Z,2023A&A...671L...7R}).
 
SMBH and NSC masses have been shown independently to scale with
several host galaxy properties, including luminosity, stellar mass and
central velocity dispersion (e.g.\
\citealt{1997AJ....114.2366C,1998AJ....116...68C,2006ApJ...644L..21F,
  2020A&ARv..28....4N}, references therein). The coexistence of NSCs
and SMBHs powering the AGN may therefore suggest that their formation
is coupled and they grow concurrently regulated by the same physical
process \citep{2008ApJ...678..116S,2015ApJ...812...72A}. Theoretical
models predict NSCs promote the formation and growth of
intermediate-mass BHs (IMBHs) and SMBHs
\citep{2014ApJ...785...71G,2015ApJ...812...72A,2017MNRAS.467.4180S,
  2020MNRAS.498.5652K,2021MNRAS.502.2682A,2022ApJ...929...84B} and
facilitate accretion on to the central SMBH by funnelling gas toward
the innermost regions \citep{2015ApJ...803...81N}.

An outstanding question is how commonly accreting SMBHs powering AGN
and dense stellar nuclei coexist at the centre of galaxies, and
whether this occurrence is connected to the mass of the host galaxy. It has been
 suggested \citep[e.g.][]{2006ApJ...644L..21F,2006ApJ...644L..17W}, in general,
massive galaxies only contain non-stellar nuclei (SMBHs) whereas low
mass galaxies contain compact stellar nuclei. In fact only $\sim20$
per cent of core-S\'ersic galaxies, which are massive
(\mbox{$M_{*,\rm gal} \ga10^{11} ~\rm M_{\sun}$}) with depleted stellar
cores, host stellar nuclei
\citep{2006ApJS..165...57C,2012ApJ...755..163D,2013ApJ...768...36D,2014MNRAS.444.2700D,
2012ApJS..203....5T,2014MNRAS.445.2385D,2017ApJ...849...55S,
2017MNRAS.471.2321D,2019ApJ...886...80D}. \citet{2010ApJ...714L.313B},  \citet{2015ApJ...812...72A} and \citet{2019MNRAS.486.5008A}
hypothesised that the low incidence of NSCs in core-S\'ersic galaxies
is due to their tidal destruction post dynamical heating by coalescing
binary SMBHs during gas-poor galaxy mergers.

However, an increasing number of galaxies, including the Milky Way,
are found to host nuclei consisting of both a stellar nucleus and an
AGN or a quiescent SMBH (e.g.\
\citealt{2008ApJ...678..116S,2008AJ....135..747G,2009MNRAS.397.2148G,
  2012MNRAS.424.2130L,2010ApJ...714...25G,2012AdAst2012E..15N,2013ARA&A..51..511K,2014MNRAS.441.3570G,2015ApJ...812...72A,2017ApJ...841...51F,2018ApJ...858..118N}). The
Milky Way itself contains a SMBH of mass
\mbox{$\sim4\times10^{6} \rm ~ M_{\sun}$}
\citep{1998ApJ...509..678G,2009ApJ...692.1075G,2022ApJ...930L..12E}
and a NSC with a mass \mbox{$\sim (2.1-4.2)\times10^{7} \rm~M_{\sun}$}
\citep{2014A&A...566A..47S,2014A&A...570A...2F, 2020A&ARv..28....4N}.
While the current technological limitations hinder dynamical BH mass
measurements at lower masses
\citep{2004ApJ...610..722G,2011Natur.470...66R}, the mass range for
hosting hybrid (NSC+AGN) nuclei also include low-mass galaxies
\citep[e.g.][]{2018ApJ...858..118N,2023MNRAS.520.5964Y}.  AGN have
been increasingly discovered in these systems, primarily with X-ray
observations, albeit at a low AGN fraction rate of $\sim 1-5$ per cent
\citep[e.g.][]{2010ApJ...714...25G,2015MNRAS.454.3722S,2018MNRAS.478.2576M,2018MNRAS.476..979P,2020MNRAS.492.2268B,2023A&A...675A.105D,2023MNRAS.522.3412D}. A
well-known low-mass
(\mbox{$M_{*,\rm gal} \sim 3.4 \times 10^{7}~\rm M_{\sun}$};
\citealt{2009MNRAS.397.2148G}) galaxy which hosts a hybrid nucleus is
NGC~4395
\citep{1989ApJ...342L..11F,2003ApJ...588L..13F,2015ApJ...809..101D,2023ApJ...950...81N}. It
is a bulgeless SA(s)m with a type 1 Seyfert nucleus.  Other low-mass,
hybrid nuclei host candidates are Pox 52, a dwarf elliptical galaxy
\citep[e.g.][]{2004ApJ...607...90B,2008ApJ...686..892T}, and the dwarf
disc galaxy RGG 118 \citep{2017ApJ...850..196B}. 

Both NSCs and AGN
emit radiation across the entire electromagnetic spectrum from radio
to X-ray.  Radio emission in NSCs is predominately driven by thermal
stellar processes \citep[e.g.][]{2020ApJ...896...84S}, whereas  active
SMBHs are associated with non-thermal processes driven by disk/corona
winds or \mbox{ (sub-)relativistic} jets \citep{2019NatAs...3..387P}. Distinguishing  between a spatially
resolved, `pure' NSC, a `pure' AGN (unresolved), and any combination
of the two in optical images  is challenging.  Nevertheless, they can all be included in a sample of `nucleated'
galaxies constructed from {\it HST} imaging.  To differentiate them and determine  the distinct origin of their nuclear emission, 
multi-band diagnostics are necessary.

Previous studies have reported that the fraction of hybrid nuclei in
(nucleated) galaxies is $\sim$10 per cent
\citep[e.g.][]{2008ApJ...678..116S,2010ApJ...714...25G,2017ApJ...841...51F}. The
useful work by \citet{2008ApJ...678..116S} investigated the presence
of AGN in a sample of 176 previously reported galaxies with NSCs using
radio, X-ray and optical spectroscopic observations. Of their 75
galaxies with available optical spectral data, seven (10 per cent) host an
optical AGN, whereas an additional  11 (15 per cent) exhibit composite, i.e.\ AGN-(star
formation)-like emission.  To detect radio emission, they used the Very
Large Array (VLA) Faint Images of the Radio Sky at Twenty cm (FIRST;
\citealt{1995ApJ...450..559B}) at 1.4 GHz, with a sensitivity limit of 1.0 mJy
and resolution of 5 arcsec. While 13 galaxies had radio detections
within 30 arcsec of the NSCs, none of them were found to host
AGN. Using heterogeneous X-ray data from {\it Chandra}, {\it ROSAT}
and XMM-Newton X-ray data, they identified 22
X-ray sources associated with their NSCs, and concluded 4/22 sources
were likely AGN.  \citet{2010ApJ...714...25G} examined the AGN
activity in 100 early-type Virgo galaxies \citep{2006ApJS..165...57C}
using {\it Chandra} observations, finding a hybrid nucleus fraction of
0.3--7 per cent for \mbox{$M_{*,\rm gal} < 10^{11}~\rm
M_{\sun}$}. \citet{2015ApJ...812...72A} used semi-analytical models and
found that the fraction of galaxies with a hybrid nucleus increases
from $\sim$5 per cent at \mbox{$M_{*,\rm gal} \sim 10^{9}~ \rm M_{\sun}$} to 30 per cent
at \mbox{$M_{*,\rm gal} \sim 10^{12}~ \rm M_{\sun}$}. Recently,
\citet{2017ApJ...841...51F} used {\it Chandra} X-ray observations and
found that $\sim$11.2 per cent of their sample of 98 galaxies with NSCs
harbour hybrid nuclei.
 
The Legacy $e$-MERLIN Multi-band Imaging of Nearby Galaxies Survey
(LeMMINGs;
\citealt{2014evn..confE..10B,2018MNRAS.476.3478B,2021MNRAS.508.2019B,2021MNRAS.500.4749B})
aims to investigate the underlying physical mechanisms that drive
nuclear emission in galaxies by combining high resolution observations
from radio ($e$-MERLIN), through optical ({\it Hubble Space Telescope,
  HST}) to X-ray ({\it Chandra}).  In this work, we use the results
from our 1D and 2D multi-component {\it HST} imaging analysis to
accurately identify the nuclei in LeMMINGs galaxies and derive
their properties (\citealt{2023A&A...675A.105D}). For a robust
characterisation of the coexistence of NSCs and AGN, our AGN
diagnostics rely on homogeneously obtained 1.5 GHz \emerlin\ radio
observations of nuclei with a sub-mJy sensitivity and resolution of
$\sim$0.15 arcsec, as well as nuclear {\it Chandra} X-ray and optical
emission-line data from  LeMMINGs
\citep{2018MNRAS.476.3478B,2021MNRAS.500.4749B,2022MNRAS.510.4909W}.
The sample covers a wide range in stellar mass, morphology and nuclear
activity, which are crucial to establish the scaling relations between
the mass and luminosity of the nuclei and their radio and X-ray
luminosities.

This study constitutes the most comprehensive multi-wavelength
investigation to date of the connection between NSCs and AGN in nearby
galaxies. The paper is organised as follows.  Section~\ref{Sec2}
provides a description of the radio and optical emission-line data and
discusses the 1D and 2D multi-component decompositions with {\it HST}
imaging that were used to characterise the nuclei.  In
Section~\ref{Sec3}, we present the relation between nucleation
fraction and several properties, including the luminosity, stellar
mass of the host galaxy and its bulge, and Hubble type.  Also
discussed in this section is the co-existence of NSCs and AGN in
LeMMINGs galaxies. Section~\ref{Sec4} presents scaling relations
between the mass/luminosity of nuclei and their radio and X-ray
luminosities and discusses the implications.  Finally, we summarise in
Section~\ref{Sec5}.  There are four appendices at the end of this
paper (Appendices \ref{App01}--\ref{AppA2}).  Our 2D decompositions
and comparison with previous fits in the literature are given in
Appendices \ref{App01} and \ref{App02}, respectively.  The global and
central properties of the sample galaxies are given in
Appendix~\ref{AppA1}, while Appendix~\ref{AppA2} provides a comparison
between our censored ({\sc asurv}) and uncensored ({\sc bces}
bisector) regression fits for the full sample of nuclei.

Throughout this paper, we use $H_{0}$
= 70 km s$^{-1}$ Mpc$^{-1}$, $\Omega_{m}$ = 0.3 and $\Omega_{\Lambda}$
= 0.7 \citep[e.g.][]{2019ApJ...882...34F}, an average of the Planck 2018
Cosmology $H_{0}$ = 67.4 $\pm$ 0.5 km s$^{-1}$ Mpc$^{-1}$
\citep{2020A&A...641A...6P} and the LMC $H_{0}$ = 74.22 $\pm$ 1.82 km
s$^{-1}$ Mpc$^{-1}$\citep{2019ApJ...876...85R}. All magnitudes are in
the Vega system, unless specified otherwise.

\begin{table} 
\centering

\setlength{\tabcolsep}{0.13029631025459268281in}
\begin {minipage}{85mm}
\caption{Multi-wavelength data.}
\label{Tab1}
\begin{tabular}{@{}llcc@{}}
\hline
Data&$N$\\
& \multicolumn{2}{c}{\centering Parent Sample/This Work}\\
 (1)&(2)&\\          
\hline  
  \\[-7.38pt]     
 1.5 GHz \emerlin\ radio data$^{\rm [1r]}$   &280/100   \\
Optical spectral classification$^{\rm [1r]}$ &280/100 \\  
{\it
  Chandra} X-ray (0.3 -- 10 keV) data$^{\rm [2r]}$&213/84\\
{\it HST} 1D   data$^{\rm [3r]}$ &173/173 \\   
{\it HST} 2D  nucleated galaxy data$^{\rm [3r]}$ &42/42 \\  
{\it HST} 2D nucleated galaxy data  (this work) &58/58 \\  
\hline      
\end{tabular}     
Note: (1): multi-wavelength data used in this work.  (2): number of
galaxies ($N$) in the parent sample and in the subsample used in this
work. The sample in this study consists of 100  nucleated
galaxies. Of this, 42 had 2D decompositions of their {\it HST} images
in \citet{2023A&A...675A.105D}, while for the remaining 58, we
performed 2D decompositions of the {\it HST} images here.  References.
1r =
\citet{2018MNRAS.476.3478B,2021MNRAS.508.2019B,2021MNRAS.500.4749B};
2r = \citet{2022MNRAS.510.4909W}; 3r = \citet{2023A&A...675A.105D}.
 \end{minipage}
\end{table}

\begin{table} 
\begin{center}
\setlength{\tabcolsep}{0.04251in}
\begin {minipage}{85mm}
\caption{Optical and radio properties. }
\label{Tab2}
\begin{tabular}{@{}llcccc@{}}
\hline
Galaxies&Number (percentage)&Radio-detected fraction\\
&(full {\it HST} sample/nucleated)&(full {\it HST} sample/nucleated)\\
 (1)&(2)&(3)&\\          
\hline       
E    &23 (13.2\%)/10 (10\%)&11/23 (47.8\%)/3/10 (30.0\%) \\
S0 & 42 (24.3\%)/25 (25\%)&  25/42 (59.5\%)/15/25 (60.0\%)\\                     
S  &102 (59.0\%)/63 (63\%)&  47/102 (46.0\%)/32/63 (50.8\%)  \\
\vspace{0.2cm}
Irr  &6 (3.5\%)/2 (2\%) &2/6 (33.3\%)/0/2 (0.0\%)   \\
Seyfert & 10 (5.8\%)/9 (9\%)& 8/10 (80.0\%)/7/9 (77.8\%)    \\
ALG &  23 (13.3\%)/9 (9\%)   &5/23 (21.7\%)/1/9 (11.1\%)    \\
LINER &71 (41\%)/48 (48\%)   &49/71 (69.0\%)/30/48 (62.5\%)\\
\vspace{0.2cm}
{\sc h ii}  &69 (39.9\%)/34 (34\%) &21/69 (30.4\%)/12/34 (35.3\%)\\
Total&173 (100\%)/100 (100\%) &83/173 (48.0\%)/50/100 (50.0\%)\\ 
\hline
\end{tabular}     
Note: The sample galaxies are first separated based on the galaxy
morphological and optical spectral classes (1--2) and then further
divided based on their radio detection (3),   see Section~\ref{Sec2.1}. The term `full {\it HST}
sample' refers to the sample of 173 LeMMINGs galaxies with {\it HST}
data \citep{2023A&A...675A.105D}, see Table~\ref{Tab1}.
 \end{minipage}
\end{center}
\end{table}

\section{Data}\label{Sec2}

In what follows, we will describe the optical, radio and X-ray data
which are used in the paper to study nucleation in  LeMMINGs
galaxies. Except for 58 per cent of the 2D decompositions of
the {\it HST} images discussed here, all other data used in this work are published
elsewhere
\citep{2018MNRAS.476.3478B,2021MNRAS.500.4749B,2021MNRAS.508.2019B,2022MNRAS.510.4909W,2023A&A...675A.105D}.

\subsection{Radio, X-ray data and optical spectral classification}\label{Sec2.1}

The LeMMINGs
\citep{2014evn..confE..10B,2018MNRAS.476.3478B,2021MNRAS.500.4749B,2021MNRAS.508.2019B,2023arXiv230308647W}
is a survey of 280 nearby galaxies, above declination
$\delta > +20^{\circ}$, see Tables~\ref{Tab1} and \ref{Tab2}. The
sample is a subset of the magnitude-limited ($B_{T} \le 12.5$ mag and
declinations $\delta > 0^{\circ}$) Palomar spectroscopic sample of 486
bright, nearby galaxies
\citep{1995ApJS...98..477H,1997ApJS..112..315H}, which in turn were
drawn from the Revised Shapley-Ames Catalog of Bright Galaxies
\citep{1981rsac.book.....S} and the Second Reference Catalogue of
Bright Galaxies \citep{1976srcb.book.....D}. The LeMMINGs declination
cutoff is implemented to ensure optimal radio visibility coverage for
the $e$-MERLIN array.  The primary goal of the survey is to provide
the deepest high-resolution radio continuum study of the local
Universe at a sub-mJy sensitivity of $\sim$ 0.08~mJy beam$^{-1}$ and
an angular resolution of $\sim$ 0.15 arcsec. The 1.5 GHz radio
continuum of the 280 galaxies were observed with \mbox{$e$-MERLIN} for
a total of 810 h \citep{2018MNRAS.476.3478B,2021MNRAS.500.4749B}.  The
radio detection and radio core luminosities for our sample are
tabulated in Table~\ref{TabA1}.

\citet{2022MNRAS.510.4909W} analysed {\it Chandra} X-ray observations
of the nuclei of a sample of 213 LeMMINGs galaxies. Of the 100
nucleated LeMMINGs galaxies targeted in this study, 84 have {\it
  Chandra} X-ray data. With an angular resolution of 0.5 arcsec and a
flux limit of 1.65 $\times$ 10$^{-14}$ erg s$^{-1}$ cm$^{-2}$ at
0.3--10 keV, the authors  detected X-ray emission in
150/213 galaxies, coincident within 2 arcsec of the optical nucleus. 

We use the optical spectral classes from
\citet{2018MNRAS.476.3478B,2021MNRAS.500.4749B}, who presented updated
spectral classifications using emission-line ratios taken mainly from
\citet[][see also
\citealt{1985ApJS...57..503F,1995ApJS...98..477H,1997ApJS..112..315H,1997ApJ...487..568H,1997ApJS..112..391H}]{1997ApJS..112..315H}.
\citet{2018MNRAS.476.3478B,2021MNRAS.500.4749B} also used new
emission-line ratio data from recent observations to refine the
spectral classification for some sample galaxies.
They applied the emission
line diagnostic diagrams by \citet{2006MNRAS.372..961K} and
\citet{2010A&A...509A...6B}. \citet{2018MNRAS.476.3478B,2021MNRAS.500.4749B} 
categorised the galaxies with nuclear emission lines as Seyfert, LINER
and \mbox{H\,{\sc ii}} galaxies, whereas LeMMINGs galaxies that
lack nuclear emission lines were referred to us `absorption line
galaxies (ALGs)'.  

 In this paper, the galaxies referred to as `optically active'  (i.e.\ AGN) are
  LINERs and Seyferts, while those referred to as `radio AGN' are, following \citet{2018MNRAS.476.3478B},   jetted objects with radio morphologies B (`one-sided
  jet'), C (`triple') and D (`doubled-lobed') as well as  radio detected LINERs and Seyferts lacking detected jets 
with $e$-MERLIN at 1.5 GHz.

\begin{figure*}
\begin{tabular}{@{}cccc@{}}
\hspace*{-.4825cm}
  \includegraphics[trim={-1.068099902769955cm 0.095cm 0cm .00965420820383603cm},clip,angle=0,scale=0.4061201]{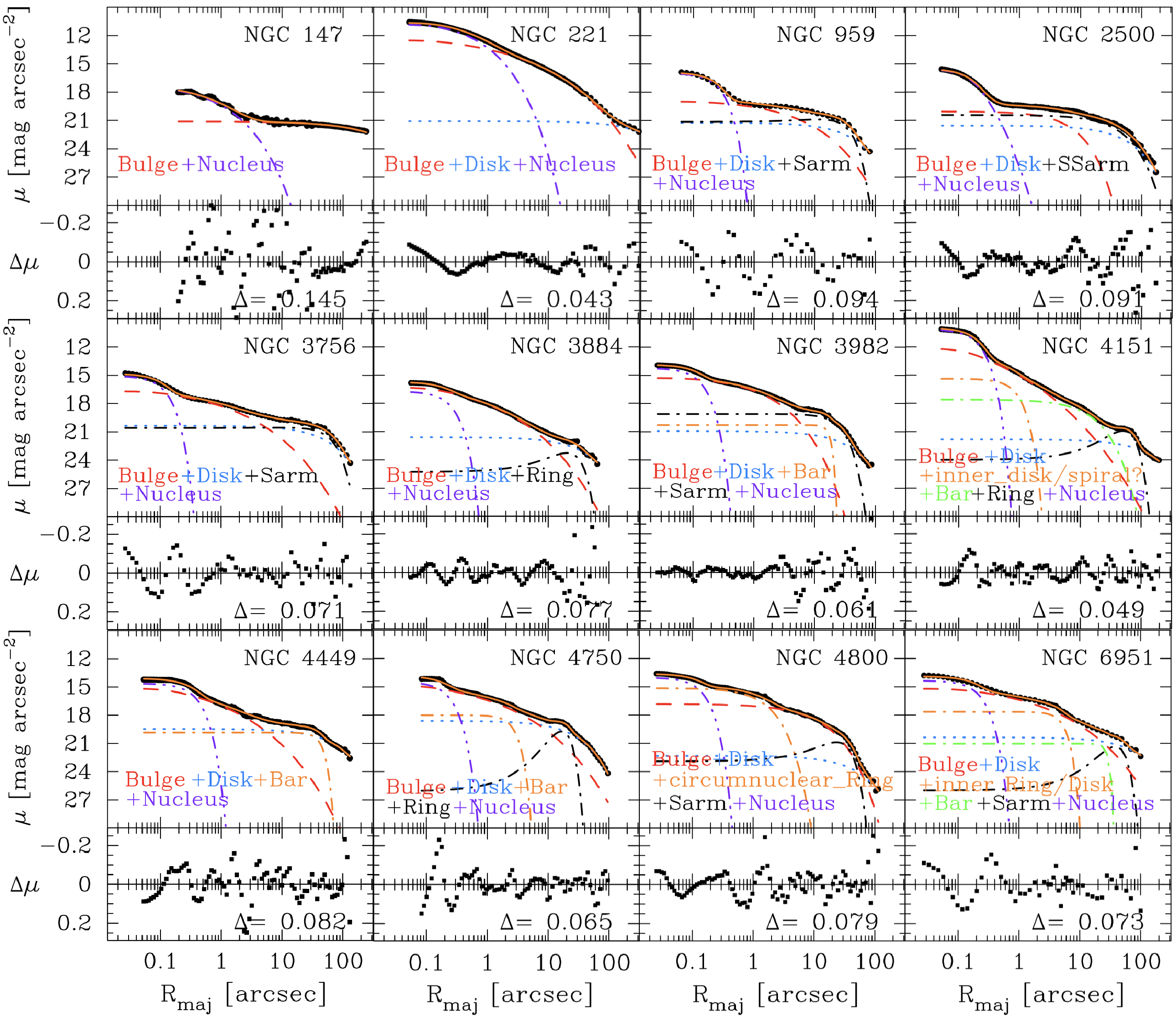}
 \vspace*{-0.3431509068830cm}
  \end{tabular}
  \caption{1D multi-component decompositions of the major-axis surface
    brightness profiles for a dozen nucleated LeMMINGs galaxies,
    selected as representative examples from the 100 nucleated
    galaxies in our sample (Table~\ref{TabA1},
    \citealt{2023A&A...675A.105D}).  The residual profiles along
    with the rms residual ($ \upDelta$) for each fit are shown. While the images for NGC~147,
    NGC~221, NGC~959 and NGC~2500 were observed  with  {\it HST} in the WFPC2 F814W
    filter,  those for NGC~3756, NGC~3982, NGC~4750, NGC~4800 and NGC~6951 were obtained
    with the {\it HST} ACS F814W filter. For the remaining three the
    {\it HST} images are from the WFPC2 and obtained in the following filters: NGC~3884 (F547M), NGC~4151 (WF791W) and
    NGC~4449 (F606W). The magnitudes are given in the Vega
    magnitude system.  We fitted the nuclei typically with a
    two-parameter Gaussian model (dash-dot-dot-dot purple curve),
    while for some galaxies, the nuclei were  described using a S\'ersic
    model (dash-dotted purple curve). The dashed red (S\'ersic) curves represent
    the bulges, while the dotted blue curve shows the large-scale
    discs which we modelled with an exponential function.  Galaxy
    components such as bars and small-scale discs, rings, spirals and
    lenses are described by S\'ersic models (i.e.\ dash-dotted  orange
    and green curves).  Large-scale ring and spiral arm (`Sarm')
    components were fitted with three-parameter Gaussian ring models
    (dash-dotted black curve). For NGC~2500, the galaxy's spiral-arm
was described using a S\'ersic model (`SSarm'). The complete fits
    (solid orange curves) match the observed galaxy profiles with a
    median rms residual $ \upDelta \sim$ 0.065 mag arcsec$^{-2}$.  We
    fit up to six model components which are summed up to a full model
    with up to 16 free parameters. } 
    \label{Fig1}
\end{figure*}

\subsection{Sample selection and identification of
  nuclei with \it HST  }\label{Sec2.2}

One of the aims of this paper is a detailed investigation  of the
structural properties of photometrically distinct nuclei observed in
the broadband {\it HST} images of nearby galaxies, paying particular
attention to NSCs and AGN (see Table~\ref{Tab1}).  We define nuclei as
distinct central light excesses with respect to the inward
extrapolation of outer S\'ersic or core-S\'ersic models which are
fitted to the underlying bulge\footnote{The term `bulge' is
  traditionally associated with the spheroidal component of disc
  galaxies but it is used here to refer to the underlying host
  spheroid in case of elliptical galaxies and the bulge for S0, spiral
  and irregular galaxies.}  profiles or to the outer disc profiles for
bulgeless galaxies
\citep[e.g.][]{2006ApJS..165...57C,2012ApJS..203....5T,2012ApJ...755..163D,2016MNRAS.462.3800D,2019ApJ...871....9D}.  NSCs are
compact with half-light radii ($R_{\rm e,nuc}$) as small as a few
parsecs \citep[e.g.][]{2004AJ....127..105B,2006ApJS..165...57C} and
for the most extended ones $R_{\rm e,nuc} \sim$ 1 arcsec
\citep[e.g.][]{2012ApJS..203....5T}. The identification of nuclei
depends strongly on the resolution of the imaging used,
high-resolution {\it HST} data being most suitable. At the mean
distance of our nearby sample galaxies ($D \sim22$ Mpc), the {\it HST}
(ACS, WFPC2, WFC3 and NICMOS) angular resolution of \mbox{0.05--0.1
  arcsec} corresponds to $\sim$ 5$-$10 pc. 
  Identification of a nucleus
and robust measurements of its luminosity/stellar mass and effective
radius also requires detailed photometric decomposition of the host
galaxy's stellar light distribution.  It is
essential to account for the effects of the PSF to reliably separate
the central light excess from the rest of the galaxy.  
In
\citet{2023A&A...675A.105D}, we revealed that the bulge mass can be
significantly overestimated when galaxy components such as bars, rings
and spirals are not included in the fits. This implies that
restricting fits to bulge-disc profiles can yield inaccurate
structural parameters for the nuclei. 

 \citet{2023A&A...675A.105D} used {\it HST} (ACS, WFPC2, WFC3 and
NICMOS) images and extracted surface brightness profiles for 173
LeMMINGs galaxies (23 E, 42 S0, 102 S and 6 Irr), see
Table~\ref{Tab1}.  In that study, we performed accurate, 1D multi-component
decompositions of the surface brightness profiles covering a large
radial extent of $R \ga 80-100$ arcsec ($ \ga 2R_{\rm e,bulge}$),
fitting up to six galaxy components (i.e.\ bulge, disc, partially
depleted core, nuclei, bar, spiral arm, and stellar halo  and ring), 
simultaneously, using S\'ersic model
\citep{1963BAAA....6...41S,1968adga.book.....S} $R^{1/n}$ and
core-S\'ersic model
\citep{2003AJ....125.2951G,2004AJ....127.1917T,2006ApJS..164..334F,2010ApJ...725.2426B,2012ApJ...755..163D,2019ApJ...886...80D}. 
To decompose a galaxy light profile in 1D, we convolved   the individual fitted components  with a Gaussian point-spread
 function (PSF) in 2D.

We note that the parent LeMMINGs sample of 280 galaxies constitute all
the Palomar galaxies \citep{1995ApJS...98..477H,1997ApJS..112..315H}
with $\delta > 20^{\circ}$.  As the Palomar sample is statistically
complete, it implies that the parent LeMMINGs sample is also
statistically complete.  As shown by \citet[][see their Figs.\ 1 and 3
and the discussion in Section~2]{2023A&A...675A.105D}, the {\it HST}
sample of 173 LeMMINGs galaxies is representative of the parent
LeMMINGs sample of 280 galaxies, and therefore is not expected to be
biased in terms of nucleation and nuclear activity (see
Tables~\ref{Tab1} and \ref{Tab2}).

The 1D decompositions presented in \citet{2023A&A...675A.105D}
  identified nucleation in 124/173 galaxies. In that analysis,  we
  revealed that a two-parameter Gaussian function (i.e.\ a special
case of the S\'ersic model when $n=0.5$) describes the light profile
of 94/124 nuclei, while we fit a three-parameter S\'ersic model with
$0.4 \la n \la 2.5$ and a median $n \sim 0.7 \pm 0.6$ to describe the
light profiles for the remaining 30 nuclei.  From this initial sample
of 124 nuclei, we have excluded 24 that do not allow reliable
determination of the NSC structural parameters because their central
source is either too large or too small, see Section~\ref{Sec2.3} for
further discussion. In what follows, we focus primarily on the
remaining sample of 100 (=124$-$24) nucleated LeMMINGs galaxies (10 E,
25 S0, 62 S and 3 Irr).

As mentioned above, 
 the identification of the 100  nuclei is based on detailed  1D
 and 2D decompositions of the host galaxies' stellar light
 distributions from the broadband {\it HST} data (Table~\ref{Tab1}).   To classify the nuclei as either  
dense star clusters, `pure' AGN or a combination of both,  we use  multi-wavelength AGN
  diagnostics  that rely on homogeneously derived, nuclear 1.5 GHz
  \emerlin\ radio, {\it Chandra} X-ray (0.3--10 keV) and optical
  emission-line data (Section~\ref{Sec3}).

\subsection{1D and 2D multi-component   decomposition}\label{Sec2.03}

Fig.~\ref{Fig1} shows the 1D multi-component decompositions of the
major-axis surface brightness profiles for a dozen nucleated LeMMINGs
galaxies.  The galaxies were selected to be representative examples
for the 100 nucleated galaxies in terms of morphology and number of
fitted galaxy structural components. Of these 12 galaxies, images of
NGC~147, NGC~221, NGC~959 and NGC~2500 were obtained with {\it HST} in
the WFPC2 F814W filter, whereas those for NGC~3756, NGC~3982,
NGC~4750, NGC~4800 and NGC~6951 were obtained with the {\it HST} ACS
F814W filter. For the remaining three objects, we used {\it HST} data
from the WFPC2 in the following filters: F547M (NGC~3884), F791W
(NGC~4151) and F606W (NGC~4449).

\citet{2023A&A...675A.105D} performed 2D decompositions of the {\it
  HST} images for 65/173 galaxies  (Table~\ref{Tab1}).  These 2D fits had the same type and
number of galaxy structural components as the corresponding 1D
fits. The 2D model images were convolved with a Moffat PSF generated
using the {\sc imfit} task {\sc makeimage}. Our findings suggested
that detailed 1D and 2D decompositions yield strong agreements,
regardless of the galaxy morphology under consideration.  Of the 100
sample nucleated galaxies (Tables~\ref{Tab1} and  \ref{TabA1} and Section~\ref{Sec2.3}),
42 had 2D decompositions in \citet{2023A&A...675A.105D}.  Here, we
  follow their fitting methodology and perform 2D decompositions of
  the {\it HST} images for the remaining 58 nucleated galaxies using
  {\sc imfit v.1.8} \citep{2015ApJ...799..226E}, see
  Table~\ref{TabA1}. We note that of  the 12 representative galaxies shown
  in Fig.~\ref{Fig1}, eight are among  these 58 galaxies newly fitted
  in 2D, while the remaining four (NGC~959, NGC~3756, NGC~3884 and
  NGC~4449) had their 2D fits published in
  \citet{2023A&A...675A.105D}.

\begin{figure*}
\begin{tabular}{@{}cccc@{}}
 \hspace*{-1.74200000699886559584068830cm}
\includegraphics[trim={-2.8047695031769955cm -5.5395cm -4.0cm .247291805cm},clip,angle=0,scale=0.3908]{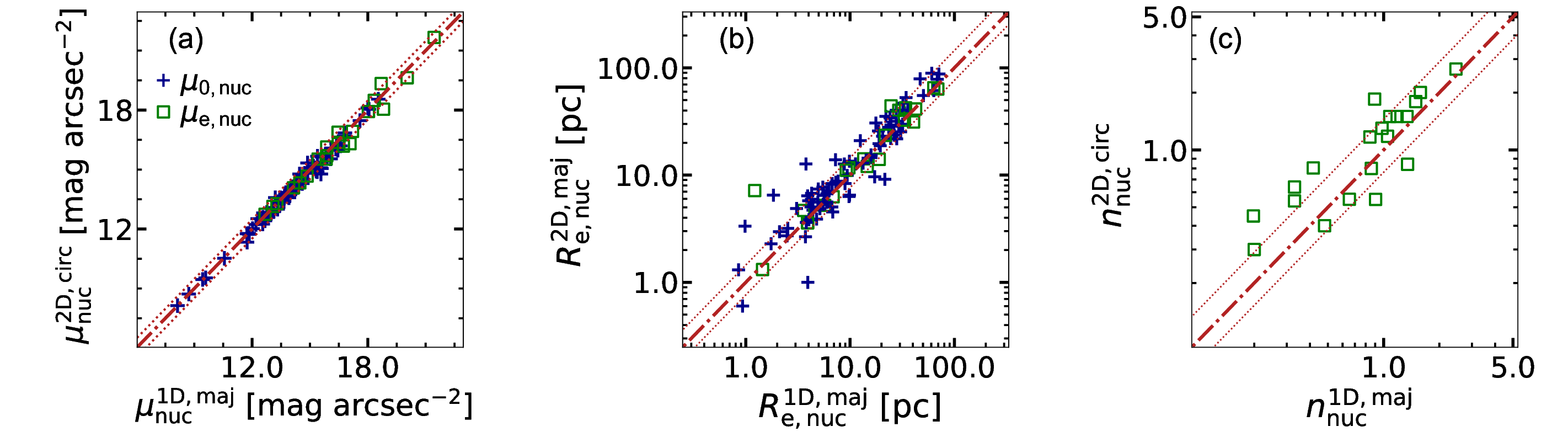}\\
 \vspace*{-2.688382084068830cm}
  \end{tabular}
  \caption{2D versus 1D structural properties of nuclei from our
    multi-component decompositions of the {\it HST} images for sample
    galaxies (Section~\ref{Sec2.2}, Table~\ref{TabA1}).  Of the 100
    nuclei in the sample, 79 were described using a Gaussian model
    (blue crosses), whereas the remaining 21 were fitted with a
    S\'ersic model (green boxes).  Comparison of 1D and 2D (a) central
    surface brightnesses of the nuclei ($\mu_{0,\rm nuc}$) or surface
    brightnesses of the nuclei at the effective radii $R_{\rm e}$
    ($\mu_{\rm e,nuc}$), (b) major-axis effective radii of the nuclei
    ($R_{\rm e,nuc}$) and (c) S\'ersic indices of the nuclei
    ($n_{\rm nuc}$).  The dash-dotted lines are the one-to-one
    relations, while the dotted lines show the 1$\sigma$
    uncertainties.}
\label{FigF3}
\end{figure*}

Fig.~\ref{FigF3} compares the 1D and 2D properties of the nuclei
including the (a) central and effective surface brightnesses, i.e.\
$\mu_{\rm 0,nuc}$ and $\mu_{\rm e,nuc}$, respectively, (b) effective
radii, $R_{\rm e,nuc}$ and (c) S\'ersic indices, $n_{\rm nuc}$. We
note that the $\mu_{\rm e,nuc}$ and $n_{\rm nuc}$ values are only for
the 21/100 galaxies whose nuclei were fitted with a S\'ersic model.
As in \citet{2023A&A...675A.105D}, we find strong agreement between
the 1D and 2D fits, where the $\mu_{\rm 0,nuc}$/$\mu_{\rm e,nuc}$,
$R_{\rm e,nuc}$ and $n_{\rm nuc}$ values from the two methods are
within the 1$\sigma$ error ranges for 89, 87 and 62 per cent of the
cases.  For 97, 92, and 76 per cent of the cases, the 1D versus 2D
measurements of $\mu_{\rm 0,nuc}$/$\mu_{\rm e,nuc}$, $R_{\rm e,nuc}$
and $n_{\rm nuc}$, respectively, are within 2$\sigma$ of perfect
agreement. Marginal discrepancies between 1D and 2D measurements are
expected, as these two methods inherently differ from each other . We
follow \citet{2012ApJS..203....5T} and \citet{2023A&A...675A.105D} and
adopt the results from our 1D decompositions in this work.  In
Appendix~\ref{App02}, we compare our data, fitting methods and sample
with those from the literature, to explain improvements and
discrepancies when they exist.

Apparent magnitudes of the nuclei are determined by integrating the
best-fitting S\'ersic and Gaussian profiles to $R = \infty$. We
applied foreground Galactic extinction corrections to the magnitudes
using the reddening values from \cite{2011ApJ...737..103S}.  For S0,
spiral and irregular galaxies, we additionally correct for internal
dust attenuation using equations from \citet{2008ApJ...678L.101D}. 
We applied the same amount of internal dust correction  to  the nuclei and their  
host bulge. The internal dust corrections, the transformation of
magnitudes from various {\it HST} filters into $V$-band magnitudes are discussed in \citet[][their
Section~3.3]{2023A&A...675A.105D}.  After correcting the nuclei 
magnitudes for dust, they were converted to stellar masses.

Table~\ref{TabA1} presents the nucleus, bulge and galaxy structural
data for the 100 nucleated galaxies. We note that the full {\it HST}
sample covers over six orders of magnitude in bulge stellar mass and
contains all Hubble types from E to Im
(\citealt{1926ApJ....64..321H,1959HDP....53..275D}),
Tables~\ref{Tab1}, \ref{Tab2} and \ref{TabA1}.  Of the 100 nuclei, 3
are hosted by bulgeless, late-type galaxies (NGC~3077, NGC~4656 and
NGC~5112). Table~\ref{TabA2}
presents the nucleus, bulge and galaxy structural data for the 24
excluded LeMMINGs galaxies with nuclei.

\subsection{Excluded   nuclei }\label{Sec2.3}

As previously noted, we use 1D and 2D decompositions and exclude
24 out of 124 nuclei in the sample, see Tables~\ref{TabA1} and
\ref{TabA2}. \citet{2004AJ....127..105B} reported a median half-light
effective radius of 3.5 pc for their sample of 39 NSCs \citep[see
also][]{2006AJ....132.2539S}.  \citet{2006ApJS..165...57C} reported
half-light radii for NSCs which are $R_{\rm e,nuc} \sim 1- 50$ pc (see
also \citealt{2004AJ....127..105B}), later work found NSCs with
\mbox{$R_{\rm e,nuc} \sim 1- 80$ }pc and stellar masses
\mbox{$M_{*,\rm nuc} \sim 10^{5}-10^{9}~ \rm M_{\sun}$}
\citep{2012ApJS..203....5T,2014MNRAS.445.2385D,2016MNRAS.457.2122G}.

Of the 24 excluded nuclei, 13 nuclei were too large for a NSC (+AGN),
with extended half-light radii of $R_{\rm e,nuc} \ga 80$ pc. In
contrast, 9/24 nuclei have sizes $R_{\rm e,nuc} < 1$ pc, whereas the
remaining 2/24 nuclei have low stellar masses of
$M_{*,\rm nuc} < 10^{5}~ \rm M_{\sun}$. The hosts of the two low-mass
nuclei (IC~2574 and NGC~5204) are optically inactive, with no
detectable nuclear radio emission at 1.5 GHz by $e$-MERLIN (see
Table~\ref{TabA2}). This suggests that these nuclei are not AGN,
instead they may be globular clusters
\citep[e.g.][]{2006ApJS..165...57C}. Among the 9/24 nuclei where the
structural analysis gives $R_{\rm e,nuc} < 1$ pc, one galaxy,
NGC~3031, hosts a resolved nucleus ($R_{\rm e,nuc} \sim$ 0.14 arcsec
$\sim 0.43$ pc).  The remaining eight were unresolved in the {\it HST}
images ($R_{\rm e,nuc} <<$ 0.05 arcsec). For these eight galaxies we
report upper-limit values of $R_{\rm e,nuc} = 0.05$ arcsec
(Table~\ref{TabA2}) given the resolution limit in the {\it HST}
images.

We consider the possibility that the 9 nuclei with
\mbox{$R_{\rm e,nuc} < 1$ pc} are `pure' AGN.  Only three of them are
hosted by optically active galaxies (NGC~3031, i.e. M81, NGC~3642 and
NGC~5273).  For NGC~3642
(\mbox{$M_{*,\rm nuc} \sim 1.3\times10^{7} \rm ~ M_{\sun}$}) and
NGC~5273
(\mbox{$M_{*,\rm nuc} \sim 9.6\times 10^{7} \rm ~ M_{\sun}$}), the
nuclei, having effective radii of $R_{\rm e,nuc} \sim$ 0.002 arcsec
$\sim 0.15$ pc, are unresolved and deep within the {\it HST} PSF. As
such, their structural properties are less secure, making it difficult
to determine whether they are AGN and/or (bright) undersized NSCs. On
the other hand, NGC~3031, which is detected both in the radio
(\citealt{2018MNRAS.476.3478B,2021MNRAS.500.4749B};
$L_{\rm R,core, ~1.5 GHz} \sim 3.2 \times 10^{35}$ erg s$^{-1}$) and
X-ray (\citealt{2022MNRAS.510.4909W};
$L_\textnormal{X} \sim 1.6 \times 10^{39}$ erg s$^{-1}$), hosts a
compact nucleus with $R_{\rm e,nuc} \sim$ 0.14 arcsec $\sim 0.43$ pc
and \mbox{$M_{*,\rm nuc} \sim 1.1\times10^{6} \rm ~M_{\sun}$}.  Our
results from the multi-band analyses seem to indicate the presence of
a genuine compact AGN in NGC~3031. We defer the analysis of the two
low-mass nuclei and the 9 nuclei with $R_{\rm e,nuc} < 1$ pc to a
future work and exclude them from our current sample.

\begin{figure}
\begin{tabular}{@{}cccc@{}}
 \hspace*{-.262882340132570cm}
\includegraphics[trim={.2890228099519955cm -8cm -3cm .204998816cm},clip,angle=0,scale=0.55604026]{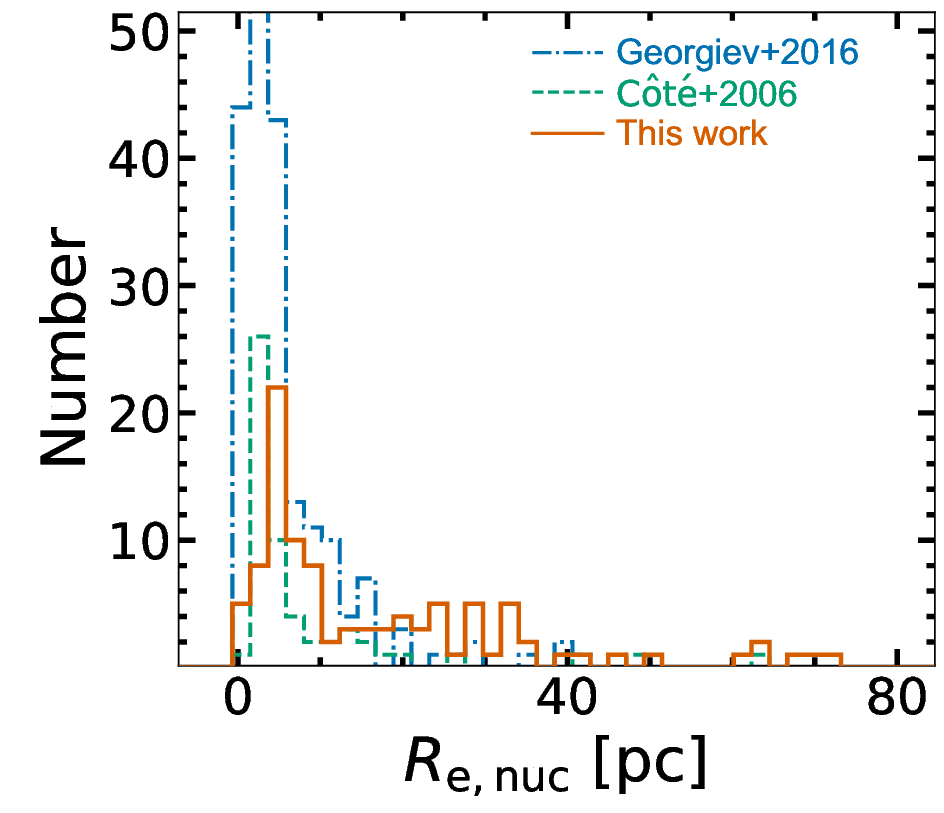}
 \vspace*{-4.570058830cm}
  \end{tabular}
  \caption{Comparison of the distribution of our 1D major-axis effective
    (half-light) radii of the nuclei (solid orange histogram) and previous
    circular effective radii of nuclei from \citet{2006ApJS..165...57C}, dashed 
    green histogram, and \citet{2016MNRAS.457.2122G}, dash-dotted blue
    histogram. }
\label{FigF4}
\end{figure}

Lastly, we explore the possibility that some nuclei in our final
sample may be due to inner stellar discs.  These inner discs tend to
be more massive and extended compared to NSCs/AGN
\citep[e.g.,][]{2007ApJ...665.1084B,2012ApJ...755..163D}.
Fig.~\ref{FigF4} compares the distribution of our major-axis effective
radii for the nuclei with the distributions of circular effective
radii of nuclei from the literature
\citep{2006ApJS..165...57C,2016MNRAS.457.2122G}.  While nine nuclei in
the final sample have masses
\mbox{$M_{*,\rm nuc} \sim 10^{9}-10^{10}~\rm M_{\sun}$}, we classify
them as NSCs/AGN rather than extended inner discs because their
effective radii are $R_{\rm e,nuc} \sim 10-70$ pc, with a mean value
of $\sim$ 38.5 pc, and they have low ellipticities ranging from 0.09
to 0.3, with a mean value of $\sim$ 0.19 (see Fig.~\ref{FigF4},
Table~\ref{TabA1}). Furthermore, \cite{2016MNRAS.457.2122G} reported
nuclei stellar masses of
\mbox{$M_{*,\rm nuc} \sim 10^{5}-10^{9}~ \rm M_{\sun}$}.  The median
stellar mass for their 228 host galaxies with nuclei is
\mbox{$M_{*,\rm gal} \sim 4 \times10^{9}~ \rm M_{\sun}$},
approximately an order magnitude smaller than the median stellar mass
for our sample galaxies
(\mbox{$ M_{*,\rm gal} \sim 5.3 \times10^{10}~ \rm M_{\sun}$}).  Given
that the stellar masses of the nuclei scale with the host galaxy
stellar masses (e.g.\
\citealt{2016MNRAS.457.2122G,2020ApJ...900...32P,2023MNRAS.520.4664H};
Dullo in prep.), the nine nuclei in our sample with
\mbox{$M_{*,\rm nuc} \sim 10^{9}-10^{10}~\rm M_{\sun}$} are expected
\citep[see][]{2012ApJS..203....5T}. Finally, we note that the 100
nuclei in this study (Table~\ref{TabA1}) should not be confused with
the extra central light components that were identified in S\'ersic
elliptical galaxies using hydrodynamical simulations \citep[][their
Fig.\ 45]{2009ApJS..181..135H}, see also
\citet{2013ARA&A..51..511K}. These latter components typically have an
extended profile with an effective radius of 100 -- 500 pc.

\begin{figure*}
\hspace{1.4208cm}
\includegraphics[trim={.006397315cm -7cm -8cm .27027293cm},clip,angle=0,scale=0.47607]{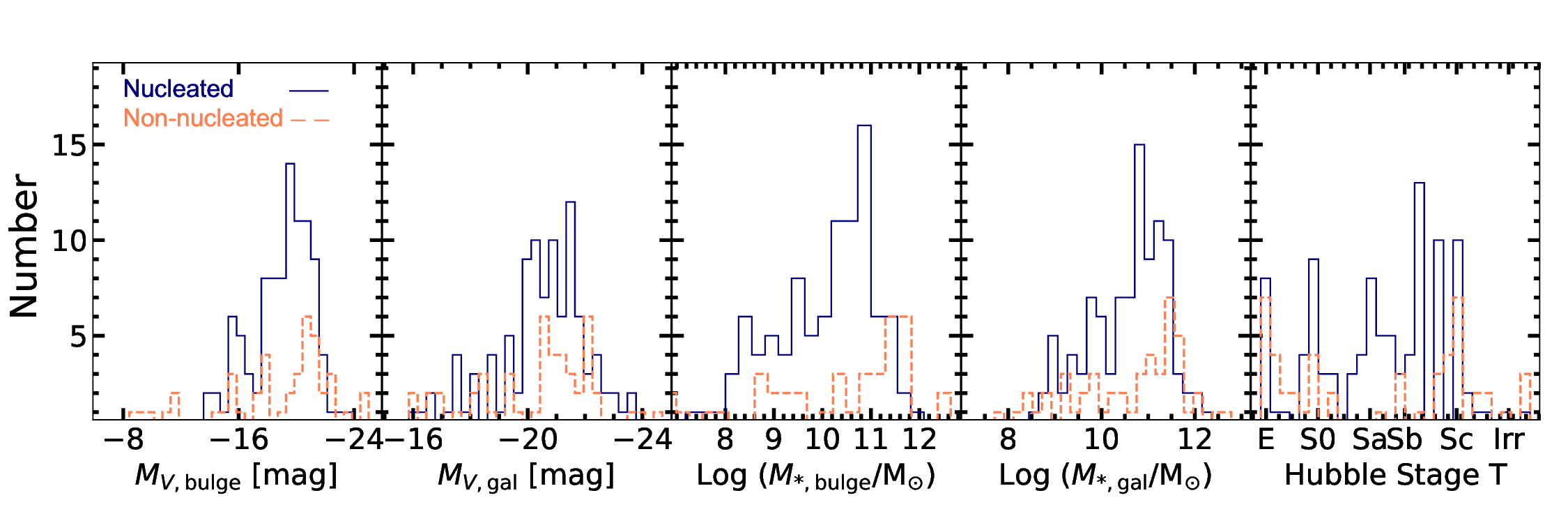}
\vspace{-4.0456299887179180cm}
\caption{Distributions of $M_{V,\rm bulge}$,
  $M_{V,\rm gal}$, $M_{*,\rm bulge}$, $M_{*,\rm gal}$ and morphology
  for nucleated (blue histograms) and non-nucleated galaxies (orange histograms) in our sample of 149
  galaxies with {\it HST} data. Our analysis excludes 24 galaxies with
  central light excesses, as explained in the text. Note that,
  including these 24 galaxies in the analysis does not significantly
  alter the observed trends. 
}
  \label{FigF5a}
\end{figure*}

\begin{figure*}
\hspace{.42045070809cm}
\includegraphics[trim={-.20505102602305397315cm -7.5cm -8.4cm .607027293cm},clip,angle=0,scale=0.4704829]{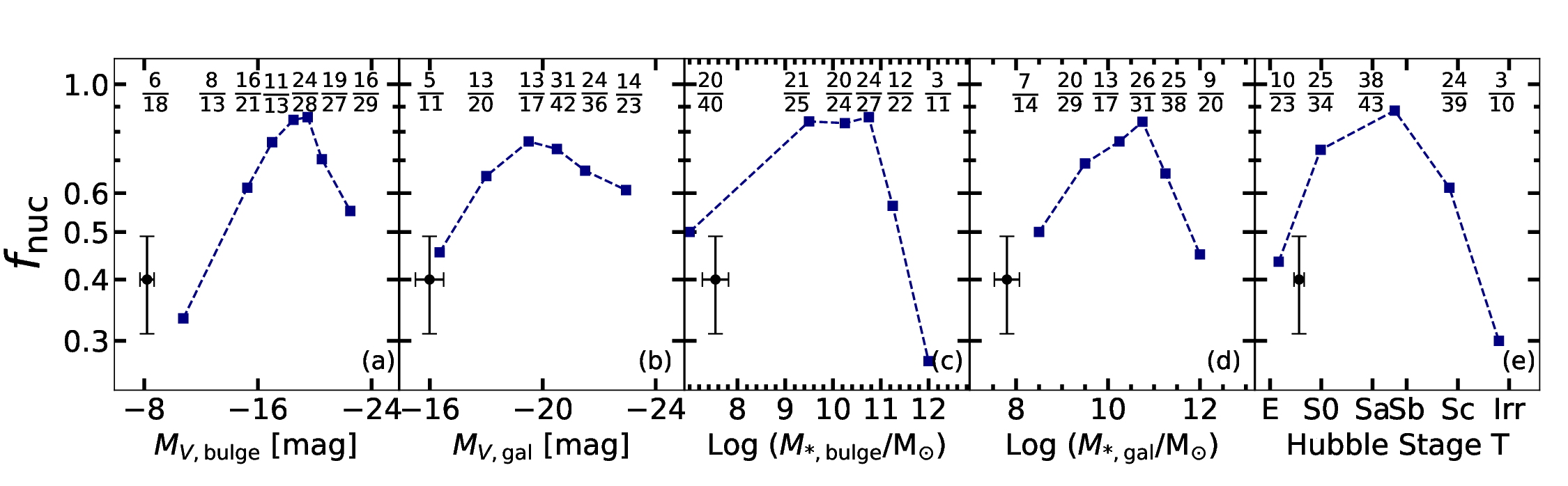}
\vspace{-04.199345629cm}
\caption{Nucleation fraction ($f_{\rm nuc}$) as a function of (a) bulge
  $V$-band absolute magnitude $M_{V,\rm bulge}$, (b) total $V$-band
  absolute galaxy magnitude $M_{V,\rm gal}$, (c) bulge stellar mass
  $M_{*,\rm bulge}$, (d) galaxy stellar mass $M_{*,\rm gal}$ and (e)
  morphology. A representative error bar is shown at the bottom left of each panel. 
}
  \label{FigF5b}
\end{figure*}

\begin{table} 
\begin{center}
\setlength{\tabcolsep}{0.04609528340533in}
\begin {minipage}{85mm}
\caption{Galaxy property intervals.  }
\label{Tab2Bin}
\begin{tabular}{@{}llcccccc@{}}
\hline
Range&Galaxies&Range&Galaxies&&Range&Galaxies\\
&per bin&&per bin&&&per bin\\
 (1)&(2)&(1)&(2)&&(1)&(2)\\          
\hline       
\multicolumn{1}{c}{$-M_{V,\rm bulge}$}&&  \multicolumn{2}{c}{${\rm log~} M_{*,\rm bulge}$}	&&  \multicolumn{2}{c}{ Hubble Type }\\
  7.0 --   14.5                     &       18      &< 9.0             &   40 &&   --5.0 -- (--4.0)	&	 23 \\
 14.5 -- 16.0             &      13       &9.0 -- 10.0	    & 	 25&&    --3.6 -- 0.5	& 34 \\
 16.0  -- 18.0	      &      21    &10.0 -- 10.5       &    24 && 0.6 -- 4.0      &   43 \\
 18.0  -- 19.0       	      &     13      &10.5 -- 11.0	    & 27 &&  4.0 -- 7.0	 	&39\\
19.0 -- 20.0	      &     28       &11.0 -- 11.5	    & 22 && 7.0 -- 9.0		&10 \\
 20.0 -- 21.0	      &      27      &11.5 -- 12.5   &  	11 \\
\vspace{0.2cm}
21.0 -- 24.0           & 	  29&& \\

\multicolumn{1}{c}{$-M_{V,\rm gal}$}  &&         \multicolumn{2}{c}{ ${\rm log~} M_{*,\rm gal}$} 	\\
15.7 -- 17.0   &  11    &   8.0 -- 9.0	& 	14 \\ 
17.0 -- 19.0  &  20   & 9.0 -- 10.0	& 	29 \\
19.0 -- 20.0  &  17   &10.0 -- 10.5   &       17 \\
20.0 -- 21.0  &  42   & 10.5 -- 11.0	&  31 \\ 			
21.0 -- 22.0  &  36   & 11.0 -- 11.5	&   38 \\			
22.0 -- 24.0  & 	 23 &11.5 -- 12.5 & 	 20 \\  
\hline
\end{tabular}     
Note: In order to calculate the nucleation fraction ($f_{\rm nuc}$), we split the sample into five to seven $M_{V,\rm bulge}$, $M_{V,\rm gal}$, $M_{*,\rm bulge}$, $M_{*,\rm gal}$ and Hubble type  bins.
 \end{minipage}
\end{center}
\end{table}

\section{ Results}\label{Sec3}

\subsection{Nucleation fraction 
  as a function of luminosity, stellar mass  and Hubble type
}\label{Sec3.1}

To investigate the relation between nucleation and host galaxy
properties, in Fig.~\ref{FigF5a} we show the distributions of bulge
absolute magnitude ($M_{V,\rm bulge}$), galaxy absolute magnitude
($M_{V,\rm gal}$), bulge stellar mass ($M_{*,\rm bulge}$), galaxy
stellar mass ($M_{*,\rm gal}$) and Hubble type for the 100 nucleated
(blue histograms) and 49(=173$-24-100$) non-nucleated (orange
histograms) LeMMINGs galaxies.  We define the nucleation fraction
  $f_{\rm nuc}$ as the ratio between the number of nucleated galaxies
  and the total number of galaxies in the bin under consideration
  (Table~\ref{Tab2Bin}). When calculating $f_{\rm nuc}$, we {\it did
  not} take into account the 24 excluded LeMMINGs galaxies
(Section~\ref{Sec2.3}). We calculate $f_{\rm nuc}$ for each bin of the
absolute magnitude, stellar mass and Hubble type distributions (see
Table~\ref{Tab2Bin}).   In Fig.~\ref{FigF5b}, we plot $f_{\rm nuc}$
against the mean $M_{V,\rm bulge}$, $M_{V,\rm gal}$,
$M_{*,\rm bulge}$, $M_{*,\rm gal}$ and Hubble type values for each bin
(Table~\ref{Tab2Bin}).  We find that $f_{\rm nuc}$ is a strong
function of $M_{V,\rm bulge}$, $M_{V,\rm gal}$, $M_{*,\rm bulge}$ and
$M_{*,\rm gal}$ (Fig.~\ref{FigF5b}a--d).

The nucleation fraction peaks at intermediate stellar mass and
luminosity ranges, declining at faint and bright end of the luminosity
function. Over the full absolute
magnitude/mass range (Fig.~\ref{FigF5a}), the nucleation fraction is
$f_{\rm nuc}=$100/149 (= 67 $\pm$ 7 per cent).  Note that the errors quoted in this
Section are Poisson errors. If we treat the 24
excluded galaxies (Section~ \ref{Sec2.3}) as non-nucleated then
$f_{\rm nuc}=$100/173 (=58 $\pm$ 6 per cent) for the full LeMMINGs
{\it HST} sample. The nucleation fraction  has a peak value of \mbox{$49/56~(= 88 \pm 13$ per
    cent)}, which occurs at bulge masses 
  \mbox{$M_{*,\rm bulge}\sim 10^{9.4}- 10^{10.8}~ \rm M_{\sun}$} (Fig.~\ref{FigF5b}). 
We find that the trends of the nucleation fraction with the
stellar mass and absolute magnitude of the host galaxy are reminiscent
of the trends with those of the bulge.

The nucleation fraction across the LeMMINGs sample agrees very well
with those reported in previous studies (typically
$f_{\rm nuc} \sim 50-80$ per cent), which are summarised in
Table~\ref{Tab3}
\citep{1997AJ....114.2366C,1998AJ....116...68C,2001AJ....122..653R,
  2001AJ....121.1385S,
  2002AJ....123..159C,2002AJ....123.1389B,2003AJ....125.2936G,2003ApJ...582L..79B,2007ApJ...665.1084B,2004ApJ...613..262L,
  2005AJ....129.2138L,2005MNRAS.363.1019G,2006ApJS..165...57C,
  2007ApJ...665.1084B,2012ApJS..203....5T,2014MNRAS.445.2385D,
  2019ApJ...878...18S,2020IAUS..351...13S,2021MNRAS.507.3246H,2021MNRAS.508..986Z,
  2022AA...664A.167S}. However, from visual inspections of
ground-based images of dwarf galaxies located in low-to-moderate
density environments, \citet{2021MNRAS.506.5494P} reported a
relatively low nucleation frequency of $f_{\rm nuc} \sim 23$ per
cent. In comparison, \citet{2001AJ....121.1385S} reported
$f_{\rm nuc} \sim 56$ per cent for dwarf ellipticals (dEs) in the
Virgo and Fornax Clusters and Leo group. Similarly,
\citet{2003AJ....125.2936G} and \citet{2014MNRAS.445.2385D} reported
$f_{\rm nuc} \sim 87$ and $80$ per cent for dwarf galaxies in the Coma
Cluster, respectively, whereas \cite{2005MNRAS.363.1019G} found
$f_{\rm nuc} \sim 61$ per cent for dEs in the Virgo cluster.

We find that nuclei are less common in core-S\'ersic galaxies, which
instead exhibit a partially depleted core
\citep{2014MNRAS.444.2700D,2017MNRAS.471.2321D,2018MNRAS.475.4670D,2019ApJ...886...80D}. This is in agreement
with \citet{2010ApJ...714L.313B} and \citet{2015ApJ...812...72A} who
suggested that coalescing SMBH binaries with individual mass of
\mbox{$M_{\rm BH} \ga 10^{8}~ \rm M_{\sun}$} can heat and lower
central stellar density, thus fully destroying the central NSC in
core-S\'ersic galaxies during gas-poor mergers.  We have identified
two nucleated core-S\'ersic galaxies (NGC~3193 and NGC~4278) out of
the 20 core-S\'ersic galaxies in LeMMINGs, i.e.\
\mbox{$f_{\rm nuc} = 10 \pm 7 $ per cent}. Combining our LeMMINGs data
with the core-S\'ersic data from our past work
\citep{2014MNRAS.444.2700D,2019ApJ...886...80D,2023A&A...675A.105D}
yields a slightly higher nucleation fraction of
\mbox{$ 10/51(20 \pm 6 $ per cent}), see also
\citet{2006ApJS..165...57C}, \citet{2012ApJ...755..163D},
\citet{2012ApJS..203....5T}, \citet{2014MNRAS.445.2385D},
\citet{2017ApJ...849...55S} and
\citet{2019ApJ...886...80D}. Conversely, nuclei are commonplace in
S\'ersic galaxies ($f_{\rm nuc} = 98/129 = 76 \pm 8$ per cent).

When separated by the Hubble type, we find that the nucleation
fractions for elliptical (Hubble $T < -4$), S0 ($-4 \le T  < 1$) and
S ($1 \le T < 9$) galaxies are \mbox{$10/23~(=44 \pm 14$ per cent)},
\mbox{$25/34~(=74 \pm 15$ per cent)} and \mbox{$63/87~(=72 \pm 9$ per cent)},
respectively. Nucleation was detected in \mbox{$2/5~(=40 \pm 28$ per cent)}
of Irr galaxies, but this figure is not conclusive due to the small
number statistics \citep[e.g.][]{2020A&ARv..28....4N,2020IAUS..351...13S,2021MNRAS.507.3246H}.

\begin{table} 
\begin{center}
\setlength{\tabcolsep}{0.09909820in}
\begin {minipage}{84mm}
\caption{Types of nuclei.}
\label{Tab4}
\begin{tabular}{@{}llccc@{}}
\hline
  Nuclei&Note&Sample &Radio&X-ray\\
&&&jetted & data/detected \\
(1)&(2)&(3)&(4)&(5)\\
  \hline
    \\[-7.38pt]    
  NSC & inactive (optical)     &43 &0&34/21 \\
      \vspace{.18cm}   
 & radio-detected ($30\%$)&\\ 
  Hybrid &AGN (optical)  &57 &13 &50/47 \\
&radio-detected   ($65\%$)     \\                          
  \hline
\end{tabular} 
Note: We have classified  the 100 nuclei in the sample
 into NSCs (43 per cent) and  hybrid nuclei  (57 per cent).
 \end{minipage}
\end{center}
\end{table}

\begin{figure}
\hspace{-.0292509cm}
\includegraphics[trim={.34027020515cm -9cm -9cm 1.59923cm},clip,angle=0,scale=0.39281]{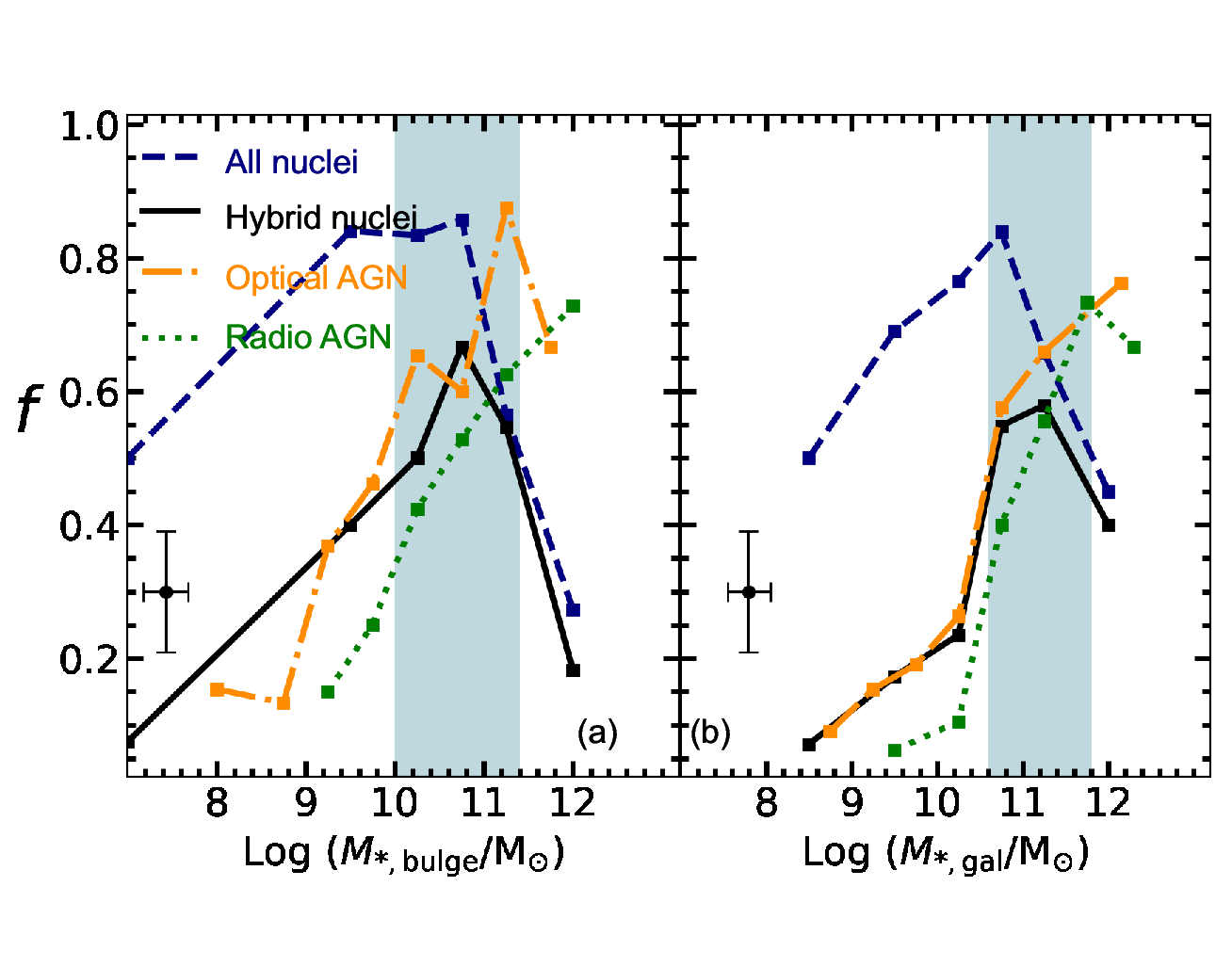}
\vspace{-4.3025cm}
\caption{The trend between the fraction of hybrid (NSC+AGN) nuclei and
  the bulge and galaxy stellar mass (solid curves). The trends of
  optical and radio AGN fractions with $M_{*,\rm bulge}$ and
  $M_{*,\rm gal}$ (dash-dotted and dotted curves, respectively) are
  taken from \citet{2023MNRAS.522.3412D} and shown here for
  comparison. The identification of radio and optical AGN is based on
  the nuclear radio continuum emission and optical emission-line data
  (Section~\ref{Sec2.1}).  The trend of all types of nuclei with
  $M_{*,\rm bulge}$ and $M_{*,\rm gal}$ (see Fig.~\ref{FigF5b}) is
  also shown for comparison. We note that the fraction $f$ for all
  types of nuclei is derived as the ratio between the number of
  nucleated galaxies and the total number of galaxies under
  consideration (Section~\ref{Sec3.1}).  The fraction of galaxies
  hosting hybrid nuclei increases with $M_{*,\rm bulge}$ and
  $M_{*,\rm gal}$, peaking at intermediate masses
  ($M_{*,\rm bulge}/M_{*,\rm gal} \sim 10^{10.0}- 10^{11.4}~\rm
  M_{\sun}$/$10^{10.6}- 10^{11.8}~\rm M_{\sun}$) before decreasing at
  higher stellar masses. Representative error bars are shown at the
  bottom of the panels.}
  \label{FigF6}
\end{figure}

\subsection{Characterisation of nuclei}\label{Sec3.2}

The sample under investigation consists of 100 nucleated
  galaxies. Using optical emission-line data, the nuclear activities
  of the galaxies were classified as either `inactive',
  ALG+\mbox{H\,{\sc ii}} for 43 galaxies or `active', Seyfert+LINER,
  for 57 galaxies \citep{2018MNRAS.476.3478B,2021MNRAS.500.4749B}, see
  also Section~\ref{Sec2.1} and Tables~\ref{Tab2} and \ref{Tab4}. We
checked for the presence of radio jets in the sample ALGs and
\mbox{H\,{\sc ii}} galaxies from weakly active SMBHs. None of the ALGs
and \mbox{H\,{\sc ii}} galaxies host radio jets that can be detected
with \mbox{$e$-MERLIN} at 1.5 GHz.

\subsubsection{NSCs}

 We classify the nuclei detected in the 43 inactive (i.e.\ 9 ALG + 34
\mbox{H\,{\sc ii}}) galaxies as `pure' NSCs, where the bulk of the central
light excess in the optical or near-IR brightness profiles of the
galaxies is of stellar origin (Tables~\ref{Tab2} and \ref{Tab4}). Our
\mbox{$e$-MERLIN} observations detected nuclear radio emission  in
thirteen of these 43 NSCs (30 per cent). We consider the possibility
that the nuclei in the 9 ALGs could be powered by non-stellar emission
of low-level AGN.  The 9 ALGs all have
$M_{*,\rm bulge} < 10^{11} ~\rm M_{\sun}$ and include three dwarf
satellites of M31, the Andromeda Galaxy (NGC~147, NGC~205 and NGC~221,
e.g.\ \citealt{2013Natur.493...62I}). All, except for one of them (NGC~2634),
lack radio detection with \mbox{$e$-MERLIN} at 1.5 GHz. Consequently,
we cannot conclude that the nuclei in the 9 ALGs are mainly due to
AGN.  Note that high-sensitivity (sub-mJy) radio continuum observations with
\mbox{$e$-MERLIN} enable detection of low-luminosity AGN at scales of
a few tens of parsecs.  Of the 43  `pure' NSC hosts in the sample, 34 have
{\it Chandra} X-ray data from LeMMINGs and we find 21/34
($\sim 62$ per cent) are X-ray detected. The majority of these X-ray
detected nuclei (13/21) have luminosities
$L_\textnormal{X}$\mbox{$\la 10^{39}$ erg s$^{-1}$}, possibly
indicating that the X-ray emission originate mainly from ULXs/XRBs
\citep{2022MNRAS.510.4909W}. The eight remaining nuclei (i.e.\ 2 ALG+6 \mbox{H\,{\sc ii}}) have
\mbox{$L_\textnormal{X} > 10^{39}$ erg s$^{-1}$}. While
only 50 per cent of these bright nuclear X-ray sources have been radio
detected with \mbox{$e$-MERLIN}, generating their core X-ray emission
may require non-thermal sources such as a weak AGN as well as
contributions from ULXs and XRBs.  Nonetheless, we classify all the 43
nuclei as `pure' NSCs, in line with their optical spectral classifications.

\begin{figure*}
\hspace{1.8509978282cm}
  \centering
\includegraphics[trim={.0952cm 0mm -4cm .03075cm},clip,angle=0,scale=.6604545]{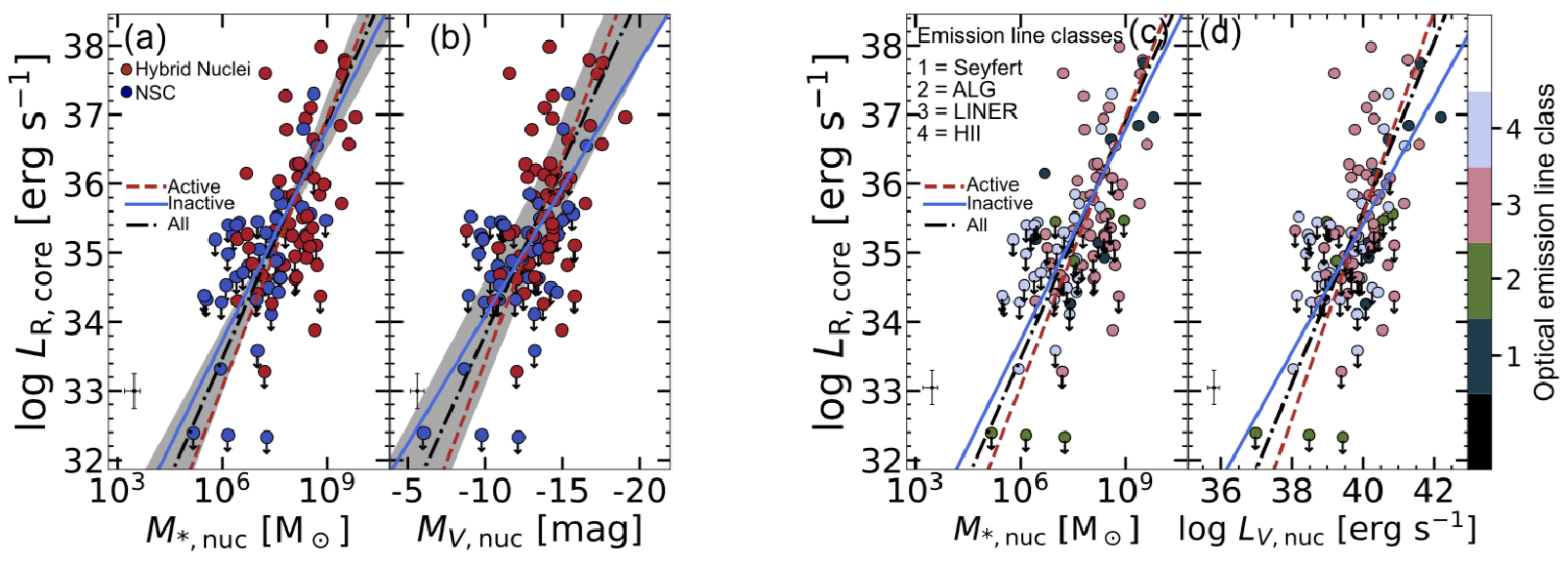}
\vspace{-.5066229978282cm}
  \caption{Radio scaling relations for our sample of 100 nucleated
    galaxies. Left panels: the 1.5 GHz radio core luminosity from
    \emerlin\ ($L_{\rm R,core}$), Table~\ref{TabA1}, is plotted as a
    function of (a) the nucleus' stellar mass ($M_{\rm *,nuc}$) and
    (b) $V$-band absolute magnitude ($M_{ V,\rm nuc}$), colour-coded
    by nucleus types. Right panels: the radio scaling
    relations by separating the galaxies based on their spectral
    classes. To allow better comparison with earlier studies on AGN,
    we converted the nuclei absolute magnitudes into $V$-band nuclei
    luminosities ($L_{V,\rm nuc}$) in units of solar luminosity
    (${\rm L}_{V,\sun}$) such that
    $L_{V,\rm nuc} = ({\rm L}_{V,\sun})10^{(0.4){({\rm M}_{V,\sun}-
        M_{V, \rm nuc} )}}$, where ${\rm M}_{V,\sun} = 4.81$
    \citep{2018ApJS..236...47W} and
    ${\rm L}_{V,\sun} = 4.4 \times 10^{32}$ erg s$^{-1}$
    \citep{2012CQGra..29n5011O}.  The nuclei that are not
    radio-detected with \emerlin\ at 1.5 GHz, and thus have $3\sigma$
    upper limit $L_{\rm R,core}$ values, are shown by downward arrows.
    The dashed, solid and dash-dotted lines are {\sc bces} bisector
    regressions for the active galaxies (i.e.\ hybrid nuclei,
    red circles), inactive
    galaxies (i.e.\ NSCs, blue circles) and all nuclei, respectively.
    A typical error bar associated with the data points is shown at
    the bottom of each panel.  }
  \label{FigF7}
\end{figure*}

\subsubsection{Hybrid nuclei}

We classify the nuclei identified in the 57 optically active,
nucleated galaxies as `hybrid = NSC+AGN' (Table~\ref{Tab4}).  
Of these
57 hybrid nuclei, 50 have {\it Chandra} X-ray data available from 
LeMMINGs \citep{2022MNRAS.510.4909W}, see Table~\ref{Tab4}. A vast
majority of these sources (47/50) are X-ray detected, with 81 per cent
of the X-ray detected, nucleated sources having X-ray luminosities 
\mbox{$L_\textnormal{X}\ga 10^{39}$ erg s$^{-1}$}. We also
find a high radio-detection fraction for these active nucleated hosts,
37/57 (65 per cent). There are only three active galaxies with nuclei
in the sample lacking detectable, nuclear radio and X-ray emission
(NGC~3486, NGC~4150 and NGC~4274, see
\citealt{2021MNRAS.500.4749B,2021MNRAS.508.2019B,2022MNRAS.510.4909W}).

We extend our efforts to investigate  hybrid nuclei within  our sample that may have most of their 
optical/near-IR   flux originating from the
AGN, thus exhibiting behaviour closely resembling that of  `pure' AGN.
Our motivation partly arises from the observation  of  massive (core-S\'ersic) galaxies
($M_{*,\rm bulge} \ga 10^{11} \rm~ M_{\sun}$), hosting massive SMBHs
($M_{\rm BH} \ga 10^{8} \rm ~M_{\sun}$). These galaxies are less likely
to contain  NSCs, which are believed to be tidally disrupted by the
inspiralling massive SMBH binary that form during galaxy mergers
\citep{2015ApJ...812...72A,2019MNRAS.486.5008A}.
Additionally, all our
nucleated galaxies with $M_{*,\rm bulge} \ga 10^{11} \rm~ M_{\sun}$ are
active.  

Therefore, we focus on radio-detected core-S\'ersic galaxies hosting hybrid  nuclei, suggesting that a significant fraction of the
optical/near-IR flux in these nuclei is likely generated by the
AGN. It should be noted that the low-luminosity AGN in massive
nearby galaxies do not always manifest themselves as detectable nuclei
in the broadband {\it HST} data
\citep{2005AJ....129.2138L,2012ApJ...755..163D,2019ApJ...886...80D}.
There are a total of 11 optically active core-S\'ersic galaxies in 
LeMMINGs sample \citep{2023MNRAS.522.3412D}. Out of these, 9/11 are
radio detected \citep{2018MNRAS.476.3478B,2021MNRAS.500.4749B} and
10/11 are X-ray detected \citep{2022MNRAS.510.4909W}.  However, only
2 (18 per cent) of these active core-S\'ersic galaxies (NGC~3193 and
NGC~4278) exhibit nuclei in our analysis of the galaxies' broadband
{\it HST} imaging data (see Table~\ref{TabA1}). Furthermore, the sizes of NGC~3193 and NGC~4278 nuclei, with 
$R_{\rm e,nuc} \sim 63$  and $5$ pc,
respectively, do not guarantee that they are compact point
sources (AGN) with sizes smaller than  any plausible NSCs. We
therefore argue that all the optical/near-IR flux for our
core-S\'ersic nuclei, seen in the broadband {\it HST} images, cannot
be explained solely by the low-luminosity AGN. In some cases, AGN
alone can account for the nuclei observed in certain core-S\'ersic
galaxies, such as M87
\citep[e.g.][]{2001AJ....122..653R,2006ApJ...644L..21F,2016MNRAS.457.3801P,2019ApJ...886...80D}.

\subsubsection{Hybrid nuclei demography }\label{Sec3.2.4}
  
Fig.~\ref{FigF6} shows the incidence of hybrid (NSC+AGN) nuclei, optical
and radio AGN, and all nuclei as a function of bulge stellar mass
($M_{*,\rm bulge}$) and galaxy stellar mass ($M_{*,\rm gal}$). We
define the hybrid nuclei fraction as the ratio between the number of
galaxies hosting a hybrid nucleus and the total number of galaxies
under consideration (i.e.\ 149 galaxies; see Section~\ref{Sec3.1}).  While  nuclei are defined as
distinct central light excesses with respect to the inward
extrapolation of outer S\'ersic or core-S\'ersic profiles (Section~\ref{Sec2.2}), we 
note that  the identification of radio and optical AGN relies on the nuclear radio 
continuum emission and optical emission-line data (Section~\ref{Sec2.1}).

The fraction of galaxies hosting hybrid nuclei
increases with increasing  $M_{*,\rm bulge}$ and $M_{*,\rm gal}$, peaking at
intermediate masses
($M_{*,\rm bulge}/M_{*,\rm gal} \sim 10^{10.0}- 10^{11.4}~\rm
M_{\sun}$/$10^{10.6}- 10^{11.8}~\rm M_{\sun}$) before declining at
higher stellar masses. We find that low mass galaxies with
$M_{*,\rm gal} \la 10^{10.6}~\rm M_{\sun}$ show  comparable frequencies of
hybrid nuclei and optical AGN. However, above this 
mass (i.e.\ $M_{*,\rm gal} \ga 10^{10.6}~\rm M_{\sun}$) galaxies
are more likely to show evidence of AGN activity rather than hosting hybrid nuclei.

As noted in the introduction, \citet{2008ApJ...678..116S} used radio,
X-ray and optical spectroscopic observations and from the optical
spectra, they reported that 10 per cent of their sample galaxies with
NSCs host both AGN and nuclear star clusters.  Using {\it Chandra}
X-ray observations for a sample of 98 (47 late-types and 51
early-types) galaxies with NSCs, \citet{2017ApJ...841...51F} reported
that $\sim$11.2 per cent harbour hybrid nuclei (see also
\citealt{2010ApJ...714...25G}). All the nuclei in our sample of
nucleated galaxies with an optical AGN are classified as hybrid
nuclei. However, we determine the lower limit for the fraction of
hybrid nuclei in the nucleated sample by adopting a criterion that
requires the use of a multi-band signature, which combines radio,
X-ray and optical spectra data. Our criterion is more stringent than
those used by \citet{2008ApJ...678..116S} and
\citet{2017ApJ...841...51F}.  We identify 30 nucleated LeMMINGs hosts
that are optically active, radio detected and X-ray luminous
($L_\textnormal{X} > 10^{39}$ erg s$^{-1}$). Relying solely on optical
and X-ray AGN diagnostics, we find that 39 nucleated LeMMINGs galaxies
are optically active and X-ray luminous ($L_\textnormal{X} > 10^{39}$
erg s$^{-1}$). We do not suggest that the sources with
$L_\textnormal{X} <10^{39}$ erg s$^{-1}$ are inactive. Instead, a high
X-ray core luminosity ($L_\textnormal{X} > 10^{39}$ erg s$^{-1}$)
likely indicate the presence of a low-luminosity AGN
\citep{2022MNRAS.510.4909W}.

To summarise, our observations suggest that at least 30 per cent of
the nucleated LeMMINGs galaxies harbour a hybrid nucleus containing
both an NSC and AGN and they exhibit a wide range in morphology
(ellipticals to late-type spirals, Sc) and in stellar mass,
$M_{*,\rm bulge} (M_{*,\rm gal}) \sim 10^{8}-10^{11.8} (10^{9}-10^{12}
\rm~ M_{\sun})$. While the hosts of 29 of the 30 hybrid nuclei
mentioned above are within 78 Mpc, the host galaxy of the remaining
hybrid nucleus is at 107 Mpc.  Excluding this furthest galaxy, we
measure a number density of $(1.5 \pm 0.4)\times 10^{-5}$ Mpc$^{-3}$
for the sample of 29 galaxies.

The hybrid nucleus fraction we measured is at least a factor of three
higher than that previously reported
\citep[e.g.][]{2008ApJ...678..116S,2017ApJ...841...51F}.  As mentioned
in the Introduction, these previous works relied on a literature
compilation of nuclei identified in {\it HST} imaging data.  In
contrast, we performed detailed 1D and 2D decompositions of the host
galaxy {\it HST} data to identify the nuclei.  Also, our AGN
diagnostics capitalise on homogeneously derived, high-quality
\emerlin\ radio, {\it Chandra} X-ray and optical emission line data.
Furthermore, with regards to the co-existence of NSCs and SMBHs, both
active and inactive, our sample of 100 nucleated galaxies includes 40
galaxies with measured SMBH masses. These 40 SMBH masses are
determined using gas dynamics, stellar dynamics, megamasers or
reverberation mapping \citep{2023A&A...675A.105D}.

\subsubsection{Do NSCs enhance AGN activity? }

Having established the AGN fraction in NSCs (i.e. in our nucleated
galaxies), we examine here whether it is different in non-nucleated
galaxies.  \citet{2015ApJ...803...81N} reported that the presence of
NSCs can facilitate the fuelling of the embedded SMBHs, by funnelling
gas toward the innermost regions, during galaxy mergers. However,
\citet{2008AJ....135..747G} did not find a trend between nucleation
and optical emission-line types. \cite{2017ApJ...841...51F} also
argued that NSCs do not enhance accretion-powered emission from the
central SMBH.  Here, we use the optical, X-ray and radio data from
LeMMINGs and re-investigate the issue. When not controlling for
bulge/galaxy mass and comparing our nucleated and the full {\it HST}
sample \citep{2023A&A...675A.105D}, we find a slight increase in the
Seyfert and LINER fractions and a slight decrease in ALG and
\mbox{H\,{\sc ii}} fractions for our sample containing only nucleated
galaxies. The optical emission-line class breakdown for the 100
nucleated galaxies is Seyferts (9 $\pm$ 3 per cent), ALGs (9 $\pm$ 3
per cent), LINERs (48 $\pm$ 7 per cent) and \mbox{H\,{\sc ii}}
galaxies (34 $\pm$ 6 per cent). In comparison, the full LeMMINGs
galaxies\footnote{For comparison the parent LeMMINGs sample of 280
  galaxies contains 6.4 per cent Seyferts, 10.0 per cent ALGs, 33.6
  per cent LINERs and 50.0 per cent \mbox{H\,{\sc ii}}.}  with {\it
  HST} data (see Sections~ \ref{Sec2.3} and \ref{Sec3.1}) consists of
Seyferts (6 $\pm$ 2 per cent), ALGs (12 $\pm$ 3 per cent), LINERs (42
$\pm$ 5 per cent) and \mbox{H\,{\sc ii}} galaxies (40 $\pm$ 5 per
cent).

Our nucleated sample also has a slightly higher radio-detection rate
than the full {\it HST} sample, but  the two samples are
indistinguishable in terms of  X-ray-detection rate.
The radio-detection fraction for the 100 nucleated
LeMMINGs galaxies is \mbox{50 $\pm$ 7}  per cent,
 compared to \mbox{45$\pm$ 5} per cent for LeMMINGs galaxies with {\it HST} data.
  We find X-ray detection fractions of \mbox{81 $\pm$ 10}
and \mbox{78 $\pm$ 7} for the 84 nucleated LeMMINGs galaxies with
X-ray data available and the LeMMINGs {\it HST} sample,
respectively. Intriguingly, while ALG and LINER hosts constitute 87
per cent of the massive ($M_{*,\rm bulge} \ga 10^{11} \rm~ M_{\sun}$)
objects in our sample of 149 galaxies, none (0/7) of the massive ALGs
contains an NSC, whereas 14/24 (58 per cent) of massive LINERs are
nucleated.

In general, our findings lend further support to the suggestion by
\citet{2015ApJ...803...81N} that NSCs may enhance accretion on to the
central SMBH, by channelling gas toward the innermost regions
\citep[see also][]{2007ApJ...668..236C}. However, the trends of AGN
activity and radio detection for the nucleated and full {\it HST}
samples are associated with large uncertainties assuming Poissonian
errors.

\begin{figure}
\hspace{-.1496882082cm}
  \centering
\includegraphics[trim={.608054090515cm -2cm -2cm 1.88394036899823cm},clip,angle=0,scale=.3969]{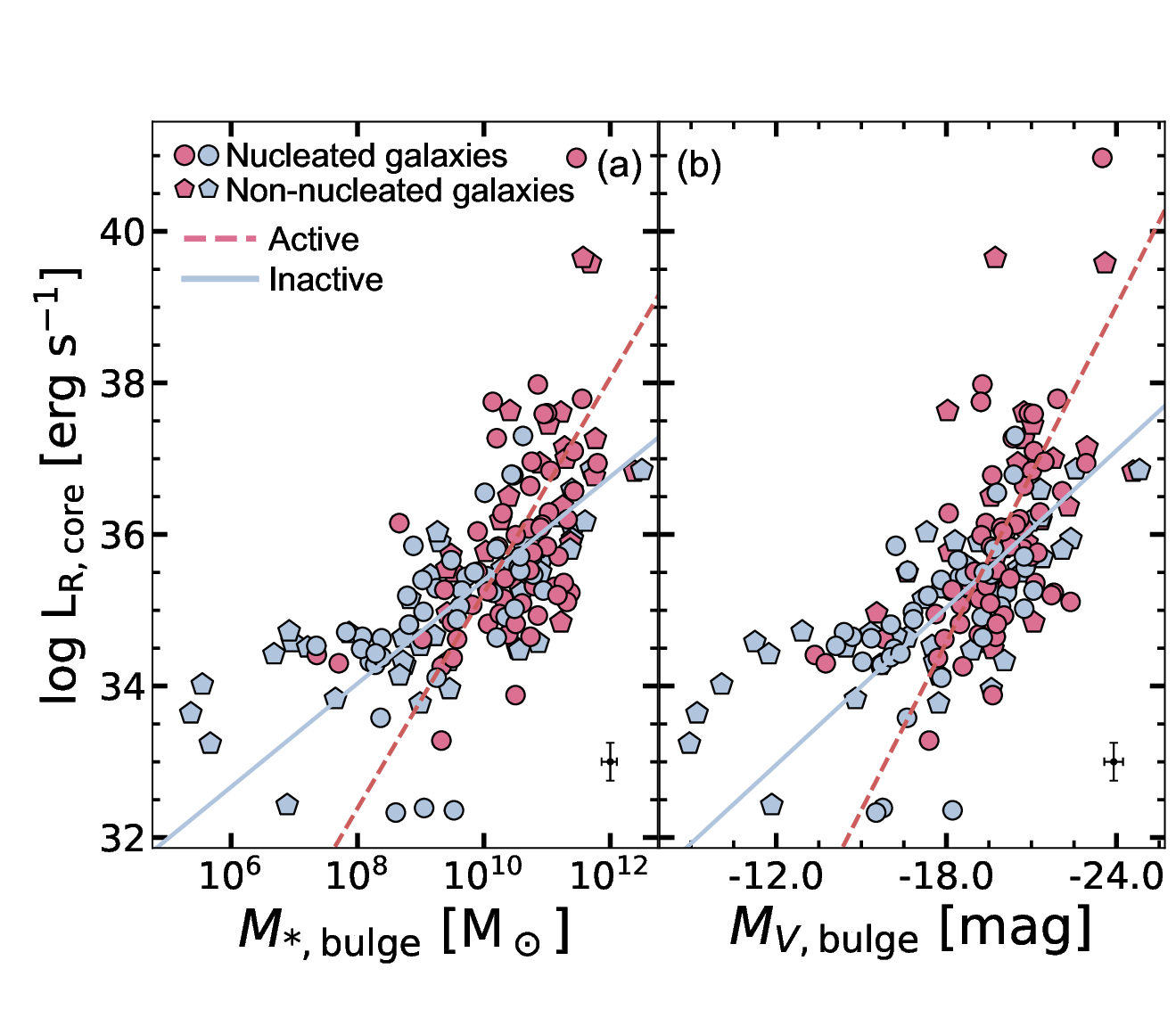}
\vspace{-1.3896533051980482cm}
\caption{1.5 GHz radio core luminosity from \emerlin\
  ($L_{\rm R,core}$) is plotted against (a) bulge stellar mass
  ($M_{*,\rm bulge}$) and (b) absolute {\it V}-band bulge magnitude
  ($M_{V,{\rm bulge}}$) for our full sample of 173 galaxies with {\it
    HST} data. The plot is similar to that in \citet[][their
  Fig.~8]{2023MNRAS.522.3412D}, but here we separated the sample into
  nucleated galaxies (circles) and non-nucleated galaxies (pentagons) to highlight the importance
  non-nucleated galaxies, which are missing in Fig.~\ref{FigF7}, in
  defining the broken $L_{\rm R,core}$--$M_{*, \rm bulge}$,
  \mbox{$L_{\rm R,core}-M_{V, \rm bulge}$} relations.   The dashed
line is our OLS bisector regression  fit to  the active (optical AGN) galaxies   (LINERs and
  Seyferts), while the solid line is the OLS bisector fit to
inactive galaxies (\mbox{H\,{\sc ii}}s and
  absorption line galaxies, ALGs).     A typical error bar associated with the data points is shown at
    the bottom of each panel.   }
  \label{FigF8}
\end{figure}

\setlength{\tabcolsep}{0.064085287390in}
\begin{table*}
\begin {minipage}{178mm}
\caption{Scaling relations between radio and X-ray core luminosities  and the mass and
  luminosity of nuclei.  } 
\label{Tab5}
\begin{tabular}{@{}lllccccccccccc@{}}
\hline
Relation &{\sc bces} bisector fit &$r_{\rm
                                              s}/P$-${\rm value}$&$r_{\rm
                                                            p}/P$-${\rm
                                                                   value}$&$\updelta_{\rm horiz}$&Sample&\\
&&&&(dex)&                                                                 \\
  \hline
               \\[-7.38pt]   
                                             \multicolumn{5}{c}{\bf  Hybrid nuclei (NSC+AGN;  active galaxies)}\\
  $L_{\rm R,core}-M_{\rm *,nuc} $
        &$\mbox{$\log$}\left(\frac{L_{\rm R,core}}{\rm erg~ s^{-1}}\right)= ( 1.30 \pm
             0.23) \mbox{log}\left(\frac{M_{\rm *,nuc}}{\mbox{$6.3\times10^{7}$ $\rm~ M_{\sun}$}}\right)$ +~($ 35.40  ~ \pm  0.17$)
 & $$0.46/$3.86\times10^{-4}$&$ $0.49/$
                                           1.42\times10^{-4}$  &  0.92& 57 \\

$L_{\rm R,core}-M_{V,\rm nuc}$
        &$\mbox{log}\left(\frac{L_{\rm R,core}}{\rm erg ~s^{-1}}\right)= ( -0.59   \pm
            0.12)\left( M_{V,\rm nuc}+13.9 \right)$ +~($ 35.73~ \pm  0.17$)
             &$-$0.46/$4.11\times10^{-4}$&$-$0.50$/
                                           7.75\times10^{-5}$&---&57 \\
  $L_\textnormal{X}-M_{\rm *,nuc} $
        &$\mbox{$\log$}\left(\frac{   L_\textnormal{X}         }{\rm erg~ s^{-1}}\right)= ( 1.44\pm
            0.18) \mbox{log}\left(\frac{M_{\rm *,nuc}}{\mbox{$1.6\times10^{8}$ $\rm~ M_{\sun}$}}\right)$ +~($ 39.98~ \pm   0.16$)
 &$$0.52/$1.25\times10^{-4}$&$$0.59/$
                                           6.70\times10^{-6}$&0.80&50 \\

$L_\textnormal{X}-M_{V,\rm nuc}$
        &$\mbox{log}\left(\frac{    L_\textnormal{X}    }{\rm erg ~s^{-1}}\right)= ( -0.68   \pm
            0.09)\left( M_{V,\rm nuc}+14.0 \right)$ +~($39.90 ~ \pm  0.22$)
             &$-$0.40/$3.66\times10^{-3}$&$-$0.51/$
                                           1.40\times10^{-4}$&---&50 \\

  \multicolumn{4}{c}{\bf  NSCs  (inactive galaxies)}\\
  $L_{\rm R,core}-M_{\rm *,nuc} $
        &$\mbox{$\log$}\left(\frac{L_{\rm R,core}}{\rm erg~ s^{-1}}\right)= (1.02 \pm
             0.16) \mbox{log}\left(\frac{M_{\rm *,nuc}}{\mbox{$1.6\times10^{7}$ $\rm~ M_{\sun}$}}\right)$ +~($ 34.96  ~ \pm  0.22$)
 &$$0.60/$2.09\times10^{-5}$&$$0.56/$
                                           9.63\times10^{-5}$&0.94&43 \\
$L_{\rm R,core}-M_{V,\rm nuc}$
        &$\mbox{log}\left(\frac{L_{\rm R,core}}{\rm erg ~s^{-1}}\right)= ( -0.37   \pm
            0.09)\left( M_{V,\rm nuc}+13.1 \right)$ +~($ 35.23~ \pm  0.16$)
             &$-$0.55/$1.26\times10^{-4}$&$-$0.58/$
                                           4.53\times10^{-5}$&---&43\\

  $L_\textnormal{X}-M_{\rm *,nuc} $
        &$\mbox{$\log$}\left(\frac{L_\textnormal{X}  }{\rm erg~ s^{-1}}\right)= (1.28\pm
            0.24) \mbox{log}\left(\frac{M_{\rm *,nuc}}{\mbox{$1.6\times10^{7}$ $\rm~ M_{\sun}$}}\right)$ +~($ 38.61~ \pm  0.19$)
 &$$0.53/$1.36\times10^{-3}$&$$0.48/$
                                           3.80\times10^{-3}$&0.84&34\\
$L_\textnormal{X}-M_{V,\rm nuc}$
        &$\mbox{log}\left(\frac{L_\textnormal{X}  }{\rm erg ~s^{-1}}\right)= (-0.63   \pm
            0.12)\left( M_{V,\rm nuc}+12.0 \right)$ +~($ 38.37~ \pm  0.22$)
             &$-$0.60/$2.06\times10^{-4}$&$-$0.47$/
                                           4.35\times10^{-3}$&---&34 \\

                                               \\[-7.38pt]   
 \hline
\end{tabular} 
Note: 1.5 GHz radio core luminosity from \emerlin\
($L_{\rm R,core}$) and (0.3--10 keV) {\it Chandra } X-ray core
luminosity ($L_\textnormal{X}$) as a function of nucleus stellar
mass ($M_{\rm *,nuc}$) and nucleus absolute magnitude
($M_{\rm V,nuc}$). We present our {\sc bces} bisector  regression fits to the
galaxy data, the Spearman and Pearson correlation coefficients
($r_{\rm s}$ and $r_{\rm p}$, respectively) and the corresponding
serendipitous correlation probabilities. The
horizontal rms scatter in the log $M_{\rm *,nuc}$ direction is denoted with $\updelta_{\rm horiz}$. 
\end{minipage}
\end{table*}

\begin{figure*}
\hspace{-.9978282cm}
  \centering
\includegraphics[trim={.0952cm 0mm -4cm .03075cm},clip,angle=0,scale=.6604545]{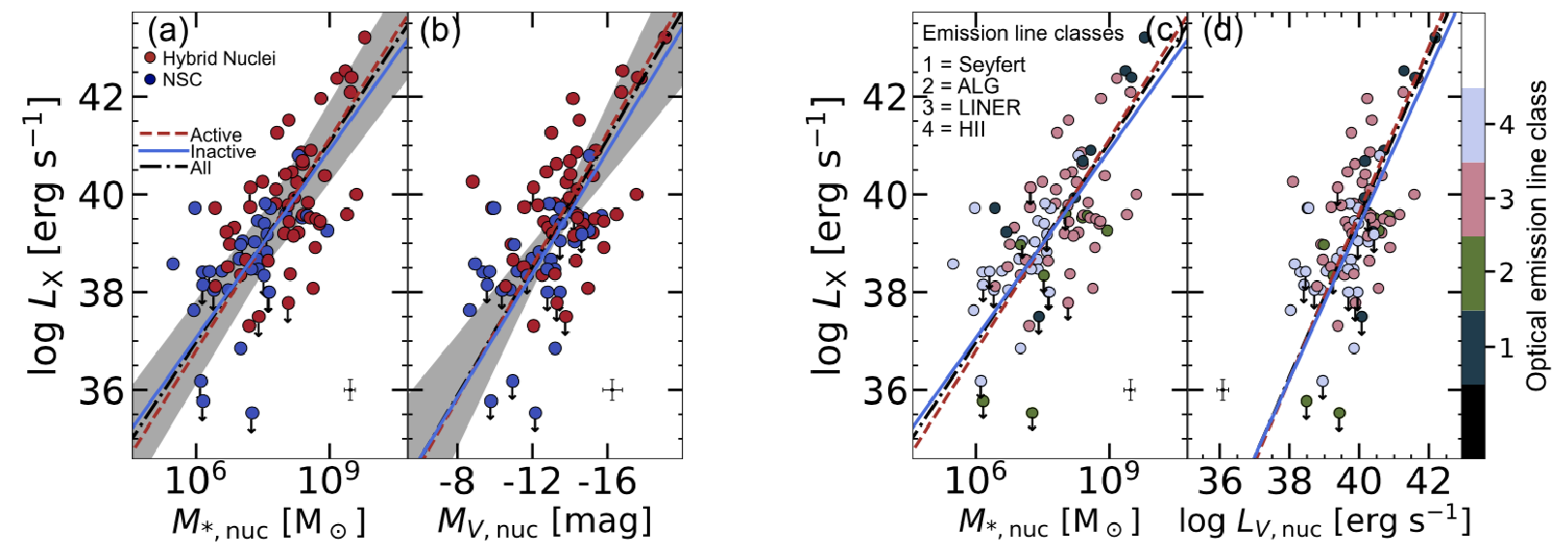}
\vspace{-.50229978282cm}
  \caption{Similar to Fig.~\ref{FigF7}, but shown here are the
    correlations between the (0.3$-$10 keV) X-ray core luminosity
    ($L_\textnormal{X}$) and $M_{*,\rm nuc}$ and
    $M_{V,\rm nuc}/L_{V,\rm nuc}$ for the 84 nucleated galaxies in our sample
    with available {\it Chandra} X-ray data. Representative error bars are shown at the bottom of the panels.}
  \label{FigF9}
\end{figure*}

\section{Radio and X-ray scaling relations for nuclei }\label{Sec4}

In \citet{2023MNRAS.522.3412D}, we showed that active and inactive
nuclei hosts follow considerably  different radio core luminosity \mbox{($L_{\rm R,core}$)--$M_{*, \rm bulge}$},
\mbox{$L_{\rm R,core}-M_{V, \rm bulge}$} and
$L_{\rm R,core}$--$\sigma$ correlations, with turnovers at
$M_{*, \rm bulge}\sim 10^{9.8 \pm 0.3} \rm~ M_{\sun}$,
$M_{V, \rm bulge}\sim -18.5 \pm 0.3$ mag and $\sigma \sim 85 \pm 5$ km
s$^{-1}$. The turnovers signify a shift from AGN-dominated nuclear
radio emission in more massive bulges to stellar-driven emission in
low-mass bulges. Similarly,
\citet{2015MNRAS.450.2317S} and \citet{2021MNRAS.508.2019B} reported broken
relations between $L_{\rm R,core}$ and the optical [\ion{O}{iii}] $\lambda$5007\,\AA\ 
luminosity ($L_{[\ion{O}{iii}]}$) and $M_{\rm BH}$ for local
galaxies. They found that below a threshold mass of
$M_{\rm BH} \sim 10^{6.5} \rm~ M_{\sun}$ the nuclear radio emission is
predominantly associated with stellar processes, whereas AGN-driven
sources are found to be dominant at higher masses.  On the other hand, \citet{2022MNRAS.510.4909W} 
found that optically active and inactive LeMMINGs galaxies together
form single $L_\textnormal{X}-\rm M_{\rm BH}$ and
$L_\textnormal{X}-L_{[\ion{O}{iii}]}$ relations across the entire
galaxy mass and luminosity ranges. Building on these earlier works, we investigate whether the
mass or luminosity   of the nucleus, derived from broadband optical and
near-IR {\it HST} images, correlates with the radio and X-ray core
luminosities. We also  examine how these relations  vary depending on the nuclear activity of
their hosts.

\subsection{$L_{\rm R,core}-M_{*,\rm nuc}$ and $L_{\rm R,core}-M_{V,\rm nuc}$  }\label{Sec4.1}

Fig.~\ref{FigF7} shows relations between $L_{\rm R,core}$ and nucleus
mass ($M_{*,\rm nuc}$) and absolute $V$-band nucleus magnitude
($M_{V,\rm nuc}$) for our sample of 100 nucleated LeMMINGs
galaxies. The data points are colour-coded based on the three classes
of nucleus introduced in Section~\ref{Sec3.2} (Figs.~\ref{FigF7}(a)
and (b)). We also divided the sample based on optical emission-line
classifications \citep{2018MNRAS.476.3478B,2021MNRAS.500.4749B} and
transformed the nuclei absolute magnitudes into $V$-band nuclei
luminosities ($L_{V,\rm nuc}$) in units of solar luminosity
(${\rm L}_{V,\sun}$) for better comparison with earlier studies on AGN
(e.g.\ \citealt{2010ApJ...725.2426B}), Figs.~\ref{FigF7}(c) and
\ref{FigF7}(d). In doing so, we use the equation
$L_{V,\rm nuc} = ({\rm L}_{V,\sun})10^{(0.4){({\rm M}_{V,\sun}- M_{V,
      \rm nuc} )}}$, where ${\rm M}_{V,\sun} = 4.81$
\citep{2018ApJS..236...47W} and
${\rm L}_{V,\sun} = 4.4 \times 10^{32}$ erg s$^{-1}$
\citep{2012CQGra..29n5011O}.

We fit the BCES bisector regressions to the
($L_{\rm R,core},M_{*,\rm nuc}$) and ($L_{\rm R,core},M_{V,\rm nuc}$)
data sets.  The BCES method allows for errors in the fitted variables
to be taken into account.  The implementation of the BCES routine
\citep{1996ApJ...470..706A} here is adapted from the python module by
\citet{2012Sci...338.1445N}. While the results from the BCES bisector
regressions are presented throughout this paper, we find good
agreement between the BCES bisector and orthogonal regression
fits. The slopes and intercepts from the two methods are consistent
within the $1\sigma$ uncertainties. We present the key radio scaling
relations in Table~\ref{Tab5}.

Furthermore, half of our sample have $L_{\rm R,core}$ upper limits,
where the sources do not have a detectable nuclear radio emission with
\emerlin\ at 1.5 GHz. We therefore performed a statistical censored
analysis with the {\sc asurv} software package
\citep{1985ApJ...293..192F,1986ApJ...306..490I}, which accounts for
upper limits (see Appendix~\ref{AppA2}). We find that the slopes and
intercepts of the $L_{\rm R,core}-M_{*,\rm nuc}$ and
$L_{\rm R,core}-M_{V,\rm nuc}$ relations from the censored and
uncensored analyses agree within the quoted errors (Table~\ref{Tab5}
and Appendix~\ref{AppA2}). However, we note that the intercepts of the
censored relations from the {\sc asurv} fits have large errors.

We find that $L_{\rm R,core}$ correlates well with $M_{*,\rm nuc}$ and
$M_{V,\rm nuc}$ (see also
\citealt{2006A&A...451...35B,2006A&A...447...97B,2010ApJ...725.2426B}). We
measure the Spearman and Pearson correlation coefficients ($r_{\rm s}$
and $r_{\rm p}$) to quantify the strength of the correlations.  For
the $L_{\rm R,core}-M_{*, \rm nuc}$ relations, we find
$r_{\rm s}/r_{\rm p} \sim 0.59/0.60$ and a very low probability $P$ of
the null hypothesis ($ \sim 10^{-11}-10^{-10}$), see
Table~\ref{Tab5}. Similarly, the $L_{\rm R,core}-M_{V,\rm nuc}$
relations are such that $-r_{\rm s}/-r_{\rm p} \sim$ $0.55/0.59$ and
$P \sim 10^{-10}-10^{-9}$.  We also derived the horizontal rms
scatters in the log $M_{\rm *,nuc}$ direction, finding
$\updelta_{\rm horiz} \sim 0.92-0.96$ dex.  NSCs and hybrid nuclei follow
$L_{\rm R,core}-L_{V,\rm nuc}$ and
\mbox{$L_{\rm R,core}-M_{*,\rm nuc}$} relations with comparable level
of scatter and strength.

Separating the sample into NSCs and hybrid nuclei (i.e.\ dividing the
full sample into active and inactive) and running the BCES regressions
yields relations where the slopes and intercepts for the two nuclei
types are consistent with each other within the $1\sigma$
errors. These relations also agree, within the errors, with those for
the full sample (Table~\ref{Tab5} and Appendix~\ref{AppA2}).  For the
full sample, the BCES bisector regression fits yield near-linear
\mbox{$L_{\rm R,core}-M_{*,\rm nuc}$} and
$L_{\rm R,core}-L_{V,\rm nuc}$ relations, see Appendix~\ref{AppA2}.

However, for nucleated galaxies of the same radio core luminosities,
NSC tend to be less massive/luminous than hybrid nuclei
(Fig.~\ref{FigF7}).  There is also a tendency in the
\mbox{$L_{\rm R,core}-M_{*, \rm nuc}$},
\mbox{$L_{\rm R,core}-M_{V,\rm nuc}$} relations where the slopes for
the hybrid nuclei are slightly steeper than those for the NSCs.  This
is reminiscent of the corresponding trend observed in the other
$L_{\rm R,core}$ scaling relations (i.e.\
$L_{\rm R,core}-M_{*, \rm bulge}$,
\mbox{$L_{\rm R,core}-M_{V, \rm bulge}$}, $L_{\rm R,core}$--$\sigma$,
$L_{\rm R,core}-M_{\rm BH}$ and $L_{\rm R,core}-L_{[\ion{O}{iii}]}$)
presented in \citet{2018A&A...616A.152S}, \citet{2021MNRAS.508.2019B}
and \citet{2023MNRAS.522.3412D}.  While our regression analysis shows
that NSCs and hybrid nuclei follow common relations within the errors
in the \mbox{$L_{\rm R,core}-M_{*, \rm nuc}$},
\mbox{$L_{\rm R,core}-M_{V,\rm nuc}$} diagrams, we believe this
apparent unity of NSCs and hybrid nuclei is artificial and driven by
the relatively low incidence of nucleation in low-mass, late-type
galaxies and massive early-type galaxies
(Figs.~\ref{FigF5a}--\ref{FigF8} and Section~\ref{Sec3.1}). These
galaxies consequently have low representation in Fig.~\ref{FigF7},
which is also evident in Fig.~\ref{FigF8}, which displays the location
of the 100 nucleated LeMMINGs galaxies and the remaining 73
LeMMINGs/{\it HST} galaxies which did not meet our criteria for having
a nucleus (Section~\ref{Sec2.3}). If large numbers of low-mass,
late-type galaxies are included in the sample, we suspect that hybrid
nuclei, similar to active bulges, define $L_{\rm R,core}$ scaling
relations with slopes that are steeper than those for the NSCs
(inactive galaxies). Note that Fig.~\ref{FigF8} is a modified version
of the \mbox{$L_{\rm R,core}-M_{*, \rm bulge}$},
\mbox{$L_{\rm R,core}-M_{V,\rm bulge}$} diagrams from \citet[][their
Fig.~8]{2023MNRAS.522.3412D}.  The dashed and solid lines
(Fig.~\ref{FigF8}) are the ordinary least-squares (OLS) bisector
regressions for the optically active (AGN) galaxies (LINER and
Seyfert) and optically inactive galaxies (\mbox{H\,{\sc ii}} and ALG),
respectively.

\subsection{$L_\textnormal{X}-M_{*,\rm nuc}$ and $L_\textnormal{X}-M_{V,\rm nuc}$ } 

In Fig.~\ref{FigF9}, we plot the (0.3$-$10 keV) X-ray core luminosity
($L_\textnormal{X}$) against $M_{*,\rm nuc}$ and $M_{V,\rm nuc}$ for
our sample of the 84 nucleated galaxies with available {\it Chandra}
X-ray data \citep{2022MNRAS.510.4909W}. Akin to Fig.~\ref{FigF7}, we
show two plots where the galaxies are separated based on nucleus types
and optical emission-line classifications (Fig.~\ref{FigF9}).  Using
our ($L_\textnormal{X},M_{*,\rm nuc}$) and
($L_\textnormal{X},M_{V,\rm nuc}$) data set, we perform BCES
regressions to derive the $L_\textnormal{X}-M_{*,\rm nuc}$ and
$L_\textnormal{X}-L_{V,\rm nuc}$ relations for NSCs and hybrid nuclei
separately (Table~\ref{Tab5}).  We also fit BCES bisector regressions
to derive $L_\textnormal{X}-M_{*, \rm nuc}$ and
$L_\textnormal{X} -M_{V,\rm nuc}$ relations for the full sample of 84
nucleated galaxies with X-ray data (see Appendix~\ref{AppA2}). We find
that the values of the slopes and intercepts for the NSCs and hybrid
nuclei and combined sample are consistent with each other within the
$1\sigma$ errors.  Out of the 84 nucleated sources, 16 have
$L_\textnormal{X}$ upper limits.  Overall, the slopes and intercepts
for our $L_\textnormal{X}$ scaling relations from the censored {\sc
  asurv} regressions and non-censored BCES regressions are in fair
agreement (Table~\ref{Tab5}).

The Spearman and Pearson correlation coefficients
and the corresponding $P$-values for the
$L_\textnormal{X}-M_{*, \rm nuc}$ relations are
$r_{\rm s} \sim 0.52-0.63$, $r_{\rm p} \sim 0.50-0.65$ and
$P \sim 10^{-11}-10^{-4}$. The $L_\textnormal{X}-M_{V,\rm nuc}$
relations are such that $-r_{\rm s} \sim$ $0.50-0.59$,
$-r_{\rm p} \sim 0.51-0.58$ and $P \sim 10^{-9}-10^{-7}$. For the
$L_\textnormal{X}-M_{*, \rm nuc}$ and
$L_\textnormal{X}-M_{V,\rm nuc}$ relations, the horizontal rms
scatters in the log $M_{\rm *,nuc}$ direction are
$\updelta_{\rm horiz} \sim 0.80-0.86$ dex.

Our analysis shows that NSCs and hybrid nuclei unite to define single
unbroken $L_\textnormal{X}-M_{*, \rm nuc}$ and
$L_\textnormal{X}-M_{V,\rm nuc}$ correlations with overlapping
uncertainties. This is in agreement with \citet{2022MNRAS.510.4909W}
who found that optically active and inactive LeMMINGs galaxies
together form single $L_\textnormal{X}-\rm M_{\rm BH}$ and
$L_\textnormal{X}-L_{[\ion{O}{iii}]}$ relations across the entire
galaxy mass and luminosity ranges.

Finally, we investigate whether nuclei are more strongly correlated
with $L_\textnormal{X}$ or $L_{\rm R,core}$, and find that
$M_{*,\rm nuc}$ and $M_{V,\rm nuc}$ are similarly correlated with both
$L_\textnormal{X}$ and $L_{\rm R,core}$.  However, nuclei are more
closely associated with $L_\textnormal{X}$ than $L_{\rm R,core}$, with
respect to the rms scatters in the log $M_{\rm *,nuc}$ direction
($\updelta_{\rm horiz}$), as the $L_\textnormal{X}$ relations have
$\updelta_{\rm horiz} \sim 0.80-0.83$ dex, which is smaller than those
for the $L_{\rm R,core}$ relations
($\updelta_{\rm horiz} \sim 0.92-0.96$ dex), see Table~\ref{Tab5} and
Appendix \ref{AppA2}.

 \section{Summary and Conclusions }\label{Sec5}
 
 The $e$-MERLIN legacy survey (LeMMINGs,
 \citealt{2014evn..confE..10B}) is the deepest, high-resolution radio
 continuum survey of a statistically complete, sample of 280 nearby
 galaxies with $e$-MERLIN at 1.25$-$1.75 and 5 GHz.  We have presented
 a study of the co-existence of nuclear star clusters (NSCs) and AGN
 in 100 nearby, nucleated LeMMINGs galaxies, including 10 elliptical,
 25 lenticular, 63 spiral, and 2 irregular galaxies.  In
 \citet{2023A&A...675A.105D}, we studied photometric and structural
 properties of nearby galaxies by selecting all LeMMINGs galaxies with
 reliable {\it HST} imaging data, yielding a subsample of 173
 galaxies, which were shown to be representative of the parent
 LeMMINGs sample.  To identify the nuclei, we used high-resolution
 optical and near-infrared {\it HST} images and performed accurate 1D
 and 2D multi-component decompositions of the stellar light
 distributions of the 173 LeMMINGs galaxies (into \ bulges, discs,
 depleted core, AGN, nuclear star clusters, bars, spiral arms, rings
 and stellar haloes) using S\'ersic and core-S\'ersic models. From
 this analysis, we identified 100 nucleated galaxies that fulfilled
 our criteria for hosting a dense star cluster, a `pure' AGN or a
 combination of both (Section~\ref{Sec2.3}). We have derived
 luminosities, stellar masses and half-light radii for the nuclei.
 
 Our detailed multi-component decompositions of 65 LT + 35 ET galaxies
 are improvements over the past literature, particularly, since past
 studies often fitted a two-component, nucleus+bulge model to the
 light distributions nucleated galaxies.  Furthermore, there are only
 a few published photometric and structural studies on the nuclei of
 late-type galaxies with {\it HST}
 \citep{2002AJ....123..159C,2002AJ....123.1389B,2014MNRAS.441.3570G,2023A&A...675A.105D},
 see Table~\ref{Tab4}. Other improvements include our AGN diagnostics,
 where we combine the {\it HST } decomposition data with homogeneously
 derived, nuclear 1.5 GHz \emerlin\ radio, {\it Chandra} X-ray (0.3 --
 10 keV) and optical emission-line data from LeMMINGs
 \citep{2018MNRAS.476.3478B,2021MNRAS.500.4749B,2021MNRAS.508.2019B,2022MNRAS.510.4909W}. The
 multi-wavelength data used in this work span wide ranges in galaxy
 stellar mass and nuclear radio and X-ray luminosities:
 \mbox{$M_{*,\rm gal} \sim 10^{8.7}-10^{12}~\rm M_{\sun}$},
 $L_{\rm R, \rm core}\sim 10^{32}-10^{40}$ erg s$^{-1}$ and
 $L_\textnormal{X} \sim 10^{35}-10^{43}$ erg s$^{-1}$. To our
 knowledge, this work represents the most comprehensive
 multi-wavelength investigation to date of the connection between NSCs
 and AGN in nearby galaxies.

Our principal conclusions are:

(1) We define the nucleation fraction ($f_{\rm nuc} $) as the relative
incidence of nuclei in the sample under consideration. Among the
sample of LeMMINGs galaxies with {\it HST} data and excluding the 24
questionable galaxies (see Section~ \ref{Sec2.3}), we find
$f_{\rm nuc}=$100/149 (= 67 $\pm$ 7 per cent), in agreement with
previous reports.  When we treat the 24 excluded galaxies as
non-nucleated then $f_{\rm nuc}=$100/173 (=58 $\pm$ 6 per cent) for
the full LeMMINGs {\it HST} sample.  We also find that $f_{\rm nuc}$
increases with increasing $M_{*,\rm bulge}$ and $M_{*,\rm gal}$,
peaking at bulge masses
\mbox{$M_{*,\rm bulge}\sim 10^{9.4}- 10^{10.8}~ \rm M_{\sun}$}
(\mbox{$f_{\rm nuc} = 49/56~=88 \pm 13$ per cent}), before decreasing
at higher stellar masses.  Core-S\'ersic galaxies exhibit a low
incidence of nucleation, \mbox{$f_{\rm nuc}$ = (10 -- 20) $ \pm$ 6 per
  cent}. This figure supports the widely accepted hypothesis of
core-S\'ersic galaxy formation, where coalescing binary SMBHs not only
scour the cores but also tidally disrupt any NSCs.  In contrast,
S\'ersic galaxies show a much higher nucleation frequency
($f_{\rm nuc} = 98/129= 76 \pm 8$ per cent).
  
(2) We classify 43 nuclei in the sample as  `pure' nuclear star clusters (NSCs),
which are optically inactive. We argue that the bulk of the central
light excesses in the optical or near-IR brightness profiles of these
nuclei are of stellar origin.  In contrast, 57 nuclei are optically
active and classified as `hybrid' (NSC+AGN).  We suggest that the
optical or near-IR flux of such nuclei likely results from
contributions of both stellar and non-stellar emissions.

(3)  In order to determine the lower
limit for the hybrid nucleus fraction in our sample, we employ AGN
diagnostics that utilise a multi-band signature. This method combines
homogeneously derived, high fidelity \emerlin\ radio, {\it Chandra}
X-ray and optical emission-line data.  We identify 30 nucleated
galaxies in our nucleated sample that are optically active, radio
detected and X-ray luminous ($L_\textnormal{X} > 10^{39}$ erg
s$^{-1}$). We conclude that our sample has a lower limit frequency for
hybrid nuclei of $\sim$ 30 per cent. This figure is at least a factor
of three higher than that previously reported
\citep[e.g.][]{2008ApJ...678..116S,2017ApJ...841...51F}. These
previous works are based on a literature compilation of nuclei
identified in {\it HST} imaging data.  In contrast, our nuclei are
identified uniformly using detailed 1D and 2D decompositions of the
host galaxy {\it HST} data and multi-wavelength data sets from
LeMMINGs. We measure a lower limit number density of
$(1.5 \pm 0.4)\times 10^{-5}$ Mpc$^{-3}$ for hybrid nuclei. The
fraction of galaxies hosting hybrid nuclei increases with
$M_{*,\rm bulge}$ and $M_{*,\rm gal}$, peaking at intermediate masses
($M_{*,\rm bulge}/M_{*,\rm gal} \sim 10^{10.0}- 10^{11.4}~\rm
M_{\sun}$/$10^{10.6}$ -- $10^{11.8}~\rm M_{\sun}$), before declining
at higher stellar masses.   For our {\it HST} sample of 149 galaxies, 
comprising   both nucleated and non-nucleated types, we find that low mass galaxies with
$M_{*,\rm gal} \la 10^{10.6}~\rm M_{\sun}$ exhibit a similar
occurrence of hybrid nuclei and optical AGN activity. However, above
this mass (i.e.\ $M_{*,\rm gal} \ga 10^{10.6}~\rm M_{\sun}$) galaxies
are more likely to exhibit evidence of optical and radio AGN activities rather
than hosting hybrid nuclei.

(4) Comparing our nucleated and the full LeMMINGs {\it HST} sample and
not controlling for bulge/galaxy mass, we find a slight increase in
the Seyfert and LINER fractions and a slight decrease in ALG and
\mbox{H\,{\sc ii}} fractions for our sample containing only nucleated
galaxies.  Our nucleated sample also has a slightly higher
radio-detection rate than our full {\it HST} sample, but the two
samples are statistically indistinguishable in terms of 
X-ray-detection rate.  Our findings suggest that NSCs may enhance
accretion on to the central SMBH, by funnelling gas toward the
innermost regions \citep[e.g.][]{2015ApJ...803...81N}.
Nevertheless, we caution
that further investigation using a larger sample of galaxies is needed
to draw firm conclusions, given the  large Poisson errors associated with these trends.

(5) Scaling relations involving carefully acquired masses/luminosities
of nuclei, observed in the broadband optical and near-IR {\it HST}
images, and the radio and X-ray core luminosities are provided in
Section~\ref{Sec4}.  Our investigation of radio core luminosities
suggests that NSCs and hybrid nuclei follow common relations in the
\mbox{$L_{\rm R,core}-M_{*, \rm nuc}$},
\mbox{$L_{\rm R,core}-M_{V,\rm nuc}$} diagrams. However, we suspect
that the apparent unity among all types of nuclei in $L_{\rm R,core}$ scaling diagrams
might be  driven by the relatively low incidence of nucleation
in low-mass late-type galaxies and massive early-type galaxies
(Figs.~\ref{FigF5a}--\ref{FigF8} and Section~\ref{Sec3.1}).  Once large
numbers of low-mass late-type galaxies are included, we argue that
hybrid nuclei, akin to active bulges, define $L_{\rm R,core}$ scaling
relations with slopes steeper than those for the `pure' NSCs (inactive
galaxies). Furthermore, our analysis reveals that NSCs and
hybrid nuclei collectively follow single unbroken (log-linear)
$L_\textnormal{X}-M_{*, \rm nuc}$ and
$L_\textnormal{X}-M_{V,\rm nuc}$ relations.

\section{ACKNOWLEDGMENTS}
We thank the referee for a timely and supportive report that improved
this manuscript.  AGdP acknowledges financial support under grant
No. PID2022-138621NB-I00.  AA acknowledges financial support from the
Spanish Ministerio de Ciencia e Innovaci\'on (grant
PID2020-117404GB-C21) and from the grant CEX2021-001131-S funded by
MCIN/AEI/ 10.13039/50110001103. JHK's research is co-funded by the
European Union. Views and opinions expressed are however those of the
author(s) only and do not necessarily reflect those of the European
Union. Neither the European Union nor the granting authority can be
held responsible for them. JHK acknowledges support from the Agencia
Estatal de Investigaci\'on del Ministerio de Ciencia, Innovaci\'on y
Universidades (MCIU/AEI) under the grant "The structure and evolution
of galaxies and their outer regions" and the European Regional
Development Fund (ERDF) with reference
PID2022-136505NB-I00/10.13039/501100011033.  Based on observations
made with the NASA/ESA {\it Hubble Space Telescope}, and obtained from
the Hubble Legacy Archive, which is a collaboration between the Space
Telescope Science Institute (STScI/NASA), the Space Telescope European
Coordinating Facility (ST-ECF/ESA) and the Canadian Astronomy Data
Centre (CADC/NRC/CSA).  We would like to acknowledge the support from
the $e$-MERLIN Legacy project `LeMMINGs', upon which this study is
based.  $e$-MERLIN, and formerly, MERLIN, is a National Facility
operated by the University of Manchester at Jodrell Bank Observatory
on behalf of the STFC. We acknowledge Jodrell Bank Centre for
Astrophysics, which is funded by the STFC.  The scientific results
reported in this article are based in part on observations made by the
{\it Chandra }X-ray Observatory and published previously in cited
articles. This research has made use of the NASA/IPAC Extragalactic
Database (NED), which is operated by the Jet Propulsion Laboratory,
California Institute of Technology, under contract with the National
Aeronautics and Space Administration.

This work has made use of {\sc numpy} \citep{2011CSE....13b..22V},
{\sc matplotlib} \citep{Hunter:2007}  and {\sc astropy}, a community-developed
core {\sc python} package for Astronomy
\citep{2013A&A...558A..33A,2018AJ....156..123A}, {\sc astroquery}
\citep{2019AJ....157...98G} {\sc cubehelix }
\citep{2011BASI...39..289G}, {\sc jupyter}
\citep{2016ppap.book...87K}, {\sc scipy} \citep{2020NatMe..17..261V},
and of {\sc topcat} (i.e.\ `Tool for Operations on Catalogues And
Tables', \citealt{2005ASPC..347...29T}).

\section{Data availability}

The data underlying this article are available in the article.

\bibliographystyle{mn2e}
\bibliography{Bil_Paps_biblo.bib}

\begin{thebibliography}{147}
\expandafter\ifx\csname natexlab\endcsname\relax\def\natexlab#1{#1}\fi

\bibitem[{{Akritas} \& {Bershady}(1996)}]{1996ApJ...470..706A}
{Akritas} M.~G., {Bershady} M.~A., 1996, \apj, 470, 706

\bibitem[{{Antonini}, {Barausse} \& {Silk}(2015){Antonini}, {Barausse}, \&
  {Silk}}]{2015ApJ...812...72A}
{Antonini} F., {Barausse} E., {Silk} J., 2015, \apj, 812, 72

\bibitem[{{Antonini}, {Gieles} \& {Gualandris}(2019){Antonini}, {Gieles}, \&
  {Gualandris}}]{2019MNRAS.486.5008A}
{Antonini} F., {Gieles} M., {Gualandris} A., 2019, \mnras, 486, 5008

\bibitem[{{Askar}, {Davies} \& {Church}(2021){Askar}, {Davies}, \&
  {Church}}]{2021MNRAS.502.2682A}
{Askar} A., {Davies} M.~B., {Church} R.~P., 2021, \mnras, 502, 2682

\bibitem[{{Astropy Collaboration} {et~al}\mbox{.}(2018){Astropy Collaboration},
  {Price-Whelan}, {Sip{\H{o}}cz}, {G{\"u}nther}, {Lim}, {Crawford}, {Conseil},
  {Shupe}, {Craig}, {Dencheva}, {Ginsburg}, {VanderPlas}, {Bradley},
  {P{\'e}rez-Su{\'a}rez}, {de Val-Borro}, {Aldcroft}, {Cruz}, {Robitaille},
  {Tollerud}, {Ardelean}, {Babej}, {Bach}, {Bachetti}, {Bakanov}, {Bamford},
  {Barentsen}, {Barmby}, {Baumbach}, {Berry}, {Biscani}, {Boquien}, {Bostroem},
  {Bouma}, {Brammer}, {Bray}, {Breytenbach}, {Buddelmeijer}, {Burke},
  {Calderone}, {Cano Rodr{\'\i}guez}, {Cara}, {Cardoso}, {Cheedella}, {Copin},
  {Corrales}, {Crichton}, {D'Avella}, {Deil}, {Depagne}, {Dietrich}, {Donath},
  {Droettboom}, {Earl}, {Erben}, {Fabbro}, {Ferreira}, {Finethy}, {Fox},
  {Garrison}, {Gibbons}, {Goldstein}, {Gommers}, {Greco}, {Greenfield},
  {Groener}, {Grollier}, {Hagen}, {Hirst}, {Homeier}, {Horton}, {Hosseinzadeh},
  {Hu}, {Hunkeler}, {Ivezi{\'c}}, {Jain}, {Jenness}, {Kanarek}, {Kendrew},
  {Kern}, {Kerzendorf}, {Khvalko}, {King}, {Kirkby}, {Kulkarni}, {Kumar},
  {Lee}, {Lenz}, {Littlefair}, {Ma}, {Macleod}, {Mastropietro}, {McCully},
  {Montagnac}, {Morris}, {Mueller}, {Mumford}, {Muna}, {Murphy}, {Nelson},
  {Nguyen}, {Ninan}, {N{\"o}the}, {Ogaz}, {Oh}, {Parejko}, {Parley}, {Pascual},
  {Patil}, {Patil}, {Plunkett}, {Prochaska}, {Rastogi}, {Reddy Janga},
  {Sabater}, {Sakurikar}, {Seifert}, {Sherbert}, {Sherwood-Taylor}, {Shih},
  {Sick}, {Silbiger}, {Singanamalla}, {Singer}, {Sladen}, {Sooley},
  {Sornarajah}, {Streicher}, {Teuben}, {Thomas}, {Tremblay}, {Turner},
  {Terr{\'o}n}, {van Kerkwijk}, {de la Vega}, {Watkins}, {Weaver}, {Whitmore},
  {Woillez}, {Zabalza}, \& {Astropy Contributors}}]{2018AJ....156..123A}
{Astropy Collaboration} {et~al.}, 2018, \aj, 156, 123

\bibitem[{{Astropy Collaboration} {et~al}\mbox{.}(2013){Astropy Collaboration},
  {Robitaille}, {Tollerud}, {Greenfield}, {Droettboom}, {Bray}, {Aldcroft},
  {Davis}, {Ginsburg}, {Price-Whelan}, {Kerzendorf}, {Conley}, {Crighton},
  {Barbary}, {Muna}, {Ferguson}, {Grollier}, {Parikh}, {Nair}, {Unther},
  {Deil}, {Woillez}, {Conseil}, {Kramer}, {Turner}, {Singer}, {Fox}, {Weaver},
  {Zabalza}, {Edwards}, {Azalee Bostroem}, {Burke}, {Casey}, {Crawford},
  {Dencheva}, {Ely}, {Jenness}, {Labrie}, {Lim}, {Pierfederici}, {Pontzen},
  {Ptak}, {Refsdal}, {Servillat}, \& {Streicher}}]{2013A&A...558A..33A}
{Astropy Collaboration} {et~al.}, 2013, \aap, 558, A33

\bibitem[{{Balcells} {et~al}\mbox{.}(2003){Balcells}, {Graham},
  {Dom{\'\i}nguez-Palmero}, \& {Peletier}}]{2003ApJ...582L..79B}
{Balcells} M., {Graham} A.~W., {Dom{\'\i}nguez-Palmero} L., {Peletier} R.~F.,
  2003, \apjl, 582, L79

\bibitem[{{Balcells}, {Graham} \& {Peletier}(2007){Balcells}, {Graham}, \&
  {Peletier}}]{2007ApJ...665.1084B}
{Balcells} M., {Graham} A.~W., {Peletier} R.~F., 2007, \apj, 665, 1084

\bibitem[{{Baldassare} {et~al}\mbox{.}(2017){Baldassare}, {Reines}, {Gallo}, \&
  {Greene}}]{2017ApJ...850..196B}
{Baldassare} V.~F., {Reines} A.~E., {Gallo} E., {Greene} J.~E., 2017, \apj,
  850, 196

\bibitem[{{Baldassare} {et~al}\mbox{.}(2022){Baldassare}, {Stone}, {Foord},
  {Gallo}, \& {Ostriker}}]{2022ApJ...929...84B}
{Baldassare} V.~F., {Stone} N.~C., {Foord} A., {Gallo} E., {Ostriker} J.~P.,
  2022, \apj, 929, 84

\bibitem[{{Baldi} {et~al}\mbox{.}(2010){Baldi}, {Chiaberge}, {Capetti},
  {Sparks}, {Macchetto}, {O'Dea}, {Axon}, {Baum}, \&
  {Quillen}}]{2010ApJ...725.2426B}
{Baldi} R.~D. {et~al.}, 2010, \apj, 725, 2426

\bibitem[{{Baldi} {et~al}\mbox{.}(2018){Baldi}, {Williams}, {McHardy},
  {Beswick}, {Argo}, {Dullo}, {Knapen}, {Brinks}, {Muxlow}, {Aalto}, {Alberdi},
  {Bendo}, {Corbel}, {Evans}, {Fenech}, {Green}, {Kl{\"o}ckner}, {K{\"o}rding},
  {Kharb}, {Maccarone}, {Mart{\'\i}-Vidal}, {Mundell}, {Panessa}, {Peck},
  {P{\'e}rez-Torres}, {Saikia}, {Saikia}, {Shankar}, {Spencer}, {Stevens},
  {Uttley}, \& {Westcott}}]{2018MNRAS.476.3478B}
{Baldi} R.~D. {et~al.}, 2018, \mnras, 476, 3478

\bibitem[{{Baldi} {et~al}\mbox{.}(2021{\natexlab{a}}){Baldi}, {Williams},
  {McHardy}, {Beswick}, {Argo}, {Dullo}, {Knapen}, {Brinks}, {Muxlow}, {Aalto},
  {Alberdi}, {Bendo}, {Corbel}, {Evans}, {Fenech}, {Green}, {Kl{\"o}ckner},
  {K{\"o}rding}, {Kharb}, {Maccarone}, {Mart{\'\i}-Vidal}, {Mundell},
  {Panessa}, {Peck}, {P{\'e}rez-Torres}, {Saikia}, {Saikia}, {Shankar},
  {Spencer}, {Stevens}, {Uttley}, \& {Westcott}}]{2021MNRAS.500.4749B}
{Baldi} R.~D. {et~al.}, 2021{\natexlab{a}}, \mnras, 500, 4749

\bibitem[{{Baldi} {et~al}\mbox{.}(2021{\natexlab{b}}){Baldi}, {Williams},
  {McHardy}, {Beswick}, {Argo}, {Dullo}, {Knapen}, {Brinks}, {Muxlow}, {Aalto},
  {Alberdi}, {Bendo}, {Corbel}, {Evans}, {Fenech}, {Green}, {Kl{\"o}ckner},
  {K{\"o}rding}, {Kharb}, {Maccarone}, {Mart{\'\i}-Vidal}, {Mundell},
  {Panessa}, {Peck}, {P{\'e}rez-Torres}, {Saikia}, {Saikia}, {Shankar},
  {Spencer}, {Stevens}, {Uttley}, \& {Westcott}}]{2021MNRAS.508.2019B}
{Baldi} R.~D. {et~al.}, 2021{\natexlab{b}}, \mnras, 508, 2019

\bibitem[{{Balmaverde} \& {Capetti}(2006)}]{2006A&A...447...97B}
{Balmaverde} B., {Capetti} A., 2006, \aap, 447, 97

\bibitem[{{Balmaverde}, {Capetti} \& {Grandi}(2006){Balmaverde}, {Capetti}, \&
  {Grandi}}]{2006A&A...451...35B}
{Balmaverde} B., {Capetti} A., {Grandi} P., 2006, \aap, 451, 35

\bibitem[{{Barth} {et~al}\mbox{.}(2004){Barth}, {Ho}, {Rutledge}, \&
  {Sargent}}]{2004ApJ...607...90B}
{Barth} A.~J., {Ho} L.~C., {Rutledge} R.~E., {Sargent} W. L.~W., 2004, \apj,
  607, 90

\bibitem[{{Becker}, {White} \& {Helfand}(1995){Becker}, {White}, \&
  {Helfand}}]{1995ApJ...450..559B}
{Becker} R.~H., {White} R.~L., {Helfand} D.~J., 1995, \apj, 450, 559

\bibitem[{{Bekki} \& {Graham}(2010)}]{2010ApJ...714L.313B}
{Bekki} K., {Graham} A.~W., 2010, \apjl, 714, L313

\bibitem[{{Beswick} {et~al}\mbox{.}(2014){Beswick}, {Argo}, {Evans}, {McHardy},
  {Williams}, \& {Westcott}}]{2014evn..confE..10B}
{Beswick} R., {Argo} M.~K., {Evans} R., {McHardy} I., {Williams} D.~R.~A.,
  {Westcott} J., 2014, in Proceedings of the 12th European VLBI Network
  Symposium and Users Meeting (EVN 2014). 7-10 October 2014. Cagliari, p.~10

\bibitem[{{Birchall}, {Watson} \& {Aird}(2020){Birchall}, {Watson}, \&
  {Aird}}]{2020MNRAS.492.2268B}
{Birchall} K.~L., {Watson} M.~G., {Aird} J., 2020, \mnras, 492, 2268

\bibitem[{{B{\"o}ker} {et~al}\mbox{.}(2002){B{\"o}ker}, {Laine}, {van der
  Marel}, {Sarzi}, {Rix}, {Ho}, \& {Shields}}]{2002AJ....123.1389B}
{B{\"o}ker} T., {Laine} S., {van der Marel} R.~P., {Sarzi} M., {Rix} H.-W.,
  {Ho} L.~C., {Shields} J.~C., 2002, \aj, 123, 1389

\bibitem[{{B{\"o}ker} {et~al}\mbox{.}(2004){B{\"o}ker}, {Sarzi}, {McLaughlin},
  {van der Marel}, {Rix}, {Ho}, \& {Shields}}]{2004AJ....127..105B}
{B{\"o}ker} T., {Sarzi} M., {McLaughlin} D.~E., {van der Marel} R.~P., {Rix}
  H.-W., {Ho} L.~C., {Shields} J.~C., 2004, \aj, 127, 105

\bibitem[{{Buttiglione} {et~al}\mbox{.}(2010){Buttiglione}, {Capetti},
  {Celotti}, {Axon}, {Chiaberge}, {Macchetto}, \&
  {Sparks}}]{2010A&A...509A...6B}
{Buttiglione} S., {Capetti} A., {Celotti} A., {Axon} D.~J., {Chiaberge} M.,
  {Macchetto} F.~D., {Sparks} W.~B., 2010, \aap, 509, A6

\bibitem[{{Carollo} {et~al}\mbox{.}(1997){Carollo}, {Stiavelli}, {de Zeeuw}, \&
  {Mack}}]{1997AJ....114.2366C}
{Carollo} C.~M., {Stiavelli} M., {de Zeeuw} P.~T., {Mack} J., 1997, \aj, 114,
  2366

\bibitem[{{Carollo}, {Stiavelli} \& {Mack}(1998){Carollo}, {Stiavelli}, \&
  {Mack}}]{1998AJ....116...68C}
{Carollo} C.~M., {Stiavelli} M., {Mack} J., 1998, \aj, 116, 68

\bibitem[{{Carollo} {et~al}\mbox{.}(2002){Carollo}, {Stiavelli}, {Seigar}, {de
  Zeeuw}, \& {Dejonghe}}]{2002AJ....123..159C}
{Carollo} C.~M., {Stiavelli} M., {Seigar} M., {de Zeeuw} P.~T., {Dejonghe} H.,
  2002, \aj, 123, 159

\bibitem[{{Chang} {et~al}\mbox{.}(2007){Chang}, {Murray-Clay}, {Chiang}, \&
  {Quataert}}]{2007ApJ...668..236C}
{Chang} P., {Murray-Clay} R., {Chiang} E., {Quataert} E., 2007, \apj, 668, 236

\bibitem[{{C{\^o}t{\'e}} {et~al}\mbox{.}(2006){C{\^o}t{\'e}}, {Piatek},
  {Ferrarese}, {Jord{\'a}n}, {Merritt}, {Peng}, {Ha{\c{s}}egan}, {Blakeslee},
  {Mei}, {West}, {Milosavljevi{\'c}}, \& {Tonry}}]{2006ApJS..165...57C}
{C{\^o}t{\'e}} P. {et~al.}, 2006, \apjs, 165, 57

\bibitem[{{de Vaucouleurs}(1959)}]{1959HDP....53..275D}
{de Vaucouleurs} G., 1959, Handbuch der Physik, 53, 275

\bibitem[{{de Vaucouleurs}, {de Vaucouleurs} \& {Corwin}(1976){de Vaucouleurs},
  {de Vaucouleurs}, \& {Corwin}}]{1976srcb.book.....D}
{de Vaucouleurs} G., {de Vaucouleurs} A., {Corwin}, H.~G. J., 1976, {Second
  reference catalogue of bright galaxies. Containing information on 4,364
  galaxies with references to papers published between 1964 and 1975.}

\bibitem[{{de Vaucouleurs} {et~al}\mbox{.}(1991){de Vaucouleurs}, {de
  Vaucouleurs}, {Corwin}, {Buta}, {Paturel}, \&
  {Fouqu{\'e}}}]{1991rc3..book.....D}
{de Vaucouleurs} G., {de Vaucouleurs} A., {Corwin}, H.~G. J., {Buta} R.~J.,
  {Paturel} G., {Fouqu{\'e}} P., 1991, {Third Reference Catalogue of Bright
  Galaxies. Volume I: Explanations and references. Volume II: Data for galaxies
  between 0$^{\rm h}$ and 12$^{\rm h}$. Volume III: Data for galaxies between
  12$^{\rm h}$ and 24$^{\rm h}$.}

\bibitem[{{den Brok} {et~al}\mbox{.}(2014){den Brok}, {Peletier}, {Seth},
  {Balcells}, {Dominguez}, {Graham}, {Carter}, {Erwin}, {Ferguson},
  {Goudfrooij}, {Guzm{\'a}n}, {Hoyos}, {Jogee}, {Lucey}, {Phillipps}, {Puzia},
  {Valentijn}, {Verdoes Kleijn}, \& {Weinzirl}}]{2014MNRAS.445.2385D}
{den Brok} M. {et~al.}, 2014, \mnras, 445, 2385

\bibitem[{{den Brok} {et~al}\mbox{.}(2015){den Brok}, {Seth}, {Barth},
  {Carson}, {Neumayer}, {Cappellari}, {Debattista}, {Ho}, {Hood}, \&
  {McDermid}}]{2015ApJ...809..101D}
{den Brok} M. {et~al.}, 2015, \apj, 809, 101

\bibitem[{{Driver} {et~al}\mbox{.}(2008){Driver}, {Popescu}, {Tuffs}, {Graham},
  {Liske}, \& {Baldry}}]{2008ApJ...678L.101D}
{Driver} S.~P., {Popescu} C.~C., {Tuffs} R.~J., {Graham} A.~W., {Liske} J.,
  {Baldry} I., 2008, \apjl, 678, L101

\bibitem[{{Dullo}(2019)}]{2019ApJ...886...80D}
{Dullo} B.~T., 2019, \apj, 886, 80

\bibitem[{{Dullo} {et~al}\mbox{.}(2019){Dullo}, {Chamorro-Cazorla}, {Gil de
  Paz}, {Castillo-Morales}, {Gallego}, {Carrasco}, {Iglesias-P{\'a}ramo},
  {Cedazo}, {Garc{\'{\i}}a-Vargas}, {Pascual}, {Cardiel}, {P{\'e}rez-Calpena},
  {G{\'o}mez-Alvarez}, {Mart{\'{\i}}nez-Delgado}, \&
  {Catal{\'a}n-Torrecilla}}]{2019ApJ...871....9D}
{Dullo} B.~T. {et~al.}, 2019, \apj, 871, 9

\bibitem[{{Dullo} \& {Graham}(2012)}]{2012ApJ...755..163D}
{Dullo} B.~T., {Graham} A.~W., 2012, \apj, 755, 163

\bibitem[{{Dullo} \& {Graham}(2013)}]{2013ApJ...768...36D}
{Dullo} B.~T., {Graham} A.~W., 2013, \apj, 768, 36

\bibitem[{{Dullo} \& {Graham}(2014)}]{2014MNRAS.444.2700D}
{Dullo} B.~T., {Graham} A.~W., 2014, \mnras, 444, 2700

\bibitem[{{Dullo}, {Graham} \& {Knapen}(2017){Dullo}, {Graham}, \&
  {Knapen}}]{2017MNRAS.471.2321D}
{Dullo} B.~T., {Graham} A.~W., {Knapen} J.~H., 2017, \mnras, 471, 2321

\bibitem[{{Dullo} {et~al}\mbox{.}(2023{\natexlab{a}}){Dullo}, {Knapen},
  {Beswick}, {Baldi}, {Williams}, {McHardy}, {Gallagher}, {Aalto}, {Argo}, {Gil
  de Paz}, {Kl{\"o}ckner}, {Marcaide}, {Mundell}, {Mutie}, \&
  {Saikia}}]{2023A&A...675A.105D}
{Dullo} B.~T. {et~al.}, 2023{\natexlab{a}}, \aap, 675, A105

\bibitem[{{Dullo} {et~al}\mbox{.}(2023{\natexlab{b}}){Dullo}, {Knapen},
  {Beswick}, {Baldi}, {Williams}, {McHardy}, {Green}, {Gil de Paz}, {Aalto},
  {Alberdi}, {Argo}, {Kl{\"o}ckner}, {Mutie}, {Saikia}, {Saikia}, \&
  {Stevens}}]{2023MNRAS.522.3412D}
{Dullo} B.~T. {et~al.}, 2023{\natexlab{b}}, \mnras, 522, 3412

\bibitem[{{Dullo} {et~al}\mbox{.}(2018){Dullo}, {Knapen}, {Williams},
  {Beswick}, {Bendo}, {Baldi}, {Argo}, {McHardy}, {Muxlow}, \&
  {Westcott}}]{2018MNRAS.475.4670D}
{Dullo} B.~T. {et~al.}, 2018, \mnras, 475, 4670

\bibitem[{{Dullo}, {Mart{\'{\i}}nez-Lombilla} \& {Knapen}(2016){Dullo},
  {Mart{\'{\i}}nez-Lombilla}, \& {Knapen}}]{2016MNRAS.462.3800D}
{Dullo} B.~T., {Mart{\'{\i}}nez-Lombilla} C., {Knapen} J.~H., 2016, \mnras,
  462, 3800

\bibitem[{{Erwin}(2015)}]{2015ApJ...799..226E}
{Erwin} P., 2015, \apj, 799, 226

\bibitem[{{Event Horizon Telescope Collaboration} {et~al}\mbox{.}(2022){Event
  Horizon Telescope Collaboration}, {Akiyama}, {Alberdi}, {Alef}, {Algaba},
  {Anantua}, {Asada}, {Azulay}, {Bach}, {Baczko}, {Ball}, {Balokovi{\'c}},
  {Barrett}, {Baub{\"o}ck}, {Benson}, {Bintley}, {Blackburn}, {Blundell},
  {Bouman}, {Bower}, {Boyce}, {Bremer}, {Brinkerink}, {Brissenden}, {Britzen},
  {Broderick}, {Broguiere}, {Bronzwaer}, {Bustamante}, {Byun}, {Carlstrom},
  {Ceccobello}, {Chael}, {Chan}, {Chatterjee}, {Chatterjee}, {Chen}, {Chen},
  {Cheng}, {Cho}, {Christian}, {Conroy}, {Conway}, {Cordes}, {Crawford},
  {Crew}, {Cruz-Osorio}, {Cui}, {Davelaar}, {De Laurentis}, {Deane}, {Dempsey},
  {Desvignes}, {Dexter}, {Dhruv}, {Doeleman}, {Dougal}, {Dzib}, {Eatough},
  {Emami}, {Falcke}, {Farah}, {Fish}, {Fomalont}, {Ford}, {Fraga-Encinas},
  {Freeman}, {Friberg}, {Fromm}, {Fuentes}, {Galison}, {Gammie}, {Garc{\'\i}a},
  {Gentaz}, {Georgiev}, {Goddi}, {Gold}, {G{\'o}mez-Ruiz}, {G{\'o}mez}, {Gu},
  {Gurwell}, {Hada}, {Haggard}, {Haworth}, {Hecht}, {Hesper}, {Heumann}, {Ho},
  {Ho}, {Honma}, {Huang}, {Huang}, {Hughes}, {Ikeda}, {Impellizzeri}, {Inoue},
  {Issaoun}, {James}, {Jannuzi}, {Janssen}, {Jeter}, {Jiang},
  {Jim{\'e}nez-Rosales}, {Johnson}, {Jorstad}, {Joshi}, {Jung}, {Karami},
  {Karuppusamy}, {Kawashima}, {Keating}, {Kettenis}, {Kim}, {Kim}, {Kim},
  {Kim}, {Kino}, {Koay}, {Kocherlakota}, {Kofuji}, {Koch}, {Koyama}, {Kramer},
  {Kramer}, {Krichbaum}, {Kuo}, {La Bella}, {Lauer}, {Lee}, {Lee}, {Leung},
  {Levis}, {Li}, {Lico}, {Lindahl}, {Lindqvist}, {Lisakov}, {Liu}, {Liu},
  {Liuzzo}, {Lo}, {Lobanov}, {Loinard}, {Lonsdale}, {Lu}, {Mao}, {Marchili},
  {Markoff}, {Marrone}, {Marscher}, {Mart{\'\i}-Vidal}, {Matsushita},
  {Matthews}, {Medeiros}, {Menten}, {Michalik}, {Mizuno}, {Mizuno}, {Moran},
  {Moriyama}, {Moscibrodzka}, {M{\"u}ller}, {Mus}, {Musoke}, {Myserlis},
  {Nadolski}, {Nagai}, {Nagar}, {Nakamura}, {Narayan}, {Narayanan},
  {Natarajan}, {Nathanail}, {Fuentes}, {Neilsen}, {Neri}, {Ni}, {Noutsos},
  {Nowak}, {Oh}, {Okino}, {Olivares}, {Ortiz-Le{\'o}n}, {Oyama}, {{\"O}zel},
  {Palumbo}, {Paraschos}, {Park}, {Parsons}, {Patel}, {Pen}, {Pesce},
  {Pi{\'e}tu}, {Plambeck}, {PopStefanija}, {Porth}, {P{\"o}tzl}, {Prather},
  {Preciado-L{\'o}pez}, {Psaltis}, {Pu}, {Ramakrishnan}, {Rao}, {Rawlings},
  {Raymond}, {Rezzolla}, {Ricarte}, {Ripperda}, {Roelofs}, {Rogers}, {Ros},
  {Romero-Ca{\~n}izales}, {Roshanineshat}, {Rottmann}, {Roy}, {Ruiz},
  {Ruszczyk}, {Rygl}, {S{\'a}nchez}, {S{\'a}nchez-Arg{\"u}elles},
  {S{\'a}nchez-Portal}, {Sasada}, {Satapathy}, {Savolainen}, {Schloerb},
  {Schonfeld}, {Schuster}, {Shao}, {Shen}, {Small}, {Sohn}, {SooHoo},
  {Souccar}, {Sun}, {Tazaki}, {Tetarenko}, {Tiede}, {Tilanus}, {Titus},
  {Torne}, {Traianou}, {Trent}, {Trippe}, {Turk}, {van Bemmel}, {van
  Langevelde}, {van Rossum}, {Vos}, {Wagner}, {Ward-Thompson}, {Wardle},
  {Weintroub}, {Wex}, {Wharton}, {Wielgus}, {Wiik}, {Witzel}, {Wondrak},
  {Wong}, {Wu}, {Yamaguchi}, {Yoon}, {Young}, {Young}, {Younsi}, {Yuan},
  {Yuan}, {Zensus}, {Zhang}, {Zhao}, {Zhao}, {Agurto}, {Allardi}, {Amestica},
  {Araneda}, {Arriagada}, {Berghuis}, {Bertarini}, {Berthold}, {Blanchard},
  {Brown}, {C{\'a}rdenas}, {Cantzler}, {Caro}, {Castillo-Dom{\'\i}nguez},
  {Chan}, {Chang}, {Chang}, {Chang}, {Chang}, {Chen}, {Chilson}, {Chuter},
  {Ciechanowicz}, {Colin-Beltran}, {Coulson}, {Crowley}, {Degenaar},
  {Dornbusch}, {Dur{\'a}n}, {Everett}, {Faber}, {Forster}, {Fuchs}, {Gale},
  {Geertsema}, {Gonz{\'a}lez}, {Graham}, {Gueth}, {Halverson}, {Han}, {Han},
  {Hasegawa}, {Hern{\'a}ndez-Rebollar}, {Herrera}, {Herrero-Illana},
  {Heyminck}, {Hirota}, {Hoge}, {Hostler Schimpf}, {Howie}, {Huang}, {Jiang},
  {Jinchi}, {John}, {Kimura}, {Klein}, {Kubo}, {Kuroda}, {Kwon}, {Lacasse},
  {Laing}, {Leitch}, {Li}, {Liu}, {Liu}, {Lin}, {Lu}, {Mac-Auliffe},
  {Martin-Cocher}, {Matulonis}, {Maute}, {Messias}, {Meyer-Zhao},
  {Monta{\~n}a}, {Montenegro-Montes}, {Montgomerie}, {Moreno Nolasco},
  {Muders}, {Nishioka}, {Norton}, {Nystrom}, {Ogawa}, {Olivares}, {Oshiro},
  {P{\'e}rez-Beaupuits}, {Parra}, {Phillips}, {Poirier}, {Pradel}, {Qiu},
  {Raffin}, {Rahlin}, {Ram{\'\i}rez}, {Ressler}, {Reynolds},
  {Rodr{\'\i}guez-Montoya}, {Saez-Madain}, {Santana}, {Shaw}, {Shirkey},
  {Silva}, {Snow}, {Sousa}, {Sridharan}, {Stahm}, {Stark}, {Test},
  {Torstensson}, {Venegas}, {Walther}, {Wei}, {White}, {Wieching}, {Wijnands},
  {Wouterloot}, {Yu}, {Yu (于威)}, \& {Zeballos}}]{2022ApJ...930L..12E}
{Event Horizon Telescope Collaboration} {et~al.}, 2022, \apjl, 930, L12

\bibitem[{{Feigelson} \& {Nelson}(1985)}]{1985ApJ...293..192F}
{Feigelson} E.~D., {Nelson} P.~I., 1985, \apj, 293, 192

\bibitem[{{Feldmeier} {et~al}\mbox{.}(2014){Feldmeier}, {Neumayer}, {Seth},
  {Sch{\"o}del}, {L{\"u}tzgendorf}, {de Zeeuw}, {Kissler-Patig}, {Nishiyama},
  \& {Walcher}}]{2014A&A...570A...2F}
{Feldmeier} A. {et~al.}, 2014, \aap, 570, A2

\bibitem[{{Ferrarese} {et~al}\mbox{.}(2006{\natexlab{a}}){Ferrarese},
  {C{\^o}t{\'e}}, {Dalla Bont{\`a}}, {Peng}, {Merritt}, {Jord{\'a}n},
  {Blakeslee}, {Ha{\c{s}}egan}, {Mei}, {Piatek}, {Tonry}, \&
  {West}}]{2006ApJ...644L..21F}
{Ferrarese} L. {et~al.}, 2006{\natexlab{a}}, \apjl, 644, L21

\bibitem[{{Ferrarese} {et~al}\mbox{.}(2006{\natexlab{b}}){Ferrarese},
  {C{\^o}t{\'e}}, {Jord{\'a}n}, {Peng}, {Blakeslee}, {Piatek}, {Mei},
  {Merritt}, {Milosavljevi{\'c}}, {Tonry}, \& {West}}]{2006ApJS..164..334F}
{Ferrarese} L. {et~al.}, 2006{\natexlab{b}}, \apjs, 164, 334

\bibitem[{{Ferrarese} {et~al}\mbox{.}(2020){Ferrarese}, {C{\^o}t{\'e}},
  {MacArthur}, {Durrell}, {Gwyn}, {Duc}, {S{\'a}nchez-Janssen}, {Santos},
  {Blakeslee}, {Boselli}, {Boyer}, {Cantiello}, {Courteau}, {Cuillandre},
  {Emsellem}, {Erben}, {Gavazzi}, {Guhathakurta}, {Huertas-Company},
  {Jord{\'a}n}, {Lan{\c{c}}on}, {Liu}, {Mei}, {Mihos}, {Peng}, {Puzia},
  {Roediger}, {Schade}, {Taylor}, {Toloba}, \& {Zhang}}]{2020ApJ...890..128F}
{Ferrarese} L. {et~al.}, 2020, \apj, 890, 128

\bibitem[{{Ferrarese} \& {Ford}(2005)}]{2005SSRv..116..523F}
{Ferrarese} L., {Ford} H., 2005, \ssr, 116, 523

\bibitem[{{Filippenko} \& {Ho}(2003)}]{2003ApJ...588L..13F}
{Filippenko} A.~V., {Ho} L.~C., 2003, \apjl, 588, L13

\bibitem[{{Filippenko} \& {Sargent}(1985)}]{1985ApJS...57..503F}
{Filippenko} A.~V., {Sargent} W.~L.~W., 1985, \apjs, 57, 503

\bibitem[{{Filippenko} \& {Sargent}(1989)}]{1989ApJ...342L..11F}
{Filippenko} A.~V., {Sargent} W. L.~W., 1989, \apjl, 342, L11

\bibitem[{{Foord} {et~al}\mbox{.}(2017){Foord}, {Gallo}, {Hodges-Kluck},
  {Miller}, {Baldassare}, {G{\"u}ltekin}, \& {Gnedin}}]{2017ApJ...841...51F}
{Foord} A., {Gallo} E., {Hodges-Kluck} E., {Miller} B.~P., {Baldassare} V.~F.,
  {G{\"u}ltekin} K., {Gnedin} O.~Y., 2017, \apj, 841, 51

\bibitem[{{Freedman} {et~al}\mbox{.}(2019){Freedman}, {Madore}, {Hatt}, {Hoyt},
  {Jang}, {Beaton}, {Burns}, {Lee}, {Monson}, {Neeley}, {Phillips}, {Rich}, \&
  {Seibert}}]{2019ApJ...882...34F}
{Freedman} W.~L. {et~al.}, 2019, \apj, 882, 34

\bibitem[{{Gallo} {et~al}\mbox{.}(2010){Gallo}, {Treu}, {Marshall}, {Woo},
  {Leipski}, \& {Antonucci}}]{2010ApJ...714...25G}
{Gallo} E., {Treu} T., {Marshall} P.~J., {Woo} J.-H., {Leipski} C., {Antonucci}
  R., 2010, \apj, 714, 25

\bibitem[{{Georgiev} \& {B{\"o}ker}(2014)}]{2014MNRAS.441.3570G}
{Georgiev} I.~Y., {B{\"o}ker} T., 2014, \mnras, 441, 3570

\bibitem[{{Georgiev} {et~al}\mbox{.}(2016){Georgiev}, {B{\"o}ker}, {Leigh},
  {L{\"u}tzgendorf}, \& {Neumayer}}]{2016MNRAS.457.2122G}
{Georgiev} I.~Y., {B{\"o}ker} T., {Leigh} N., {L{\"u}tzgendorf} N., {Neumayer}
  N., 2016, \mnras, 457, 2122

\bibitem[{{Ghez} {et~al}\mbox{.}(1998){Ghez}, {Klein}, {Morris}, \&
  {Becklin}}]{1998ApJ...509..678G}
{Ghez} A.~M., {Klein} B.~L., {Morris} M., {Becklin} E.~E., 1998, \apj, 509, 678

\bibitem[{{Gillessen} {et~al}\mbox{.}(2009){Gillessen}, {Eisenhauer}, {Trippe},
  {Alexander}, {Genzel}, {Martins}, \& {Ott}}]{2009ApJ...692.1075G}
{Gillessen} S., {Eisenhauer} F., {Trippe} S., {Alexander} T., {Genzel} R.,
  {Martins} F., {Ott} T., 2009, \apj, 692, 1075

\bibitem[{{Ginsburg} {et~al}\mbox{.}(2019){Ginsburg}, {Sip{\H{o}}cz},
  {Brasseur}, {Cowperthwaite}, {Craig}, {Deil}, {Guillochon}, {Guzman},
  {Liedtke}, {Lian Lim}, {Lockhart}, {Mommert}, {Morris}, {Norman}, {Parikh},
  {Persson}, {Robitaille}, {Segovia}, {Singer}, {Tollerud}, {de Val-Borro},
  {Valtchanov}, {Woillez}, {Astroquery Collaboration}, \& {a subset of astropy
  Collaboration}}]{2019AJ....157...98G}
{Ginsburg} A. {et~al.}, 2019, \aj, 157, 98

\bibitem[{{Gnedin}, {Ostriker} \& {Tremaine}(2014){Gnedin}, {Ostriker}, \&
  {Tremaine}}]{2014ApJ...785...71G}
{Gnedin} O.~Y., {Ostriker} J.~P., {Tremaine} S., 2014, \apj, 785, 71

\bibitem[{{Gonz{\'a}lez Delgado} {et~al}\mbox{.}(2008){Gonz{\'a}lez Delgado},
  {P{\'e}rez}, {Cid Fernandes}, \& {Schmitt}}]{2008AJ....135..747G}
{Gonz{\'a}lez Delgado} R.~M., {P{\'e}rez} E., {Cid Fernandes} R., {Schmitt} H.,
  2008, \aj, 135, 747

\bibitem[{{Graham}(2016)}]{2016ASSL..418..263G}
{Graham} A.~W., 2016, in Astrophysics and Space Science Library, Vol. 418,
  Galactic Bulges, {Laurikainen} E., {Peletier} R., {Gadotti} D., eds., p. 263

\bibitem[{{Graham} {et~al}\mbox{.}(2003){Graham}, {Erwin}, {Trujillo}, \&
  {Asensio Ramos}}]{2003AJ....125.2951G}
{Graham} A.~W., {Erwin} P., {Trujillo} I., {Asensio Ramos} A., 2003, \aj, 125,
  2951

\bibitem[{{Graham} \& {Guzm{\'a}n}(2003)}]{2003AJ....125.2936G}
{Graham} A.~W., {Guzm{\'a}n} R., 2003, \aj, 125, 2936

\bibitem[{{Graham} \& {Spitler}(2009)}]{2009MNRAS.397.2148G}
{Graham} A.~W., {Spitler} L.~R., 2009, \mnras, 397, 2148

\bibitem[{{Grant}, {Kuipers} \& {Phillipps}(2005){Grant}, {Kuipers}, \&
  {Phillipps}}]{2005MNRAS.363.1019G}
{Grant} N.~I., {Kuipers} J.~A., {Phillipps} S., 2005, \mnras, 363, 1019

\bibitem[{{Green}(2011)}]{2011BASI...39..289G}
{Green} D.~A., 2011, Bulletin of the Astronomical Society of India, 39, 289

\bibitem[{{Greene} \& {Ho}(2004)}]{2004ApJ...610..722G}
{Greene} J.~E., {Ho} L.~C., 2004, \apj, 610, 722

\bibitem[{{Ho}, {Filippenko} \& {Sargent}(1995){Ho}, {Filippenko}, \&
  {Sargent}}]{1995ApJS...98..477H}
{Ho} L.~C., {Filippenko} A.~V., {Sargent} W.~L., 1995, \apjs, 98, 477

\bibitem[{{Ho}, {Filippenko} \& {Sargent}(1997{\natexlab{a}}){Ho},
  {Filippenko}, \& {Sargent}}]{1997ApJS..112..315H}
{Ho} L.~C., {Filippenko} A.~V., {Sargent} W. L.~W., 1997{\natexlab{a}}, \apjs,
  112, 315

\bibitem[{{Ho}, {Filippenko} \& {Sargent}(1997{\natexlab{b}}){Ho},
  {Filippenko}, \& {Sargent}}]{1997ApJ...487..568H}
{Ho} L.~C., {Filippenko} A.~V., {Sargent} W. L.~W., 1997{\natexlab{b}}, \apj,
  487, 568

\bibitem[{{Ho} {et~al}\mbox{.}(1997){Ho}, {Filippenko}, {Sargent}, \&
  {Peng}}]{1997ApJS..112..391H}
{Ho} L.~C., {Filippenko} A.~V., {Sargent} W. L.~W., {Peng} C.~Y., 1997, \apjs,
  112, 391

\bibitem[{{Hopkins} {et~al}\mbox{.}(2009){Hopkins}, {Cox}, {Dutta},
  {Hernquist}, {Kormendy}, \& {Lauer}}]{2009ApJS..181..135H}
{Hopkins} P.~F., {Cox} T.~J., {Dutta} S.~N., {Hernquist} L., {Kormendy} J.,
  {Lauer} T.~R., 2009, \apjs, 181, 135

\bibitem[{{Hoyer} {et~al}\mbox{.}(2021){Hoyer}, {Neumayer}, {Georgiev}, {Seth},
  \& {Greene}}]{2021MNRAS.507.3246H}
{Hoyer} N., {Neumayer} N., {Georgiev} I.~Y., {Seth} A.~C., {Greene} J.~E.,
  2021, \mnras, 507, 3246

\bibitem[{{Hoyer} {et~al}\mbox{.}(2023){Hoyer}, {Neumayer}, {Seth}, {Georgiev},
  \& {Greene}}]{2023MNRAS.520.4664H}
{Hoyer} N., {Neumayer} N., {Seth} A.~C., {Georgiev} I.~Y., {Greene} J.~E.,
  2023, \mnras, 520, 4664

\bibitem[{{Hubble}(1926)}]{1926ApJ....64..321H}
{Hubble} E.~P., 1926, \apj, 64, 321

\bibitem[{Hunter(2007)}]{Hunter:2007}
Hunter J.~D., 2007, Computing in Science \& Engineering, 9, 90

\bibitem[{{Ibata} {et~al}\mbox{.}(2013){Ibata}, {Lewis}, {Conn}, {Irwin},
  {McConnachie}, {Chapman}, {Collins}, {Fardal}, {Ferguson}, {Ibata}, {Mackey},
  {Martin}, {Navarro}, {Rich}, {Valls-Gabaud}, \&
  {Widrow}}]{2013Natur.493...62I}
{Ibata} R.~A. {et~al.}, 2013, \nat, 493, 62

\bibitem[{{Isobe}, {Feigelson} \& {Nelson}(1986){Isobe}, {Feigelson}, \&
  {Nelson}}]{1986ApJ...306..490I}
{Isobe} T., {Feigelson} E.~D., {Nelson} P.~I., 1986, \apj, 306, 490

\bibitem[{{Karachentsev} {et~al}\mbox{.}(2004){Karachentsev}, {Karachentseva},
  {Huchtmeier}, \& {Makarov}}]{2004AJ....127.2031K}
{Karachentsev} I.~D., {Karachentseva} V.~E., {Huchtmeier} W.~K., {Makarov}
  D.~I., 2004, \aj, 127, 2031

\bibitem[{{Kewley} {et~al}\mbox{.}(2006){Kewley}, {Groves}, {Kauffmann}, \&
  {Heckman}}]{2006MNRAS.372..961K}
{Kewley} L.~J., {Groves} B., {Kauffmann} G., {Heckman} T., 2006, \mnras, 372,
  961

\bibitem[{{King}(1962)}]{1962AJ.....67..471K}
{King} I., 1962, \aj, 67, 471

\bibitem[{{King}(1966)}]{1966AJ.....71...64K}
{King} I.~R., 1966, \aj, 71, 64

\bibitem[{Kluyver {et~al}\mbox{.}(2016)Kluyver, Ragan-Kelley, P{\'e}rez,
  Granger, Bussonnier, Frederic, Kelley, Hamrick, Grout, Corlay, Ivanov, Avila,
  Abdalla, \& Willing}]{2016ppap.book...87K}
Kluyver T. {et~al.}, 2016, in Positioning and Power in Academic Publishing:
  Players, Agents and Agendas, Loizides F., Schmidt B., eds., Amsterdam: IOS
  Press, pp. 87 -- 90

\bibitem[{{Kormendy} \& {Ho}(2013)}]{2013ARA&A..51..511K}
{Kormendy} J., {Ho} L.~C., 2013, \araa, 51, 511

\bibitem[{{Kroupa} {et~al}\mbox{.}(2020){Kroupa}, {Subr}, {Jerabkova}, \&
  {Wang}}]{2020MNRAS.498.5652K}
{Kroupa} P., {Subr} L., {Jerabkova} T., {Wang} L., 2020, \mnras, 498, 5652

\bibitem[{{Larsen}(1999)}]{1999A&AS..139..393L}
{Larsen} S.~S., 1999, \aaps, 139, 393

\bibitem[{{Lauer} {et~al}\mbox{.}(1995){Lauer}, {Ajhar}, {Byun}, {Dressler},
  {Faber}, {Grillmair}, {Kormendy}, {Richstone}, \&
  {Tremaine}}]{1995AJ....110.2622L}
{Lauer} T.~R. {et~al.}, 1995, \aj, 110, 2622

\bibitem[{{Lauer} {et~al}\mbox{.}(2005){Lauer}, {Faber}, {Gebhardt},
  {Richstone}, {Tremaine}, {Ajhar}, {Aller}, {Bender}, {Dressler},
  {Filippenko}, {Green}, {Grillmair}, {Ho}, {Kormendy}, {Magorrian}, {Pinkney},
  \& {Siopis}}]{2005AJ....129.2138L}
{Lauer} T.~R. {et~al.}, 2005, \aj, 129, 2138

\bibitem[{{Leigh}, {B{\"o}ker} \& {Knigge}(2012){Leigh}, {B{\"o}ker}, \&
  {Knigge}}]{2012MNRAS.424.2130L}
{Leigh} N., {B{\"o}ker} T., {Knigge} C., 2012, \mnras, 424, 2130

\bibitem[{{Lotz}, {Miller} \& {Ferguson}(2004){Lotz}, {Miller}, \&
  {Ferguson}}]{2004ApJ...613..262L}
{Lotz} J.~M., {Miller} B.~W., {Ferguson} H.~C., 2004, \apj, 613, 262

\bibitem[{{Magorrian} {et~al}\mbox{.}(1998){Magorrian}, {Tremaine},
  {Richstone}, {Bender}, {Bower}, {Dressler}, {Faber}, {Gebhardt}, {Green},
  {Grillmair}, {Kormendy}, \& {Lauer}}]{1998AJ....115.2285M}
{Magorrian} J. {et~al.}, 1998, \aj, 115, 2285

\bibitem[{{Mezcua} {et~al}\mbox{.}(2018){Mezcua}, {Civano}, {Marchesi}, {Suh},
  {Fabbiano}, \& {Volonteri}}]{2018MNRAS.478.2576M}
{Mezcua} M., {Civano} F., {Marchesi} S., {Suh} H., {Fabbiano} G., {Volonteri}
  M., 2018, \mnras, 478, 2576

\bibitem[{{Naiman} {et~al}\mbox{.}(2015){Naiman}, {Ramirez-Ruiz}, {Debuhr}, \&
  {Ma}}]{2015ApJ...803...81N}
{Naiman} J.~P., {Ramirez-Ruiz} E., {Debuhr} J., {Ma} C.~P., 2015, \apj, 803, 81

\bibitem[{{Nandi} {et~al}\mbox{.}(2023){Nandi}, {Stalin}, {Saikia}, {Muneer},
  {Mountrichas}, {Wylezalek}, {Sagar}, \&
  {Kissler-Patig}}]{2023ApJ...950...81N}
{Nandi} P., {Stalin} C.~S., {Saikia} D.~J., {Muneer} S., {Mountrichas} G.,
  {Wylezalek} D., {Sagar} R., {Kissler-Patig} M., 2023, \apj, 950, 81

\bibitem[{{Nemmen} {et~al}\mbox{.}(2012){Nemmen}, {Georganopoulos}, {Guiriec},
  {Meyer}, {Gehrels}, \& {Sambruna}}]{2012Sci...338.1445N}
{Nemmen} R.~S., {Georganopoulos} M., {Guiriec} S., {Meyer} E.~T., {Gehrels} N.,
  {Sambruna} R.~M., 2012, Science, 338, 1445

\bibitem[{{Neumayer}, {Seth} \& {B{\"o}ker}(2020){Neumayer}, {Seth}, \&
  {B{\"o}ker}}]{2020A&ARv..28....4N}
{Neumayer} N., {Seth} A., {B{\"o}ker} T., 2020, \aapr, 28, 4

\bibitem[{{Neumayer} \& {Walcher}(2012)}]{2012AdAst2012E..15N}
{Neumayer} N., {Walcher} C.~J., 2012, Advances in Astronomy, 2012, 709038

\bibitem[{{Nguyen} {et~al}\mbox{.}(2018){Nguyen}, {Seth}, {Neumayer}, {Kamann},
  {Voggel}, {Cappellari}, {Picotti}, {Nguyen}, {B{\"o}ker}, {Debattista},
  {Caldwell}, {McDermid}, {Bastian}, {Ahn}, \&
  {Pechetti}}]{2018ApJ...858..118N}
{Nguyen} D.~D. {et~al.}, 2018, \apj, 858, 118

\bibitem[{{O'Shaughnessy}, {Kopparapu} \& {Belczynski}(2012){O'Shaughnessy},
  {Kopparapu}, \& {Belczynski}}]{2012CQGra..29n5011O}
{O'Shaughnessy} R., {Kopparapu} R.~K., {Belczynski} K., 2012, Classical and
  Quantum Gravity, 29, 145011

\bibitem[{{Panessa} {et~al}\mbox{.}(2019){Panessa}, {Baldi}, {Laor},
  {Padovani}, {Behar}, \& {McHardy}}]{2019NatAs...3..387P}
{Panessa} F., {Baldi} R.~D., {Laor} A., {Padovani} P., {Behar} E., {McHardy}
  I., 2019, Nature Astronomy, 3, 387

\bibitem[{{Pechetti} {et~al}\mbox{.}(2020){Pechetti}, {Seth}, {Neumayer},
  {Georgiev}, {Kacharov}, \& {den Brok}}]{2020ApJ...900...32P}
{Pechetti} R., {Seth} A., {Neumayer} N., {Georgiev} I., {Kacharov} N., {den
  Brok} M., 2020, \apj, 900, 32

\bibitem[{{Penny} {et~al}\mbox{.}(2018){Penny}, {Masters}, {Smethurst},
  {Nichol}, {Krawczyk}, {Bizyaev}, {Greene}, {Liu}, {Marinelli}, {Rembold},
  {Riffel}, {Ilha}, {Wylezalek}, {Andrews}, {Bundy}, {Drory}, {Oravetz}, \&
  {Pan}}]{2018MNRAS.476..979P}
{Penny} S.~J. {et~al.}, 2018, \mnras, 476, 979

\bibitem[{{Phillips} {et~al}\mbox{.}(1996){Phillips}, {Illingworth},
  {MacKenty}, \& {Franx}}]{1996AJ....111.1566P}
{Phillips} A.~C., {Illingworth} G.~D., {MacKenty} J.~W., {Franx} M., 1996, \aj,
  111, 1566

\bibitem[{{Planck Collaboration} {et~al}\mbox{.}(2020){Planck Collaboration},
  {Aghanim}, {Akrami}, {Ashdown}, {Aumont}, {Baccigalupi}, {Ballardini},
  {Banday}, {Barreiro}, {Bartolo}, {Basak}, {Battye}, {Benabed}, {Bernard},
  {Bersanelli}, {Bielewicz}, {Bock}, {Bond}, {Borrill}, {Bouchet}, {Boulanger},
  {Bucher}, {Burigana}, {Butler}, {Calabrese}, {Cardoso}, {Carron},
  {Challinor}, {Chiang}, {Chluba}, {Colombo}, {Combet}, {Contreras}, {Crill},
  {Cuttaia}, {de Bernardis}, {de Zotti}, {Delabrouille}, {Delouis}, {Di
  Valentino}, {Diego}, {Dor{\'e}}, {Douspis}, {Ducout}, {Dupac}, {Dusini},
  {Efstathiou}, {Elsner}, {En{\ss}lin}, {Eriksen}, {Fantaye}, {Farhang},
  {Fergusson}, {Fernandez-Cobos}, {Finelli}, {Forastieri}, {Frailis},
  {Fraisse}, {Franceschi}, {Frolov}, {Galeotta}, {Galli}, {Ganga},
  {G{\'e}nova-Santos}, {Gerbino}, {Ghosh}, {Gonz{\'a}lez-Nuevo}, {G{\'o}rski},
  {Gratton}, {Gruppuso}, {Gudmundsson}, {Hamann}, {Handley}, {Hansen},
  {Herranz}, {Hildebrandt}, {Hivon}, {Huang}, {Jaffe}, {Jones}, {Karakci},
  {Keih{\"a}nen}, {Keskitalo}, {Kiiveri}, {Kim}, {Kisner}, {Knox},
  {Krachmalnicoff}, {Kunz}, {Kurki-Suonio}, {Lagache}, {Lamarre}, {Lasenby},
  {Lattanzi}, {Lawrence}, {Le Jeune}, {Lemos}, {Lesgourgues}, {Levrier},
  {Lewis}, {Liguori}, {Lilje}, {Lilley}, {Lindholm}, {L{\'o}pez-Caniego},
  {Lubin}, {Ma}, {Mac{\'\i}as-P{\'e}rez}, {Maggio}, {Maino}, {Mandolesi},
  {Mangilli}, {Marcos-Caballero}, {Maris}, {Martin}, {Martinelli},
  {Mart{\'\i}nez-Gonz{\'a}lez}, {Matarrese}, {Mauri}, {McEwen}, {Meinhold},
  {Melchiorri}, {Mennella}, {Migliaccio}, {Millea}, {Mitra},
  {Miville-Desch{\^e}nes}, {Molinari}, {Montier}, {Morgante}, {Moss}, {Natoli},
  {N{\o}rgaard-Nielsen}, {Pagano}, {Paoletti}, {Partridge}, {Patanchon},
  {Peiris}, {Perrotta}, {Pettorino}, {Piacentini}, {Polastri}, {Polenta},
  {Puget}, {Rachen}, {Reinecke}, {Remazeilles}, {Renzi}, {Rocha}, {Rosset},
  {Roudier}, {Rubi{\~n}o-Mart{\'\i}n}, {Ruiz-Granados}, {Salvati}, {Sandri},
  {Savelainen}, {Scott}, {Shellard}, {Sirignano}, {Sirri}, {Spencer},
  {Sunyaev}, {Suur-Uski}, {Tauber}, {Tavagnacco}, {Tenti}, {Toffolatti},
  {Tomasi}, {Trombetti}, {Valenziano}, {Valiviita}, {Van Tent}, {Vibert},
  {Vielva}, {Villa}, {Vittorio}, {Wand elt}, {Wehus}, {White}, {White},
  {Zacchei}, \& {Zonca}}]{2020A&A...641A...6P}
{Planck Collaboration} {et~al.}, 2020, \aap, 641, A6

\bibitem[{{Poulain} {et~al}\mbox{.}(2021){Poulain}, {Marleau}, {Habas}, {Duc},
  {S{\'a}nchez-Janssen}, {Durrell}, {Paudel}, {Ahad}, {Chougule}, {M{\"u}ller},
  {Lim}, {B{\'\i}lek}, \& {Fensch}}]{2021MNRAS.506.5494P}
{Poulain} M. {et~al.}, 2021, \mnras, 506, 5494

\bibitem[{{Prieto} {et~al}\mbox{.}(2016){Prieto}, {Fern{\'a}ndez-Ontiveros},
  {Markoff}, {Espada}, \& {Gonz{\'a}lez-Mart{\'\i}n}}]{2016MNRAS.457.3801P}
{Prieto} M.~A., {Fern{\'a}ndez-Ontiveros} J.~A., {Markoff} S., {Espada} D.,
  {Gonz{\'a}lez-Mart{\'\i}n} O., 2016, \mnras, 457, 3801

\bibitem[{{Ravindranath} {et~al}\mbox{.}(2001){Ravindranath}, {Ho}, {Peng},
  {Filippenko}, \& {Sargent}}]{2001AJ....122..653R}
{Ravindranath} S., {Ho} L.~C., {Peng} C.~Y., {Filippenko} A.~V., {Sargent} W.
  L.~W., 2001, \aj, 122, 653

\bibitem[{{Reines} {et~al}\mbox{.}(2011){Reines}, {Sivakoff}, {Johnson}, \&
  {Brogan}}]{2011Natur.470...66R}
{Reines} A.~E., {Sivakoff} G.~R., {Johnson} K.~E., {Brogan} C.~L., 2011, \nat,
  470, 66

\bibitem[{{Richstone} {et~al}\mbox{.}(1998){Richstone}, {Ajhar}, {Bender},
  {Bower}, {Dressler}, {Faber}, {Filippenko}, {Gebhardt}, {Green}, {Ho},
  {Kormendy}, {Lauer}, {Magorrian}, \& {Tremaine}}]{1998Natur.395A..14R}
{Richstone} D. {et~al.}, 1998, \nat, 395, A14

\bibitem[{{Riess} {et~al}\mbox{.}(2019){Riess}, {Casertano}, {Yuan}, {Macri},
  \& {Scolnic}}]{2019ApJ...876...85R}
{Riess} A.~G., {Casertano} S., {Yuan} W., {Macri} L.~M., {Scolnic} D., 2019,
  \apj, 876, 85

\bibitem[{{Rom{\'a}n} {et~al}\mbox{.}(2023){Rom{\'a}n},
  {S{\'a}nchez-Alarc{\'o}n}, {Knapen}, \& {Peletier}}]{2023A&A...671L...7R}
{Rom{\'a}n} J., {S{\'a}nchez-Alarc{\'o}n} P.~M., {Knapen} J.~H., {Peletier} R.,
  2023, \aap, 671, L7

\bibitem[{{Saikia} {et~al}\mbox{.}(2018){Saikia}, {K{\"o}rding}, {Coppejans},
  {Falcke}, {Williams}, {Baldi}, {Mchardy}, \& {Beswick}}]{2018A&A...616A.152S}
{Saikia} P., {K{\"o}rding} E., {Coppejans} D.~L., {Falcke} H., {Williams} D.,
  {Baldi} R.~D., {Mchardy} I., {Beswick} R., 2018, \aap, 616, A152

\bibitem[{{Saikia}, {K{\"o}rding} \& {Falcke}(2015){Saikia}, {K{\"o}rding}, \&
  {Falcke}}]{2015MNRAS.450.2317S}
{Saikia} P., {K{\"o}rding} E., {Falcke} H., 2015, \mnras, 450, 2317

\bibitem[{{S{\'a}nchez-Janssen} {et~al}\mbox{.}(2019){S{\'a}nchez-Janssen},
  {C{\^o}t{\'e}}, {Ferrarese}, {Peng}, {Roediger}, {Blakeslee}, {Emsellem},
  {Puzia}, {Spengler}, {Taylor}, {{\'A}lamo-Mart{\'\i}nez}, {Boselli},
  {Cantiello}, {Cuillandre}, {Duc}, {Durrell}, {Gwyn}, {MacArthur},
  {Lan{\c{c}}on}, {Lim}, {Liu}, {Mei}, {Miller}, {Mu{\~n}oz}, {Mihos},
  {Paudel}, {Powalka}, \& {Toloba}}]{2019ApJ...878...18S}
{S{\'a}nchez-Janssen} R. {et~al.}, 2019, \apj, 878, 18

\bibitem[{{Sandage} \& {Tammann}(1981)}]{1981rsac.book.....S}
{Sandage} A., {Tammann} G.~A., 1981, {A Revised Shapley-Ames Catalog of Bright
  Galaxies (Carnegie Inst. of Washington)}

\bibitem[{{Sartori} {et~al}\mbox{.}(2015){Sartori}, {Schawinski}, {Treister},
  {Trakhtenbrot}, {Koss}, {Shirazi}, \& {Oh}}]{2015MNRAS.454.3722S}
{Sartori} L.~F., {Schawinski} K., {Treister} E., {Trakhtenbrot} B., {Koss} M.,
  {Shirazi} M., {Oh} K., 2015, \mnras, 454, 3722

\bibitem[{{Scarlata} {et~al}\mbox{.}(2004){Scarlata}, {Stiavelli}, {Hughes},
  {Axon}, {Alonso-Herrero}, {Atkinson}, {Batcheldor}, {Binney}, {Capetti},
  {Carollo}, {Dressel}, {Gerssen}, {Macchetto}, {Maciejewski}, {Marconi},
  {Merrifield}, {Ruiz}, {Sparks}, {Tsvetanov}, \& {van der
  Marel}}]{2004AJ....128.1124S}
{Scarlata} C. {et~al.}, 2004, \aj, 128, 1124

\bibitem[{{Schlafly} \& {Finkbeiner}(2011)}]{2011ApJ...737..103S}
{Schlafly} E.~F., {Finkbeiner} D.~P., 2011, \apj, 737, 103

\bibitem[{{Sch{\"o}del} {et~al}\mbox{.}(2014){Sch{\"o}del}, {Feldmeier},
  {Kunneriath}, {Stolovy}, {Neumayer}, {Amaro-Seoane}, \&
  {Nishiyama}}]{2014A&A...566A..47S}
{Sch{\"o}del} R., {Feldmeier} A., {Kunneriath} D., {Stolovy} S., {Neumayer} N.,
  {Amaro-Seoane} P., {Nishiyama} S., 2014, \aap, 566, A47

\bibitem[{{S\'ersic}(1963)}]{1963BAAA....6...41S}
{S\'ersic} J.~L., 1963, Boletin de la Asociacion Argentina de Astronomia La
  Plata Argentina, 6, 41

\bibitem[{{S\'ersic}(1968)}]{1968adga.book.....S}
{S\'ersic} J.~L., 1968, {Atlas de Galaxias Australes (C\'ordoba, Argentina:
  Observatorio Astron\'omico)}

\bibitem[{{Seth} {et~al}\mbox{.}(2008){Seth}, {Ag{\"u}eros}, {Lee}, \&
  {Basu-Zych}}]{2008ApJ...678..116S}
{Seth} A., {Ag{\"u}eros} M., {Lee} D., {Basu-Zych} A., 2008, \apj, 678, 116

\bibitem[{{Seth} {et~al}\mbox{.}(2006){Seth}, {Dalcanton}, {Hodge}, \&
  {Debattista}}]{2006AJ....132.2539S}
{Seth} A.~C., {Dalcanton} J.~J., {Hodge} P.~W., {Debattista} V.~P., 2006, \aj,
  132, 2539

\bibitem[{{Seth}, {Neumayer} \& {B{\"o}ker}(2020){Seth}, {Neumayer}, \&
  {B{\"o}ker}}]{2020IAUS..351...13S}
{Seth} A.~C., {Neumayer} N., {B{\"o}ker} T., 2020, in Star Clusters: From the
  Milky Way to the Early Universe, {Bragaglia} A., {Davies} M., {Sills} A.,
  {Vesperini} E., eds., Vol. 351, pp. 13--18

\bibitem[{{Smith} {et~al}\mbox{.}(2020){Smith}, {Bajaj}, {Ryon}, \&
  {Sabbi}}]{2020ApJ...896...84S}
{Smith} L.~J., {Bajaj} V., {Ryon} J., {Sabbi} E., 2020, \apj, 896, 84

\bibitem[{{Spengler} {et~al}\mbox{.}(2017){Spengler}, {C{\^o}t{\'e}},
  {Roediger}, {Ferrarese}, {S{\'a}nchez-Janssen}, {Toloba}, {Liu},
  {Guhathakurta}, {Cuillandre}, {Gwyn}, {Zirm}, {Mu{\~n}oz}, {Puzia},
  {Lan{\c{c}}on}, {Peng}, {Mei}, \& {Powalka}}]{2017ApJ...849...55S}
{Spengler} C. {et~al.}, 2017, \apj, 849, 55

\bibitem[{{Stiavelli} {et~al}\mbox{.}(2001){Stiavelli}, {Miller}, {Ferguson},
  {Mack}, {Whitmore}, \& {Lotz}}]{2001AJ....121.1385S}
{Stiavelli} M., {Miller} B.~W., {Ferguson} H.~C., {Mack} J., {Whitmore} B.~C.,
  {Lotz} J.~M., 2001, \aj, 121, 1385

\bibitem[{{Stone}, {K{\"u}pper} \& {Ostriker}(2017){Stone}, {K{\"u}pper}, \&
  {Ostriker}}]{2017MNRAS.467.4180S}
{Stone} N.~C., {K{\"u}pper} A. H.~W., {Ostriker} J.~P., 2017, \mnras, 467, 4180

\bibitem[{{Su} {et~al}\mbox{.}(2022){Su}, {Salo}, {Janz}, {Venhola}, \&
  {Peletier}}]{2022AA...664A.167S}
{Su} A.~H., {Salo} H., {Janz} J., {Venhola} A., {Peletier} R.~F., 2022, \aap,
  664, A167

\bibitem[{{Taylor}(2005)}]{2005ASPC..347...29T}
{Taylor} M.~B., 2005, in Astronomical Society of the Pacific Conference Series,
  Vol. 347, Astronomical Data Analysis Software and Systems XIV, {Shopbell} P.,
  {Britton} M., {Ebert} R., eds., p.~29

\bibitem[{{Thornton} {et~al}\mbox{.}(2008){Thornton}, {Barth}, {Ho},
  {Rutledge}, \& {Greene}}]{2008ApJ...686..892T}
{Thornton} C.~E., {Barth} A.~J., {Ho} L.~C., {Rutledge} R.~E., {Greene} J.~E.,
  2008, \apj, 686, 892

\bibitem[{{Trujillo} {et~al}\mbox{.}(2004){Trujillo}, {Erwin}, {Asensio Ramos},
  \& {Graham}}]{2004AJ....127.1917T}
{Trujillo} I., {Erwin} P., {Asensio Ramos} A., {Graham} A.~W., 2004, \aj, 127,
  1917

\bibitem[{{Turner} {et~al}\mbox{.}(2012){Turner}, {C{\^o}t{\'e}}, {Ferrarese},
  {Jord{\'a}n}, {Blakeslee}, {Mei}, {Peng}, \& {West}}]{2012ApJS..203....5T}
{Turner} M.~L., {C{\^o}t{\'e}} P., {Ferrarese} L., {Jord{\'a}n} A., {Blakeslee}
  J.~P., {Mei} S., {Peng} E.~W., {West} M.~J., 2012, \apjs, 203, 5

\bibitem[{{van der Walt}, {Colbert} \& {Varoquaux}(2011){van der Walt},
  {Colbert}, \& {Varoquaux}}]{2011CSE....13b..22V}
{van der Walt} S., {Colbert} S.~C., {Varoquaux} G., 2011, Computing in Science
  and Engineering, 13, 22

\bibitem[{{Virtanen} {et~al}\mbox{.}(2020){Virtanen}, {Gommers}, {Oliphant},
  {Haberland}, {Reddy}, {Cournapeau}, {Burovski}, {Peterson}, {Weckesser},
  {Bright}, {van der Walt}, {Brett}, {Wilson}, {Millman}, {Mayorov}, {Nelson},
  {Jones}, {Kern}, {Larson}, {Carey}, {Polat}, {Feng}, {Moore}, {VanderPlas},
  {Laxalde}, {Perktold}, {Cimrman}, {Henriksen}, {Quintero}, {Harris},
  {Archibald}, {Ribeiro}, {Pedregosa}, {van Mulbregt}, \& {SciPy 1. 0
  Contributors}}]{2020NatMe..17..261V}
{Virtanen} P. {et~al.}, 2020, Nature Methods, 17, 261

\bibitem[{{Wehner} \& {Harris}(2006)}]{2006ApJ...644L..17W}
{Wehner} E.~H., {Harris} W.~E., 2006, \apjl, 644, L17

\bibitem[{{Williams} {et~al}\mbox{.}(2022){Williams}, {Pahari}, {Baldi},
  {McHardy}, {Mathur}, {Beswick}, {Beri}, {Boorman}, {Aalto}, {Alberdi},
  {Argo}, {Dullo}, {Fenech}, {Green}, {Knapen}, {Mart{\'\i}-Vidal}, {Moldon},
  {Mundell}, {Muxlow}, {Panessa}, {P{\'e}rez-Torres}, {Saikia}, {Shankar},
  {Stevens}, \& {Uttley}}]{2022MNRAS.510.4909W}
{Williams} D.~R.~A. {et~al.}, 2022, \mnras, 510, 4909

\bibitem[{{Williams} {et~al}\mbox{.}(2023){Williams}, {Pahari}, {Baldi},
  {McHardy}, {Mathur}, {Beswick}, {Beri}, {Boorman}, {Aalto}, {Alberdi},
  {Argo}, {Dullo}, {Fenech}, {Green}, {Knapen}, {Mart{\'\i}-Vidal}, {Moldon},
  {Mundell}, {Muxlow}, {Panessa}, {P{\'e}rez-Torres}, {Saikia}, {Shankar},
  {Stevens}, \& {Uttley}}]{2023arXiv230308647W}
{Williams} D.~R.~A. {et~al.}, 2023, arXiv e-prints, arXiv:2303.08647

\bibitem[{{Willmer}(2018)}]{2018ApJS..236...47W}
{Willmer} C. N.~A., 2018, \apjs, 236, 47

\bibitem[{{Yang} {et~al}\mbox{.}(2023){Yang}, {Paragi}, {Frey}, {Gurvits},
  {Liao}, {Liu}, {Cui}, {Yang}, {Chen}, {Varenius}, {Conway}, {Chen}, \&
  {Chang}}]{2023MNRAS.520.5964Y}
{Yang} J. {et~al.}, 2023, \mnras, 520, 5964

\bibitem[{{Zanatta} {et~al}\mbox{.}(2021){Zanatta}, {S{\'a}nchez-Janssen},
  {Chies-Santos}, {de Souza}, \& {Blakeslee}}]{2021MNRAS.508..986Z}
{Zanatta} E., {S{\'a}nchez-Janssen} R., {Chies-Santos} A.~L., {de Souza} R.~S.,
  {Blakeslee} J.~P., 2021, \mnras, 508, 986

\end{thebibliography}

\setcounter{section}{0}
\renewcommand{\thesection}{A\arabic{section}}

\setcounter{table}{0}
\renewcommand{\thetable}{A\arabic{table}} 

\setlength{\tabcolsep}{0.010922134in}
\begin{table*}
\begin {minipage}{177mm}
\caption{Comparison with previous studies of nucleation in nearby galaxies.  } 
\label{Tab3}
\begin{tabular}{@{}lllccccc@{}}
  \hline
  Study &Data used&Type&Method&$f_{\rm nuc } $&Full sample\\
  \hline
  {\color{blue}This work} &{\it HST}&ETs+LTs&multicompt (1D,2D)&58--67\%/core (10 -- 20\%)&173&\\
\citet{2022AA...664A.167S}&ground-based  &dwarf+massive galaxies &multicompt (2D)&---& 557\\
 &(seeing  $\sim$ 1.1--1.2 arcsec)&\\
 \citet{2021MNRAS.506.5494P}&ground-based&dwarfs&PSFNuc+S\'ersic(2D)&23\%&2210&\\
 & (seeing $\sim$ 0.5--1.6 arcsec)&\\
  \citet{ 2021MNRAS.508..986Z}&{\it HST} &low-mass, quiescent &PSFNuc+S\'ersic(2D)&50\%&66 \\
  \citet{2021MNRAS.507.3246H}&compilation& local volume galaxies&SNuc+flatbg (2D)&---&601& \\
  \citet{2019ApJ...878...18S}&ground-based  &quiescent Virgo&SNuc+S\'ersic (1D)&$90\%$&400 \\
  &(seeing  $\sim$ 0.6 arcsec)&\\
  \citet{2014MNRAS.441.3570G}&{\it HST} &Spirals&KingNuc(2D)&80\%&228\\
  \citet{2014MNRAS.445.2385D}&{\it HST} &ET dwarfs&PSFNuc+S\'ersic&80\%&22\\
  \citet{2012ApJS..203....5T}&{\it HST}&ETs&multicompt (1D,2D)&72\%& 43\\
 \citet{2007ApJ...665.1084B}&{\it     HST}   & S0–Sbc galaxies&multicompt (1D)&84\%         & 19\\
  \citet{2006ApJS..165...57C}&{\it     HST}&ETs Virgo&KingNuc+core-/S\'ersic(1D)&($66-82\%$)&100 \\
  \citet{2005MNRAS.363.1019G}&ground-based &dEs&PSFNuc+S\'ersic(1D+2D)&61\%&181 \\
  &(seeing $\sim$ 1.7--1.9 arcsec)&\\
  \citet{2005AJ....129.2138L}&{\it HST}&ETs&Nuc\_excluded+Nuker(1D)&core (29\%)/power-law (60\%)&77 \\
   \citet{2003AJ....125.2936G}&{\it     HST}   &dEs&PSFNuc+S\'ersic(1D)&87\%&     15\\
   \citet{2002AJ....123..159C}&{\it     HST} &Spirals&PSFNuc(1D)&65\%&69\\
  \citet{2002AJ....123.1389B}&{\it     HST} & LT spirals  &Galaxy inc\_Nuc, Nuker(1D)&77\%&77 \\
  \citet{2001AJ....121.1385S}&{\it     HST}  &dEs&Nuc\_excluded+Nuk/Ser(1D)&56\%&25& \\
  \citet{2001AJ....122..653R}&{\it HST} &ETs&PSFNuc+Nuker(1D,2D)&50\%&33\\
  \citet{1998AJ....116...68C}&{\it   HST}   &Spirals&PSFNuc(1D)&50\%&75 \\
  \hline
\end{tabular} 
Note: Ground-based and {\it HST} data have been used to study
nucleation in early- and late-type (i.e.\ ET and LT) and quiescent
galaxies.  Previous studies in the literature which performed
multi-component decompositions for photometric and structural analysis
of nuclei include e.g.\ \citet{2022AA...664A.167S}.  Several studies
fitted the nuclei with a PSF model (e.g.\
\citealt{2021MNRAS.506.5494P}), whereas others fitted a S\'ersic model
or a King profile \citep{1962AJ.....67..471K,1966AJ.....71...64K},
see\ \citet{2006ApJS..165...57C}.
\citet{2001AJ....121.1385S} and \citet{2005AJ....129.2138L} fit the surface
brightness profiles of their galaxies by excluding the
nuclei. \citet{2002AJ....123.1389B} fitted the light profiles of their
late-type spiral galaxies with a Nuker (nucleus+bulge) model. See
Section~\ref{App02} for further details.
\end{minipage}
\end{table*}

 \setcounter{figure}{0}
\renewcommand{\thefigure}{A\arabic{figure}} 

\setcounter{section}{0}
\renewcommand{\thesection}{A\arabic{section}}

\section{2D decomposition}\label{App01}
 
 Fig.~\ref{2Dfits} displays the {\it HST} images and the residual images
from the 2D multi-component decompositions for the dozen
representative nucleated galaxies whose 1D profiles are shown in Fig.~\ref{Fig1}. The
residual images are generated after subtracting the best-fitting {\sc
  imfit} model images from the galaxies' {\it HST} images.

 \section{Literature  comparison}\label{App02}
 
\begin{figure*}
	\includegraphics[trim={-1.1cm 0mm -4cm 0cm},clip, width=01.03765099506\linewidth]{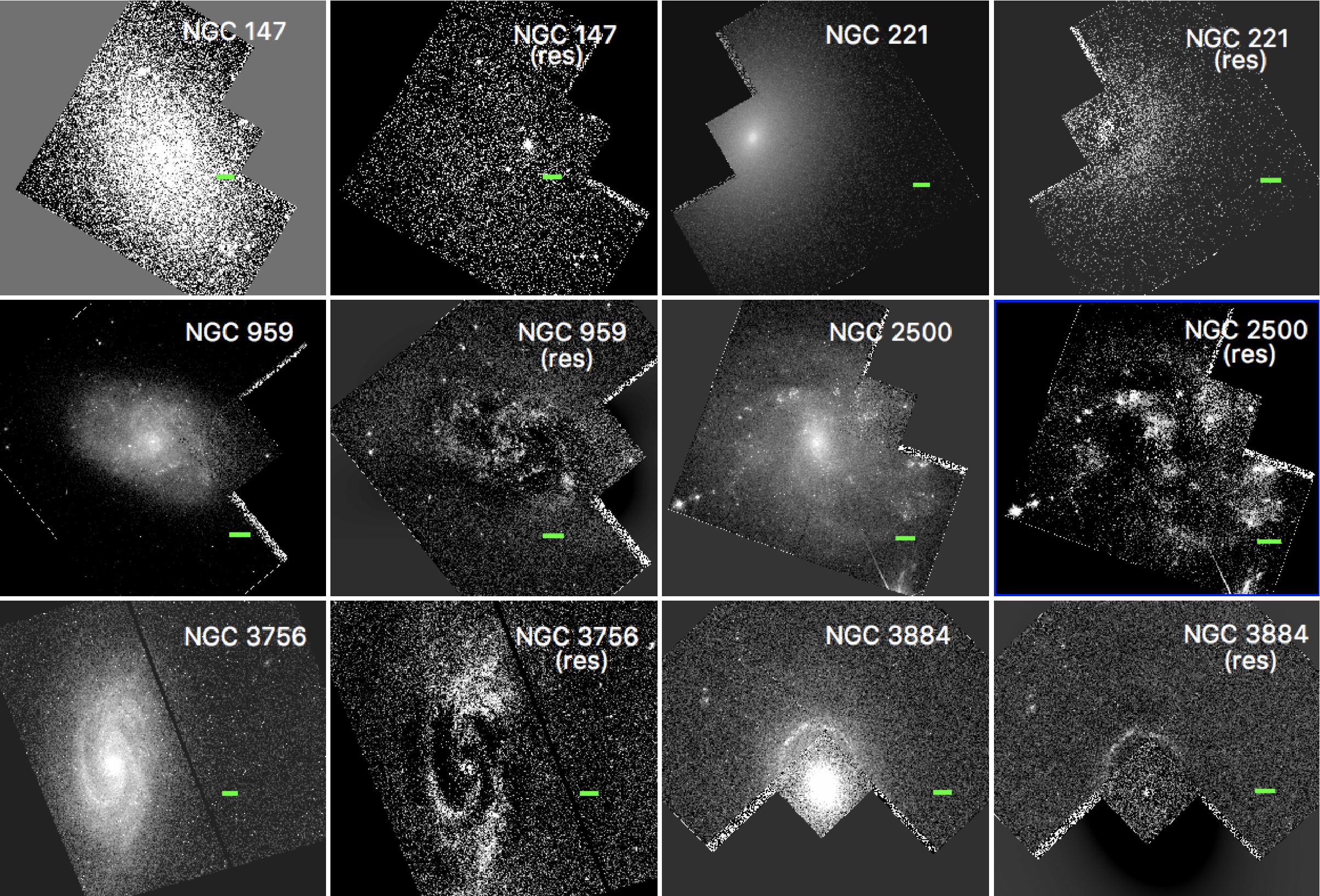}\\
	\includegraphics[trim={-1.1cm 0mm -4cm 0cm},clip, width=01.035099506\linewidth]{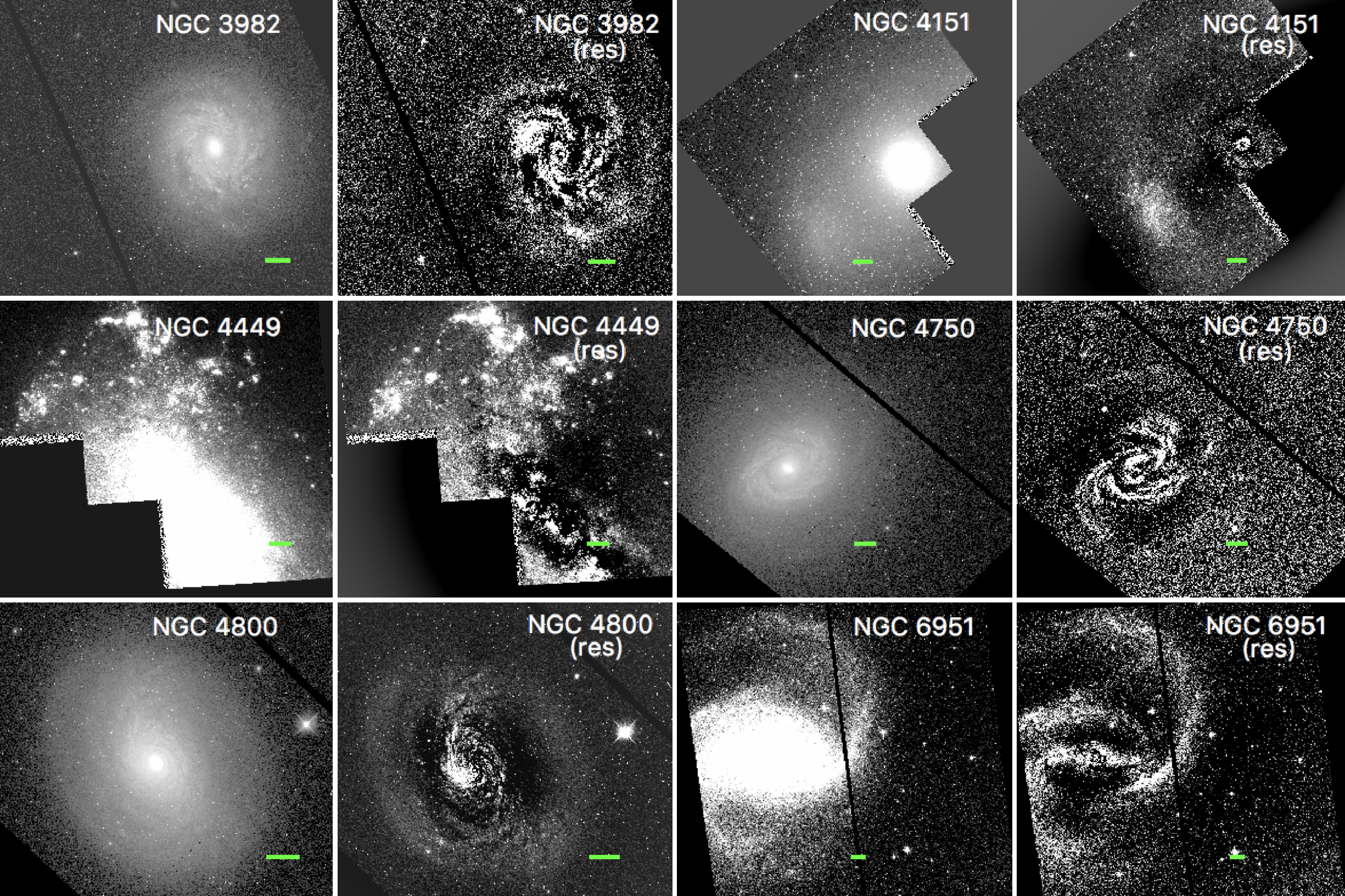}
\vspace{-0.328114cm}
\caption{ {\it HST} images and the corresponding residual images from
  our 2D multi-component decompositions with {\sc imfit} for the
  sample of 12 LeMMINGs galaxies shown in Fig.~\ref{Fig1}.  The
    {\it HST} instruments and filters used for imaging the galaxies
    are as in Fig.~\ref{Fig1} (see Section~\ref{Sec2.03}).  The 2D
  fits have the same type and number of galaxy structural components
  as the corresponding 1D modelling (Fig.~\ref{Fig1}).  We find good
  agreement between the 1D and 2D fits for the sample galaxies.  The
  green scale bar is 10 arcsec in length. }
 \label{2Dfits}
\end{figure*}

 To identify the 100 nuclei in this work, we have decomposed the full
 extent of the {\it HST} stellar light distributions of the 173 (23 E,
 42 S0, 102 S and 6 Irr) galaxies (Table~\ref{Tab1}), and, when
 necessary, fitted galaxy components beyond our multi-component,
 bulge+disc+nucleus model (Figs.~\ref{Fig1}, \ref{2Dfits} and see also
 \citealt{2023A&A...675A.105D}). Table~\ref{Tab3} presents previous
 studies of nucleation in nearby \mbox{($D \la 100$ Mpc)} galaxies,
 the pertaining data, the adopted galaxy fitting method and sample
 size. Given the common presence of structural components such as
 bars, discs and rings in nearby galaxies, employing a multi-component
 galaxy decomposition that extends beyond the traditional bulge+disk
 profiles is essential. It enables us to properly investigate the
 scaling relations between nuclei and host bulge properties and SMBH
 masses.

 Previous studies which performed multi-component decompositions for
 photometric and structural analysis of nuclei include
 \citet{2003ApJ...582L..79B}, \citet{2007ApJ...665.1084B},
 \citet{2012ApJS..203....5T}, \citet{2017ApJ...849...55S},
 \citet{2022AA...664A.167S} and \citet{2023A&A...671L...7R}.  However,
 our study of nuclei in 35 early-type and 65 late-type galaxies with
 {\it HST} is based on the most detailed decompositions for the
 largest sample of galaxies to date (see Table~\ref{Tab3}).

 A common approach for describing the stellar light distribution in
 nucleated galaxies is to limit the fit to a two-component model that
 describes the nucleus and the underlying galaxy
 \citep[e.g.][]{2020ApJ...900...32P,2021MNRAS.508..986Z,2021MNRAS.506.5494P,2023MNRAS.520.4664H}.
 Several studies modelled nucleated dwarf galaxies using a PSF (i.e.\
 Gaussian) nucleus + S\'ersic bulge model
 \citep{2003AJ....125.2936G,2005MNRAS.363.1019G,2014MNRAS.445.2385D,2021MNRAS.508..986Z,2021MNRAS.506.5494P}.
 \citet{1997AJ....114.2366C} and \citet{1998AJ....116...68C} fitted
 the nuclei in spiral galaxies with a PSF model. To separate the
 nuclei, these latter studies described underlying galaxy light with
 Gaussian wings and also attempted modelling the galaxy light in 2D,
 which was then subtracted from the galaxy
 images. \citet{2001AJ....122..653R} performed PSF nucleus + Nuker
 bulge model fits to the {\it HST} images of their early-type
 galaxies. \citet{2001AJ....121.1385S} and \citet{2005AJ....129.2138L}
 excluded the nuclei and performed Nuker model
 \citep{1995AJ....110.2622L} fits to the host galaxy {\it HST}
 profiles.  \citet{2001AJ....121.1385S} also fitted $R^{1/4}$,
 exponential, and S\'ersic models to the bugle profiles.
 \citet{2002AJ....123.1389B} fitted the light profiles of their
 late-type spiral galaxies with a Nuker (nucleus + bulge) model.  We
 note that \citet{1998AJ....116...68C}, \citet{2001AJ....122..653R},
 \citet{2001AJ....121.1385S}, \citet{2002AJ....123.1389B},
 \citet{2002AJ....123..159C} and \citet{2005AJ....129.2138L} used
 inner 10 arcsec {\it HST} images or light profiles to model the
 stellar light distributions of their galaxies.
 \citet{2006ApJS..165...57C} fitted King models to the nuclei and
 S\'ersic or core-S\'ersic models to the bulges of their early-type
 galaxies.

 Recently, \citet{2014MNRAS.441.3570G} and \citet{2016MNRAS.457.2122G}
 identified the nuclei in their spiral galaxy sample through visual
 inspection of the galaxies' {\it HST} images. To derive effective
 radii and luminosities for the nuclei, they used the {\sc ishape}
 task in {\sc baolab} \citep{1999A&AS..139..393L} and fitted King
 profiles \citep{1962AJ.....67..471K,1966AJ.....71...64K} to the
 nuclei, restricting the fitted radial range typical to 0.5 arcsec.
 \citet{2019ApJ...878...18S} and \citet{2020ApJ...890..128F} modelled
 the ground-based data of their nucleated galaxies using a S\'ersic
 nucleus + S\'ersic bulge model.
 \citet{2021MNRAS.507.3246H,2023MNRAS.520.4664H} fitted a S\'ersic
 model to the nuclei and a flat background model to the underlying
 light distributions of their galaxies.

\section{Data Tables}\label{AppA1}

Table~\ref{TabA1} presents central and global properties for our
sample of 100 nucleated galaxies, which includes distance,
morphological classification, plus luminosities, stellar masses and
effective radii of the nuclei.  Additionally, the table provides bulge
and galaxy magnitudes and stellar masses, as well as radio and X-ray
core luminosities.  Table~\ref{TabA2} is similar to Table~\ref{TabA1}
but presents properties for our sample of 24 nucleated LeMMINGs
galaxies excluded from our analysis. Of the 24 excluded nuclei, 13
nuclei, with extended half-light radii of $R_{\rm e,nuc} \ga 80$ pc,
were too large for a NSC (+AGN). In contrast, 9/24 nuclei have sizes
$R_{\rm e,nuc} < 1$ pc based on our structural decomposition, whereas
the remaining 2/24 nuclei have low stellar masses of
$M_{*,\rm nuc} < 10^{5}~ \rm M_{\sun}$.

\begin{center}
\begin{table*} 
\caption{Global and central properties for our sample of 100 nucleated galaxies.}
\label{TabA1}
\setlength{\tabcolsep}{0.015in}
\begin {minipage}{178mm}
\begin{tabular}{@{}llcccccccccccccccccccccc@{}}
\hline
\hline
Galaxy&Type& $D$&Class&$M_{\rm V}$&$M_{\rm V}$&log$M_{\rm *}$&log$M_{\rm *}$&$M_{\rm V,nuc}$&log$M_{\rm *,nuc}$&$R_{\rm e,nuc}$  &$R_{\rm e,nuc}$&$\mu_{\rm 0,nuc}(\mu_{\rm e,nuc})$&log$L_{\rm R,core}$ & Det &log$L_\textnormal{X}   $&Det&Type\\
&&&&(bulge)&(gal)&(bulge)&(gal)&&&&(1D/2D)&(1D/2D)&&&\\
&&(Mpc)&&(mag)&(mag)&($\rm M_{\sun}$)&($\rm M_{\sun}$)&(mag)&($\rm M_{\sun}$)&(arcsec)&(pc)&&(erg s$^{-1}$)&(radio)&(erg s$^{-1}$)&(X-ray)&(nuclei)\\
(1)&(2)&(3)&(4)&(5)&(6)&(7)&(8)&(9)&(10)&(11)&(12)&(13)&(14)&(15)&(16)&(17)&(18)\\   
\hline 
I0239	&	SAB	    &	9.9	&	H{\sc{ii}}	&	--16.82	&	--18.80	&	9.1	&	9.8	&	--9.60	&	6.2	&	0.09	&	4.4/4.7	&	16.34/16.04 	&	<35.0	&	N	&	<38.2	&	N	&	NSC	\\
I0356	&	SA	    &	11.7	&	H{\sc{ii}}	&	--16.63	&	--19.03	&	10.6	&	11.5	&	--9.10	&	7.6	&	0.02	&	0.8/1.3	&	11.73/11.77	 &	35.5		&	Y	&	---		&	---	&	NSC	\\
I0520	&	SAB	    &	50.9	&	L		&	--20.74	&	--22.62	&	10.7	&	11.5	&	--15.48	&	8.6	&	0.10	&	23.5/27.8	&	15.31/15.09	&	<36.1	&	N	&	---		&	---	&	HYB		\\
N0147	&	E5pec  &	0.8$^{\rm k}$ &ALG &	--15.76	&	--15.76	&	9.1	&	9.1	&	--6.04	&	5.2	&	1.43	&	5.7/5.2	&	(20.04/19.64)	&	<32.4	&	N	&	---		&	---	&	NSC	\\
N0205	&	E5pec  &	0.8$^{\rm k}$&	ALG	&	--18.22	&	--18.22	&	9.5	&	9.5	&	--9.80	&	6.2	&	0.36	&	1.5/1.3	&	(15.87/15.70)	&	<32.4	&	N	&	<35.8	&	N	&	NSC	\\
N0221	&	E2	    &	0.8$^{\rm k}$&ALG	&	--15.52	&	--16.48	&	8.6	&	9.0	&	--12.17	&	7.3	&	0.9	&	3.6/4.6	&	(13.10/13.13) 	&	<32.3	&	N	&	<35.5	&	N	&	NSC	\\
N0266	&	SB	    &	62.9	&	L		&	--22.93	&	--23.36	&	11.8	&	12.0	&	--14.36	&	8.4	&	0.08	&	24.5/21.9  &	16.07/15.54	&	36.9		&	Y	&	40.9		&	Y	&	HYB		\\
N0278	&	SAB	    &	5.3	&	H		&	--16.04	&	--17.65	&	8.8	&	9.5	&	--10.82	&	6.7	&	0.07	&	1.7/1.9  	&	12.66/12.42 	&	34.8		&	Y	&	38.1		&	Y	&	NSC	\\
N0404	&	E	    &	3.1$^{\rm k}$	&L	&	--17.39	&	--17.40	&	9.3	&	9.3	&	--12.06	&	7.2	&	0.30	&	4.2/5.0	&	13.53/13.38	&	<33.3	&	N	&	37.3		&	Y	&	HYB		\\
N0959	&	Sdm?   &	5.4	&	H{\sc{ii}}	&	--14.68	&	--16.12	&	8.1	&	8.7	&	--10.38	&	6.4	&	0.16	&	4.0/3.8	&	15.59/15.08	&	<34.7	&	N	&	<38.0	&	N	&	NSC	\\
N1058	&	SA	    &	4.5	&	L		&	--13.36	&	--16.34	&	7.4	&	8.6	&	--12.53	&	7.0	&	0.87	&	19.0/14.0  &	(14.85/14.67) 	&	<34.4	&	N	&	38.4		&	Y	&	HYB		\\
N1161	&	S0	    &	25.6	&	L/RL		&	--22.08	&	--22.25	&	11.4	&	11.5	&	--17.55	&	9.6	&	0.57	&	69.3/63.4	&	(17.21/16.93)	&	36.6		&	Y	&	40.0		&	Y	&	HYB		\\
N1275	&	cDpec  &	73.9	&	L/RL		&	--23.50	&	--23.77	&	11.5	&	11.6	&	--17.77	&	9.2	&	0.08	&	27.7/31.8	&	12.86/12.78	&	41.0		&	Y	&	42.4		&	Y	&	HYB		\\
N2273	&	SB	    &	26.9	&	S		&	--20.74	&	--21.27	&	10.7	&	10.9	&	--15.37	&	8.6	&	0.26	&	33.3	/38.0	&	15.39/15.67	&	36.6		&	Y	&	40.9		&	Y	&	HYB		\\
N2342	&	Spec	&	78.6	&	H{\sc{ii}}	&	--19.78	&	--21.52	&	10.0	&	10.7	&	--16.54	&	8.7	&	0.13	&	46.9/78.0	&	15.59/15.34	&	36.5		&	Y	&	---		&	---	&	NSC	\\
N2500	&	SB	&	9.0	&	H{\sc{ii}}	&	--14.39	&	--18.09	&	7.8	&	9.3	&	--11.21	&	6.6	&	0.09	&	3.9/3.6	&	(15.84/15.53)	&	<34.7	&	N	&	38.4		&	Y	&	NSC	\\
N2549	&	SA0 &16.6&ALG			&	--21.07	&	--21.74	&	10.9	&	11.2	&	--14.70	&	8.4	&	0.10	&	7.7/8.7	&	13.00/13.93	&	<35.3	&	N	&	39.5		&	Y	&	NSC	\\
N2634	&	E1?	&	33.0	&	ALG		&	--19.68	&	--20.53	&	10.2	&	10.5	&	--14.23	&	8.0	&	0.10	&	15.9/14.8	&	14.31/14.35	&	35.8		&	Y	&	<39.6	&	N	&	NSC	\\
N2655	&	SAB0	&	20.3	&	L/RL	&	--21.08	&	--21.72	&	11.0	&	11.2	&	--17.28	&	9.4	&	0.60	&	60.7/89.0  &	15.18/15.18	&	37.6		&	Y	&	---		&	---	&	HYB		\\
N2681	&	SAB0	&	12.1	& L		&	--19.0	&	--20.22	&	9.9	&	10.3	&	--15.2	&	8.3	&	0.19 &	11.0/	9.2    &      11.74/11.54	&	35.5		&	Y	&	39.2		&	Y	&	HYB		\\
N2685	&	SB0	&14.4&L				&	--18.48	&	--19.88	&	10.1	&	10.6	&	--12.71	&	7.8	&	0.10	&	7.3/13.0	&	13.96/13.73	&	<34.8	&	N	&	---		&	---	&	HYB		\\
N2748	&	SAbc	&	21.6	&H{\sc{ii}}	&	--18.47	&	--21.18	&	9.7	&	10.8	&	--10.36	&	6.5	&	0.02	&	2.1/2.9	&	8.71/8.72          &	<35.4	&	N	&	38.4		&	Y	&	NSC	\\
N2768	&	E6?	&	21.6	&	L/RL		&	--21.09	&	--21.27	&	11.4	&	11.5	&	--13.84	&	8.5	&	0.32	&	31.9/29.8	&	14.37/14.41	&	37.1		&	Y	&	39.8		&	Y	&	HYB		\\
N2782	&	SAB	&	39.7	&	H{\sc{ii}}	&	--20.38	&	--21.51	&	10.5	&	10.9	&	--15.05	&	8.3	&	0.15	&	27.7/40.2	&	14.62/14.51	&	36.8		&	Y	&	40.8		&	Y	&	NSC	\\
N2787	&	SB0	&11.0&L				&	--18.08	&	--19.34	&	10.3	&	10.8	&	--12.62	&	8.1	&	0.09	&	5.0/7.4	&	12.99/12.78	&	36.3		&	Y	&	39.4		&	Y	&	HYB		\\
N2859	&	SB0	&27.8&L				&	--20.15	&	--20.88	&	11.1	&	11.4	&	--11.89	&	7.8	&	0.02	&	1.0/3.0	&	9.58/9.54		&	<35.3	&	N	&	---		&	---	&	HYB		\\
N2903	&	SAB	&	12.1	&	H{\sc{ii}}	&	--15.04	&	--22.03	&	8.2	&	11.0	&	--11.98	&	7.0	&	0.05	&	3.1/4.9	&	13.65/13.39 	&	<34.3	&	N	&	38.7		&	Y	&	NSC	\\
N2950	&	SB0 &20.9&ALG			&	--19.57	&	--20.46	&	10.8	&	11.1	&	--15.01	&	8.9	&	0.27	&	26.5/23.6	&	13.91/13.58	&	<35.5	&	N	&	39.3		&	Y	&	NSC	\\
N2985	&	SA	&	19.8	& L			&	--20.70	&	--21.13	&	10.8	&	11.0	&	--13.20	&	7.8	&	0.05	&	4.9/5.6	&	12.75/12.68	&	35.8		&	Y	&	39.8		&	Y	&	HYB		\\
N3073	&	SAB0&19.1& H{\sc{ii}}		&	--18.14	&	--18.70 	&	9.7	&	9.9	&	--9.73	&	6.3	&	0.04	&	3.9/1.0	&	8.13/8.14		&	<35.3	&	N	&	<38.4	&	N	&	NSC	\\
N3077\textsuperscript{$\dagger$}	& I0pec	&	1.4 		& H{\sc{ii}}	&	---	&	--16.24	&	---	&	9.0	&	--8.66	&	6.0		&	0.36			&	2.5/2.7	&	15.13/15.33	&	33.3		&	Y	&	37.6		&	Y	&	NSC	\\
N3079	&	SB&	18.3	&	L			&	--20.34	&	--22.33	&	10.2	&	11.0	&	--14.38	&	7.8	&	0.02	&	1.2/7.2	&	(18.82/18.05)	&	37.3		&	Y	&	40.1		&	Y	&	HYB		\\
N3185	&	SB&	22.0	&	S			&	--19.16	&	--19.94	&	10.3	&	10.6	&	--13.98	&	8.2	&	0.20	&	20.5/23.1	&	15.33/14.92	&	<35.2	&	N	&	39.9		&	Y	&	HYB		\\
N3190	&	SA	&	21.8	&	L		&	--19.42	&	--21.79	&	10.8	&	11.7	&	--8.82		&	7.5	&	0.27	&	29.3/40.2	&	(18.69/19.34)	&	<35.3	&	N	&	40.3		&	Y	&HYB		\\
N3193	&	E2	&	24.3	&	L		&	--21.79	&	--21.93	&	11.4	&	11.4	&	--14.40	&	8.4	&	0.54	&	63.5/61.3	&	17.59/17.47	&	<35.2	&	N	&	39.6		&	Y	&	HYB		\\
N3198	&	SB	&	12.6	&	H{\sc{ii}}	&	--20.74	&	--21.51	&	10.5	&	10.8	&	--12.98	&	7.4	&	0.21	&	12.5/20.9	&	12.55/12.26	&	35.0		&	Y	&	38.7		&	Y	&	NSC	\\
N3245	&	SA0 &	22.3&H{\sc{ii}}		&	--20.74	&	--21.39	&	10.6	&	10.8	&	--13.50	&	7.7	&	0.04	&	4.3/6.8 	&	12.20/12.23	&	35.7		&	Y	&	39.7		&	Y	&	NSC	\\
N3319	&	SB	&	14.1	&	L	        &	--19.44	&	--20.56	&	9.5	&	9.9	&	--12.85	&	6.9	&	0.13	&	8.9/8.3 	&	15.83/15.77	&	<34.8	&	N	&	39.3		&	Y	&	HYB		\\
N3344	&	SAB	&	12.3	&	H{\sc{ii}}	&	--17.83	&	--20.82	&	9.3	&	10.5	&	--13.19	&	7.4	&	0.18	&	11.3/12.7	&	14.35/14.09	&	<34.1	&	N	&	38.6		&	Y	&	NSC	\\
N3414	&	S0	&	24.4	&	L		&	--20.59	&	--20.77	&	11.3	&	11.4	&	--13.28	&	8.4	&	0.23	&	28.0/21.8 &	15.56/15.09	&	36.2		&	Y	&	40.6		&	Y	&	HYB		\\
N3486	&	SAB	&	14.1	&	S		&	--18.58	&	--20.27	&	9.3	&	10.0	&	--13.75	&	7.4	&	0.15	&	9.9/13.5	&	13.75/13.84	&	<34.3	&	N	&	<37.5	&	N	&	HYB		\\
N3504	&	SAB &	26.2	&	H{\sc{ii}}	&	--20.43	&	--20.90 	&	10.6	&	10.8	&	--15.36	&	8.6	&	0.26	&	32.1/46.9 &	12.66/12.53	&	37.3		&	Y	&	---		&	---	&	NSC	\\
N3516	&	SB0	&37.5&S				&	--21.01	&	--22.36	&	11.1	&	11.6	&	--16.80	&	9.4	&	0.19	&	35.7/40.4	&	12.94/12.78	&	36.8		&	Y	&	42.5		&	Y	&	HYB		\\
N3610	&	E5?	&	25.6	&	ALG		&	--20.79	&	--20.88	&	10.6	&	10.6	&	--15.70	&	8.5	&	0.33	&	42.5/41.3	&	(13.29/13.28)	&	<35.6	&	N	&	39.6		&	Y	&	NSC	\\
\hline
Error   &---&---&---&   0.30&0.33& 0.15 dex&    0.17 dex    &0.35 & 0.24 dex& 15\%& 15\%&0.32& 1 dex&---&1 dex& ---&--- \\
\hline
\end{tabular} 
Note: (1) galaxy name. Bulgeless galaxies are indicated by
the superscript `$\dagger$'. (2) morphological classification from RC3
\citep{1991rc3..book.....D}.  (3) distance ($D$) are primarily from
the NASA/IPAC Extragalactic Database (NED;
\url{http://nedwww.ipac.caltech.edu}), other source is
\citet[][k]{2004AJ....127.2031K}.  (4) optical spectral class from
\citet{2021MNRAS.500.4749B}: \mbox{H {\sc{ii}}}, L = LINER, S =
Seyfert and ALG = Absorption Line Galaxy. Radio-loud galaxies (RL) are
based on classifications by \citet{2023MNRAS.522.3412D}.
(5)--(6) $V$-band bulge and galaxy absolute magnitudes calculated by
integrating the best-fitting  S\'ersic or core-S\'ersic functions. The
magnitudes are given in the Vega magnitude system.  (7)--(8) logarithm
of the stellar masses of the bulge and galaxy.  (9) $V$-band absolute
magnitude of the nucleus calculated by integrating the best-fitting 
Gaussian or S\'ersic function.  (10) logarithm of the stellar mass of
the nucleus.  (11) angular 1D major-axis effective (half-light) radius of the
nucleus in arcseconds. (12) 1D and 2D, major-axis effective radii of the
nucleus in pc. (13) central and effective  surface brightnesses of the nuclei, 
$\mu_{\rm 0,nuc}$ and $\mu_{\rm e,nuc}$, respectively. $\mu_{\rm 0,nuc}$
 and $\mu_{\rm e,nuc}$ are in units of mag
arcsec$^{-2}$ and values given inside a  parenthesis `()' are $\mu_{\rm e,nuc}$.
The {\it HST} filters in which $\mu_{\rm 0,nuc}$ and $\mu_{\rm e,nuc}$
are obtained are given in \citet{2023A&A...675A.105D}.  (14) logarithm
of the \emerlin\ 1.5 GHz radio core luminosity ($L_{\rm R,core}$).
(15) radio detection of the galaxies based on
\citet{2021MNRAS.500.4749B}: `Y = Yes' = detected; `N = No' =
undetected. (16) logarithm of the (0.3$-$10 keV) X-ray core luminosity
($L_\textnormal{X}$).  (17) X-ray detection of the galaxies are from
\citet{2022MNRAS.510.4909W}: `Y = Yes' = detected; `N = No' =
undetected.   (18) our classification of nuclei based on
multi-wavelength data: NSCs and hybrid nuclei (HYB).

\end {minipage}
\end{table*}
\end{center}

\setcounter{table}{1}
\begin{center}
\begin{table*} 
\setlength{\tabcolsep}{0.0172240304059833in}
\begin {minipage}{178mm}
\caption{(\it continued)}
\begin{tabular}{@{}llcccccccccccccccccccccc@{}}
\hline
\hline
Galaxy&Type& $D$&Class&$M_{\rm V}$&$M_{\rm V}$&log$M_{\rm *}$&log$M_{\rm *}$&$M_{\rm V,nuc}$&log$M_{\rm *,nuc}$&$R_{\rm e,nuc}$  &$R_{\rm e,nuc}$&$\mu_{\rm 0,nuc}(\mu_{\rm e,nuc})$&log$L_{\rm R,core}$ & Det &log$L_\textnormal{X} $&Det&Type\\
&&&&(bulge)&(gal)&(bulge)&(gal)&&&&(1D/2D)&(1D/2D)&&&\\
&&(Mpc)&&(mag)&(mag)&($\rm M_{\sun}$)&($\rm M_{\sun}$)&(mag)&($\rm M_{\sun}$)&(arcsec)&(pc)&&(erg s$^{-1}$)&(radio)&(erg s$^{-1}$)&(X-ray)&(nuclei)\\
(1)&(2)&(3)&(4)&(5)&(6)&(7)&(8)&(9)&(10)&(11)&(12)&(13)&(14)&(15)&(16)&(17)&(18)\\   
\hline   
N3631	&	SA	&	19.2	&	H{\sc{ii}}	&	--20.13	&	--20.33	&	10.2	&	10.2	&	--13.24	&	7.4	&	0.24	&	22.1/35.2		&	16.45/15.97	&	<35.2	&	N	&	39.5		&	Y	&	NSC	\\
N3718	&	SB	&	16.9	&	L/RL		&	--19.62	&	--20.15	&	10.5	&	10.7	&	--13.01	&	7.9	&	0.09	&	7.3/8.5		&	13.37/13.27	&	36.8		&	Y	&	41.3		&	Y	&	HYB	\\
N3729	&	SB	&	17.8	&	H{\sc{ii}}	&	--16.22	&	--18.86	&	8.9	&	9.9	&	--12.89	&	7.6	&	0.25	&	21.5/9.1		&	16.12/16.08	&	35.8		&	Y	&	39.8		&	Y	&	NSC	\\
N3756	&	SAB	&	21.4	&	H{\sc{ii}}	&	--17.81	&	--20.57	&	9.3	&	10.4	&	--12.79	&	7.2	&	0.06	&	5.5/5.3  		&	14.20/14.02	&	<35.3	&	N	&	<38.5	&	N	&	NSC	\\
N3884	&	SA0	&	107.0	&	L	&	--21.91	&	--23.12	&	11.6	&	12.0	&	--16.74	&	9.5	&	0.14	&	71.0/87.5		&	16.38/16.31	&	37.8		&	Y	&	42.1		&	Y	&	HYB	\\
N3898	&	SA	&	19.2	&	L		&	--20.25	&	--20.59	&	11.0	&	11.1	&	--14.71	&	8.8	&	0.38	&	34.4/52.8		&	14.46/14.77	&	35.8		&	Y	&	39.4		&	Y	&	HYB	\\
N3949	&	SA	&	14.5	&	H{\sc{ii}}	&	--19.92	&	--20.43	&	9.6	&	9.8	&	--13.49	&	7.1	&	0.08	&	5.5/7.7		&	14.32/14.23	&	<35.0	&	N	&	<39.1	&	N	&	NSC	\\
N3982	&	SAB	&	18.3	&	S		&	--19.39	&	--20.77	&	8.7	&	9.2	&	--14.55	&	6.7	&	0.06	&	5.6/6.5 		&	13.62/13.42	&	36.2		&	Y	&	39.2		&	Y	&	HYB	\\
N3992	&	SB	&	17.6	&	L		&	--18.50	&	--21.12	&	9.8	&	10.9	&	--10.87	&	6.8	&	0.02	&	1.8/6.5		&	14.60/14.30	&	<35.1	&	N	&	39.0		&	Y	&	HYB		\\
N3998	&	SA0	&	17.4	&	L/RL		&	--19.27	&	--20.04	&	10.9	&	11.2	&	--14.16	&	8.8	&	0.12	&	9.7/11.6		&	(12.67/12.74)	&	38.0		&	Y	&	42.0		&	Y	&	HYB		\\
N4026	&	S0	&	16.9	&	L		&	--22.38	&	--23.12	&	11.3	&	11.6	&	--15.80		&	8.7	&	0.03	&	2.5/3.2		&	10.56/10.53	&	<35.1	&	N	&	38.9		&	Y	&HYB		\\
N4036	&	S0	&	21.7	&	L		&	--19.23	&	--21.45	&	10.5	&	11.4	&	--15.23	&	8.9	&	0.39	&	40.6/31.5		&	(15.87/16.14)	&	36.0		&	Y	&	40.4		&	Y	&	HYB		\\
N4041	&	SA	&	19.5	&	H{\sc{ii}}	&	--19.32	&	--20.14	&	9.9	&	10.2	&	--13.54	&	7.6	&	0.15	&	13.6/14.1		&	(15.42/15.53)	&	35.5		&	Y	&	39.4		&	Y	&	NSC	\\
N4062	&	SA	&	14.9	&	H{\sc{ii}}	&	--15.35	&	--20.31	&	8.4	&	10.4	&	--13.48	&	7.6	&	0.7	&	50.5/55.1		&	18.54/18.55	&	<34.6	&	N	&	<38.0	&	N	&	NSC	\\
N4125	&	E6pec	&	21.0	&	L	&	--21.14	&	--21.14	&	11.3	&	11.3	&	--14.39	&	8.6	&	0.68	&	68.5/78.1		&	15.63/15.63	&	<35.4	&	N	&	39.5		&	Y	&	HYB		\\
N4143	&	SAB0	&	16.9	&	L	&	--20.45	&	--21.22	&	10.9	&	11.2	&	--13.82	&	8.3	&	0.08	&	6.7/5.3		&	14.32/14.33	&	36.1		&	Y	&	40.3		&	Y	&	HYB		\\
N4150	&	SA0	&	7.1	&	L		&	--15.80	&	--17.76	&	9.0	&	9.8	&	--13.31	&	8.1	&	0.52	&	17.7/30.0		&	14.88/14.93	&	<34.6	&	N	&	<37.8	&	N	&	HYB		\\
N4151	&	SAB	&	17.8	&	S		&	--19.22	&	--20.85	&	10.2	&	10.8	&	--17.62	&	9.5	&	0.1	&	8.8/12.4		&	9.43/	9.46		&	37.8		&	Y	&	42.4		&	Y	&	HYB		\\
N4203	&	SAB0	&	19.5	&	L	&	--19.59	&	--20.44	&	10.9	&	11.2	&	--12.75	&	8.2	&	0.05	&	4.3/5.9		&	11.82/11.78	&	36.1		&	Y	&	40.5		&	Y	&	HYB		\\
N4220	&	SA0	&	16.0	&	L		&	--18.26	&	--19.96	&	10.1	&	10.7	&	--13.08	&	8.0	&	0.83	&	63.5/65.1		&	(18.03/17.93)	&	35.2		&	Y	&	39.2		&	Y	&	HYB		\\
N4245	&	SB0	&	16.7	&	H{\sc{ii}}	&	--19.28	&	--20.21	&	10.2	&	10.6	&	--12.81	&	7.7	&	0.18	&	14.6/12.0		&	(17.08/16.30)	&	<34.6	&	N	&	<38.0	&	N	&	NSC	\\
N4258	&	SAB	&	9.4	&	S		&	--20.03	&	--21.72	&	10.8	&	11.5	&	--14.00		&	8.4	&	0.15	&	6.7/6.2		&	(14.15/14.08)	&	34.9		&	Y	&	40.7		&	Y	&HYB		\\
N4274	&	SB	&	17.4	&	L		&	--17.93	&	--20.10	&	9.6	&	10.4	&	--12.06	&	7.2	&	0.29	&	24.7/44.1		&	(18.34/18.49)	&	<34.6	&	N	&	<40.1	&	N	&	HYB		\\
N4278	&	E1-2	&	15.6	&	L/RL		&	--20.91	&	--20.91	&	11.0	&	11.0	&	--11.57	&	7.2	&	0.08	&	5.0/4.8		&      13.32/13.00	&	37.6		&	Y	&	39.7		&	Y	&	HYB		\\
N4314	&	SB	&	17.8	&	L		&	--19.74	&	--20.69	&	10.8	&	11.1	&	--13.20	&	8.1	&	0.22	&	18.7/25.7		&	14.73/14.81	&	<34.7	&	N	&	38.4		&	Y	&	HYB		\\
N4414	&	SA	&	14.2	&	L		&	--19.14	&	--20.11	&	10.4	&	10.8	&	--10.98	&	7.1	&	0.06	&	3.9/6.3		&	13.47/13.58	&	<34.7	&	N	&	38.7		&	Y	&	HYB		\\
N4449	&	IBm	&	6.1	&	H{\sc{ii}}	&	--16.61	&	--20.10	&	8.4	&	9.8	&	--13.21	&	7.0	&	0.24	&	6.7/8.3		&	14.46/14.43	&	<33.6	&	N	&	36.9		&	Y	&	NSC	\\
N4559	&	SAB	&	15.6	&	H{\sc{ii}}	&	--16.16	&	--19.64	&	8.1	&	9.5	&	--14.25	&	7.3	&	0.1	&	6.3/7.0		&	13.87/13.85	&	<34.5	&	N	&	39.0		&	Y	&	NSC	\\
N4565	&	SA	&	21.8	&	S		&	--21.70	&	--23.62	&	11.2	&	11.9	&	--9.86		&	6.4	&	0.05	&	4.8/3.9		&	16.81/16.85	&	35.2		&	Y	&	39.7		&	Y	&HYB		\\
N4648	&	E3	&	21.0	&	ALG		&	--18.66	&	--19.58	&	10.0	&	10.4	&	--11.04	&	7.0	&	0.04	&	4.0/5.5		&	13.13/12.90	&	<35.5	&	N	&	39.0		&	Y	&	NSC	\\
N4656\textsuperscript{$\dagger$}	&	SBm	&	13.0	&	H{\sc{ii}}	&	---		&	--21.48	&	---	&	9.8	&	--10.88	&	5.6	&	0.54	&	34.0/42.3		&	(21.44/21.67)	&	<34.3	&	N	&	---		&	---	&	NSC	\\
N4736	&	SA	&	7.6	&	L		&	--19.78	&	--21.35	&	10.5	&	11.1	&	--15.33	&	8.7	&	0.1	&	3.8/16.8		&	12.57/12.59	&	34.8		&	Y	&	39.5		&	Y	&	HYB		\\
N4750	&	SA	&	24.1	&	L		&	--19.57	&	--20.89	&	10.2	&	10.8	&	--14.03	&	8.0	&	0.14	&	15.9/15.3		&	14.23/14.16	&	35.8		&	Y	&	40.4		&	Y	&	HYB		\\
N4800	&	SA	&	15.6	&	H{\sc{ii}}	&	--19.24	&	--19.41	&	10.4	&	10.4	&	--12.37	&	7.6	&	0.08	&	5.9/6.6		&	13.53/13.54	&	<34.9	&	N	&	38.8		&	Y	&	NSC	\\
N4826	&	SA	&	10.0	&	L		&	--19.63	&	--21.45	&	10.5	&	11.2	&	--14.96	&	8.6	&	0.36	&	17.2/9.6		&	13.35/13.38	&	33.9		&	Y	&	38.1		&	Y	&	HYB		\\
N5005	&	SAB	&	16.8	&	L		&	--21.30	&	--21.47	&	11.0	&	11.1	&	--14.22	&	8.2	&	0.12	&	9.8/6.3		&	13.20/13.58	&	36.3		&	Y	&	39.2		&	Y	&	HYB		\\
N5033	&	SA	&	15.8	&	L		&	--20.00	&	--22.31	&	9.9	&	10.8	&	--14.31	&	7.6	&	0.13	&	9.4/9.9		&	13.79/13.67	&	36.0		&	Y	&	38.6		&	Y	&	HYB		\\
N5055	&	SA	&	9.9	&	L		&	$-$17.70	&	--21.56	&	9.5	&	11.1	&	--15.8		&	8.8	&	0.61	&	29.3/29.1		&	13.73/13.77	&	<34.4	&	N	&	39.5		&	Y	&HYB		\\
N5112\textsuperscript{$\dagger$}	&	SB	&	17.0	&	H{\sc{ii}}	&	---		&	--19.51	&	---	&	9.5	&	--11.15	&	6.2	&	0.23	&	19.0/19.6		&	17.99/18.11	&	<35.4	&	N	&	---		&	---	&	NSC	\\
N5377	&	SB	&	28.0	&	L		&	$-$20.99	&	--21.78	&	11.2	&	11.5	&	--16.48	&	9.4	&	0.18	&	24.7/29.3		&	13.11/12.88	&	35.7		&	Y	&	39.6		&	Y	&	HYB		\\
N5457	&	SAB	&	5.2	&	H{\sc{ii}}	&	$-$15.72	&	--17.42	&	8.3	&	9.0	&	--9.96		&	6.0	&	0.21	&	5.1/4.9		&	16.59/16.19	&	<34.3	&	N	&	39.7		&	Y	&	NSC	\\
N5474	&	SA	&	5.7	&	H{\sc{ii}}	&	$-$16.04	&	--17.59	&	8.4	&	9.0	&	--8.95		&	5.5	&	0.25	&	6.7/8.0		&	18.06/18.04	&	<34.4	&	N	&	38.6		&	Y	&	NSC	\\
N5548	&	SA0	&	77.6	&	S		&	$-$21.45	&	--22.31	&	10.8	&	11.1	&	--19.06	&	9.8	&	0.07	&	24.1/34.1		&	12.02/12.02	&	37.0		&	Y	&	43.2		&	Y	&	HYB		\\
N5585	&	SAB	&	5.6	&	H{\sc{ii}}	&	$-$14.11	&	--17.20	&	7.4	&	8.6	&	--10.96	&	6.1	&	0.15	&	3.9/3.6		&	(16.73/16.18)	&	<34.5	&	N	&	<36.2	&	N	&	NSC	\\
N5879	&	SA	&	12.0	&	L		&	$-$18.20	&	--20.29	&	9.4	&	10.2	&	--11.54	&	6.7	&	0.06	&	3.7/2.8		&	14.87/15.33	&	35.3		&	Y	&	38.5		&	Y	&	HYB		\\
N5985	&	SAB	&	36.7	&	L		&	$-$21.22	&	--22.70	&	10.8	&	11.4	&	--14.49	&	8.1	&	0.11	&	19.5/18.9		&	16.10/16.05	&	35.8		&	Y	&	41.5		&	Y	&	HYB		\\
N6207	&	SA	&	12.3	&	H{\sc{ii}}	&	$-$17.36	&	--20.20	&	8.8	&	9.9	&	--9.86		&	5.8	&	0.06	&	3.8/3.6		&	16.77/16.70	&	<35.2	&	N	&	---		&	---	&	NSC	\\
N6217	&	SB	&	19.1	&	H{\sc{ii}}	&	$-$18.40	&	--19.85	&	9.5	&	10.1	&	--15.37	&	8.3	&	0.11	&	9.9/6.9		&	12.29/12.52	&	35.7		&	Y	&	---		&	---	&	NSC	\\
N6340	&	SA0	&	16.6	&	L		&	$-$19.78	&	--20.20	&	10.7	&	10.9	&	--14.02	&	8.4	&	0.38	&	30.5/25.3		&	15.38/14.89	&	35.5		&	Y	&	---		&	---	&	HYB		\\
N6412	&	SA	&	18.1	&	H{\sc{ii}}	&	$-$17.82	&	--19.35	&	9.0	&	9.6	&	--11.25	&	6.4	&	0.08	&	6.0/5.8		&	15.57/14.77	&	<35.4	&	N	&	---		&	---	&	NSC	\\
N6503	&	SA	&	4.9	&	L		&	$-$13.75	&	--19.20	&	7.7	&	9.9	&	--10.56	&	6.4	&	0.04	&	1.0/0.6		&	12.76/12.83	&	<34.3	&	N	&	38.1		&	Y	&	HYB		\\
N6946	&	SAB	&	5.0	&	H{\sc{ii}}	&	$-$16.38	&	--18.79	&	8.3	&	9.3	&	--14.64	&	7.6	&	0.91	&	21.8/29.8		&	(16.46/16.88)	&	34.4		&	Y	&	<39.2	&	N	&	NSC	\\
N6951	&	SAB	&	18.2	&	L		&	$-$20.25	&	--21.51	&	10.3	&	10.8	&	--14.34	&	7.9	&	0.15	&	13.3/12.9		&	14.01/14.07	&	35.4		&	Y	&	---		&	---	&	HYB		\\
N7217	&	SA	&	9.0	&	L		&	$-$19.55	&	--19.86	&	10.6	&	10.7	&	--14.15	&	8.5	&	0.77	&	33.2/32.5		&	(16.47/16.37)	&	35.1		&	Y	&	---		&	---	&	HYB		\\
N7331	&	SA	&	7.0	&	L		&	$-$17.62	&	--19.29	&	10.3	&	10.9	&	--12.30		&	8.1	&	0.27	&	9.18/11.1		&	(14.47/14.33)	&	<35.0	&	N	&	39.8		&	Y	&HYB		\\
N7457	&	SA0	&	7.2	&	ALG		&	$-$16.84	&	--18.11	&	9.6	&	10.1	&	--11.72	&	7.5	&	0.05	&	1.75/2.2		&	11.85/11.74	&	<34.9	&	N	&	<38.3	&	N	&	NSC	\\
\hline
Error   &---&---&---&   0.30&0.33& 0.15 dex&    0.17 dex    &0.35 & 0.24 dex& 15\%& 15\%&0.32& 1 dex&---&1 dex& ---&--- \\
\hline
\end{tabular} 
\end {minipage}
\end{table*}
\end{center}

\renewcommand{\thetable}{A\arabic{table}} 
\begin{center}
\begin{table*} 
\caption{Excluded  nuclei.}
\label{TabA2}
\setlength{\tabcolsep}{0.04553in}
\begin {minipage}{178mm}
\begin{tabular}{@{}llcccccccccccccccccccccc@{}}
\hline
\hline
Galaxy&Type& $D$&Class&$M_{\rm V}$&$M_{\rm V}$&log$M_{\rm *}$&log$M_{\rm *}$&$M_{\rm V,nuc}$&log$M_{\rm *,nuc}$&$R_{\rm e,nuc}$  &$R_{\rm e,nuc}$&log$L_{\rm R,core}$ & Det &log$L_\textnormal{X}$&Det\\
&&&&(bulge)&(gal)&(bulge)&(gal)&&&&&&&&\\
&&(Mpc)&&(mag)&(mag)&($\rm M_{\sun}$)&($\rm M_{\sun}$)&(mag)&($\rm M_{\sun}$)&(arcsec)&(pc)&(erg s$^{-1}$)&(radio)&(erg s$^{-1}$)&(X-ray)\\
(1)&(2)&(3)&(4)&(5)&(6)&(7)&(8)&(9)&(10)&(11)&(12)&(13)&(14)&(15)&(16)\\   
\hline 
I2574\textsuperscript{$\dagger$}	& 	SAB &     2.1&H{\sc{ii}}& 	   ---             &    --19.87            &           ---     & 	9.5   		       &	--4.83                         &	3.5        	&	0.242	    	&	2.4   		        &      <33.7        &	N  	&	< 36.55	&	N \\
N0841	&	SAB		    &	   62.2	& 	   L       		&	 --21.34	   &	  --22.23	     & 10.3		&      10.6			&	--20.17   			&	9.8	  	&	1.498		& 	438.9		&	<36.2	&	N  	&	---		&	---\\
N1023	&	SB0		    &	   6.2		&	   ALG	    	&	 --19.36	   &	  --19.94	     &	10.9		& 	11.1			&	--10.69			&	7.4		&	<0.1		&	--			&	<34.6	&	N  	&	39.05	&	Y\\
N1961	&	SAB(rs	    &	   56.4	&         L	    		&   	 --22.96	   & 	  --23.50	     &	11.3		& 	11.5			&	--16.93			&	8.9  		&      0.300		&	79.9		        &	37.2		&	Y  	&	40.43	&	Y \\
N2403	&	SAB(s)cd	    &	   2.6		& 	   H{\sc{ii}}		&	 --19.58  	   &	  --20.45	     &	9.5		&  	9.8			&	--16.19			&	8.1	 	&	21.443		&	278.8		&	<34.0	&	N  	&	38.58	&	Y  \\
N2541	&	SA(s)cd	    &	   9.8	   	&	   H{\sc{ii}}	         &  	 --12.92 	   &	  --19.15	     & 6.9		& 	9.4			&	--8.21			&	5.0		&	<0.1		&	--			&	<34.7	&	N  	&	38.10	&	Y\\
N2639	&	(R)SA           &    50.3  	&	   L			&	 --20.74 	   &	  --21.97        &	11.2		&  	11.7			&	--16.80			&	9.6		&	0.427		&	101.6		&	37.6		&	Y  	&	41.05	&	Y\\
N2683	&	SA(rs)	    &	   9.1	  	&	   L			&	 --19.64 	   &	  --21.51        &	10.5		&  	11.3			&	--16.07			&	9.1		&	2.503		&	110.2		&	34.5		&	Y  	&	39.39	&	Y\\
N2770	&	SA(s)c?	    &	   31.4 	&	   H{\sc{ii}}		&	 --18.62	   &	  --22.46	     &	9.2		&  	10.7			&	--14.24			&	7.4		&	0.710		&	106.5		&	<35.5	&	N  	&	<38.34	&	N\\
N2964	&	SAB(r)	    &	   22.9 	&	   H{\sc{ii}}		&	 --17.30	   &     --20.85        &	9.3		& 	10.7			&	--17.04			&	9.2		&	0.823		&	90.5		        &	36.0		&	Y 	&	--- 		&	---\\
N3031	&	SA(s)a	    &	   0.7		&	   L			&	 --16.62        &	  --17.63	     &	9.5		&  	 9.9			&	--7.92			&	6.0		&	0.142		&	0.4			&	35.5		&	Y 	&	39.21	&	Y\\
N3184	&	SAB(rs)cd	    &	   11.4  	& 	   H{\sc{ii}}		&	 --16.62        &	  --20.24	     &	8.7		&	10.2			&	--13.55			&	7.5		&	1.355		&	77.2		         &	<34.6	&	N 	&	38.47	&	Y\\
N3642	&	SA(r)		    &	   24.9	&	   L			&	 --18.86	   &	  --20.63	     &	9.4		&	10.1			&	--13.20			&	7.2		&	<0.1 		&	--			&	<35.6	&	N 	&	40.19	&	Y\\
N3838	&	SA0a?	    &	   21.0	& 	   ALG		&	 --18.69        &	  --22.11	     & 10.0		&	11.4			&	--13.01			&	7.7		&	<0.1 		&	--			&	35.4		&	Y 	&	39.29	&	Y\\
N3900	&	SA0		    &	   30.2	&	   ALG		&	 --20.83 	   &     --21.61	     &	10.7		&	11.1			&	--14.00			&	8.0		&	<0.1		&	--			&	<35.6	&	N 	&	---		&	---\\
N4102	&	SAB		    &	  14.7	&	   H{\sc{ii}}		&	 --18.30 	   &	  --18.77	     & 10.6		&	10.8			&	--16.33			&	9.8		&	1.443		&	102.5		&	35.9		&	Y 	&	40.83	&	Y\\
N4138	&	SA0		    &	  16.0     	&	   L			&	 --20.01 	   &	  --20.58        &	10.6		&      10.8			&	--14.93			&	8.5		&	1.101		&	84.8			&	<35.3	&	N 	&	41.21	&	Y\\
N4448	&	SB		    &	  13.5    	&	   H{\sc{ii}}		&	 --18.89 	   &	  --19.38	     & 10.6		&	10.8			&	--9.23			&	6.7		&	<0.1		&	--		&	<34.5	&	N 	&	<38.48	&	N\\
N5204	&	SA(s)	    &	  4.6	    	&	   H{\sc{ii}}		&	 --17.75    	   &     --18.31        &	8.7		&	 8.9			&	--6.93			&	4.3		&	0.236		&	5.2   			&	<34.1	&	N  	&	<36.18	&	N\\
N5273	&	SA0		    &	  18.6   	&	   S			&	 --18.86  	   &	  --20.24	     &	10.1		&	10.6			&	--13.73			&	8.0		&	<0.1 		&	--		&	35.3		&	Y  	&	40.96 	&	Y\\
N5308	&	S0		    &	  30.1   	&	   ALG		&	 --19.99 	   &	  --21.08	     &	10.8		&	11.3			&	--17.52			&	9.8		&	0.645		&	92.9			&	<35.8	&	N  	&	41.43	&	Y\\
N5354	&	S0ed		    &    39.8   	&	   L			&	 --20.52    	   &	  --21.34	     &	10.9		&	11.2			&	--15.82			&	9.0		&	0.603		&	114.0		&	36.9		&	Y   	&	40.10	&	Y\\
N5448	&	(R)SAB	    &	  31.1  	&	   L			&	 --19.77    	   &	  --23.39	     &	9.5		&	10.9			&	--16.53			&	8.2		&	0.556		&	82.8			&	35.7		&	Y   	&	39.36 	&	Y\\
N6654	&	(R')SB0	    &	  25.0  	&	   ALG		&	 --18.61  	   &	  --20.19	     &	10.5		&	11.1			&	--10.83			&	7.3		&	<0.1		& 	--			&	<35.6	&	N   	& 	---		&	---\\
\hline
Error   &---&---&---&   0.30&0.33& 0.15 dex&    0.17 dex    &0.35 & 0.24 dex& 15\%& 15\%&1 dex&---&1 dex&---& \\
\hline
\end{tabular} 
Note: Similar to  Table~\ref{TabA1}, but here for sample of nucleated LeMMINGs galaxies  excluded  from our analyses. Of the 24 excluded nuclei, 13 
nuclei were too large for a NSC (+AGN) with extended half-light radii of $R_{\rm e,nuc} \ga 80$
pc. In contrast,  9/24 nuclei have sizes  $R_{\rm e,nuc} < 1$ pc based on our decomposition, whereas the remaining 2/24 nuclei have low stellar masses of $M_{*,\rm nuc} < 10^{5}~ \rm M_{\sun}$. For  eight nuclei  unresolved in the {\it HST} images, we report upper-limit angular $R_{\rm e,nuc} $ values.
\end {minipage}
\end{table*}
\end{center}

\section{Additional relations for nuclei  }\label{AppA2}

Table~\ref{TabA3} presents radio and X-ray scaling relations for
nuclei derived by fitting censored linear regressions using {\sc
  asurv} to account for upper limits and to help illustrate the good
agreement between our censored ({\sc asurv}) and uncensored ({\sc
  bces} bisector) regression fits.  The radio and X-ray scaling
relations here are derived from the full sample of nuclei, without
making a distinction between NSCs and hybrid nuclei, whereas in
Section~\ref{Sec4}, NSCs and hybrid nuclei were fitted separately.

\setlength{\tabcolsep}{0.08in}
\begin{table*}
\begin {minipage}{178mm}
\caption{ Radio and X-ray scaling  relations for nuclei obtained using  censored {\sc asurv} and uncensored {\sc bces} bisector regressions.} 
\label{TabA3}
\begin{tabular}{@{}lllccccccccccc@{}}
\hline

 Relation &{\sc asurv} fit &$r_{\rm
                                              s}/P$-${\rm value}$&$r_{\rm
                                                            p}/P$-${\rm
                                                                   value}$&$\updelta_{\rm horiz}$&Sample&\\
  \hline
 
 \multicolumn{4}{c}{\bf  Censored analysis}\\
  \\[-9.608pt]   
$L_{\rm R,core}-M_{\rm *,nuc} $
        &$\mbox{$\log$}\left(\frac{L_{\rm R,core}}{\rm erg~ s^{-1}}\right)= (1.25 \pm
             0.22) \mbox{log}\left(\frac{M_{\rm
          *,nuc}}{\mbox{$5\times10^{7}$ $\rm M_{\sun}$}}\right)$
          +~($34.60  ~ \pm  0.88$)
           & ---&---&---&100 \\
  
$L_{\rm R,core}-M_{V,\rm nuc}$
        &$\mbox{log}\left(\frac{L_{\rm R,core}}{\rm erg ~s^{-1}}\right)= (-0.53   \pm
            0.17)\left( M_{V,\rm nuc}+13.2 \right)$ +~($34.66 ~ \pm  0.76$)
             & ---&---&---&100 \\
$L_\textnormal{X}-M_{\rm *,nuc} $
        &$\mbox{$\log$}\left(\frac{ L_\textnormal{X}   }{\rm erg~ s^{-1}}\right)= (1.09\pm
            0.16) \mbox{log}\left(\frac{M_{\rm *,nuc}}{\mbox{$4\times10^{7}$ $\rm M_{\sun}$}}\right)$ +~($ 39.09~ \pm  0.28$)
& ---&---&---&84 \\
$L_\textnormal{X}-M_{V,\rm nuc}$
        &$\mbox{log}\left(\frac{L_\textnormal{X}        }{\rm erg ~s^{-1}}\right)= (-0.48   \pm
            0.10)\left( M_{V,\rm nuc}+13.2 \right)$ +~($39.22 ~ \pm  0.20$)
             &---&---&---&84 \\
               \\[-11.8838pt]

  \multicolumn{5}{c}{}\\
 \multicolumn{4}{c}{\bf  Uncensored analysis }\\
 Relation &{\sc bces} bisector fit &$r_{\rm
                                              s}/P$-${\rm value}$&$r_{\rm
                                                            p}/P$-${\rm
                                                                   value}$&$\updelta_{\rm horiz}$&Sample&\\
&&&&(dex)&                                                                 \\
  $L_{\rm R,core}-M_{\rm *,nuc} $
        &$\mbox{$\log$}\left(\frac{L_{\rm R,core}}{\rm erg~ s^{-1}}\right)= (  1.15 \pm
             0.22) \mbox{log}\left(\frac{M_{\rm *,nuc}}{\mbox{$5\times10^{7}$ $\rm~ M_{\sun}$}}\right)$ +~($ 35.40  ~ \pm  0.10$)
 &0.60/$6.7\times10^{-11}$&0.59/$ 1.5\times10^{-10}$&0.96&100 \\

$L_{\rm R,core}-M_{V,\rm nuc}$
        &$\mbox{log}\left(\frac{L_{\rm R,core}}{\rm erg ~s^{-1}}\right)= (-0.49   \pm
            0.07)\left( M_{V,\rm nuc}+13.2 \right)$ +~($35.41 ~ \pm  0.55$)
             &$-$0.55/$2.6\times10^{-9}$&$-$0.60/$ 1.2\times10^{-10}$&---&100 \\
  $L_\textnormal{X}-M_{\rm *,nuc} $
        &$\mbox{$\log$}\left(\frac{L_\textnormal{X} }{\rm erg~ s^{-1}}\right)= (1.36\pm
            0.16) \mbox{log}\left(\frac{M_{\rm *,nuc}}{\mbox{$4\times10^{7}$ $\rm~ M_{\sun}$}}\right)$ +~($ 39.16~ \pm  0.60$)
 &0.63/$1.4\times10^{-10}$&0.65/$ 1.6\times10^{-11}$&0.83&84 \\

$L_\textnormal{X}-M_{V,\rm nuc}$
        &$\mbox{log}\left(\frac{L_\textnormal{X}    }{\rm erg ~s^{-1}}\right)= (-0.66   \pm
            0.07)\left( M_{V,\rm nuc}+13.2 \right)$ +~($39.30 ~ \pm  0.28$)
             &$-$0.53/$2.06\times10^{-7}$&$-$0.58/$
                                           5.24\times10^{-9}$&---&84 \\                                                                   
       \\[-6.82080608pt]      
 \hline
\end{tabular} 
Note:  Radio and X-ray scaling  relations  are based on  the full sample of nuclei, without separating  NSCs and hybrid nuclei. 1.5 GHz radio core luminosity from \emerlin\
($L_{\rm R,core}$) and (0.3--10 keV) {\it Chandra } X-ray core
luminosity ($L_\textnormal{X}$) as a function of nucleus stellar
mass ($M_{\rm *,nuc}$) and nucleus absolute magnitude
($M_{\rm V,nuc}$). We performed the censored linear regressions
using {\sc asurv} to account for upper limits. We present our {\sc bces} bisector  regression fits to the
galaxy data, the Spearman and Pearson correlation coefficients
($r_{\rm s}$ and $r_{\rm p}$, respectively) and the corresponding
serendipitous correlation probabilities. The
horizontal rms scatter in the log $M_{\rm *,nuc}$ direction is denoted with $\updelta_{\rm horiz}$. 
\end{minipage}
\end{table*}

\label{lastpage}
\end{document}